\documentclass[11pt,letterpaper]{article}
\usepackage[T1]{fontenc}
\usepackage[utf8]{inputenc}
\usepackage{lmodern}
\usepackage[margin=1in]{geometry}
\usepackage{amsmath,amssymb,amsthm,mathtools}
\usepackage{booktabs,array,longtable,enumitem}
\usepackage{needspace}
\usepackage{tikz}
\usepackage{microtype,xurl,hyperref}
\hypersetup{hidelinks,pdftitle={Sharp Bounds for Truthful-in-Expectation Scheduling},pdfauthor={Yunpeng Li}}
\setlist{itemsep=2pt,topsep=4pt}
\newtheorem{theorem}{Theorem}[section]
\newtheorem{lemma}[theorem]{Lemma}
\newtheorem{proposition}[theorem]{Proposition}
\newtheorem{corollary}[theorem]{Corollary}
\theoremstyle{definition}

\theoremstyle{remark}

\newcommand{\E}{\mathop{\mathbb E}\nolimits}
\newcommand{\Prb}{\mathop{\mathbb P}\nolimits}
\newcommand{\OPT}{\operatorname{OPT}}
\newcommand{\ind}{\mathbf 1}
\newcommand{\tnorm}{\lVert\theta\rVert_2}
\DeclareMathOperator{\loadwidth}{width}
\newcommand{\R}{\mathbb R}

\numberwithin{equation}{section}
\begin{document}
\hypersetup{pageanchor=false}
\begin{titlepage}
\centering
\vspace*{0.55in}
{\LARGE\bfseries Sharp Bounds for Truthful-in-Expectation Scheduling\par}
\vspace{0.45in}
\renewcommand{\thefootnote}{\fnsymbol{footnote}}
{\large Yunpeng Li\footnote{The Chinese University of Hong Kong, Shenzhen.
Email: \href{mailto:liyunpeng@cuhk.edu.cn}{\texttt{liyunpeng@cuhk.edu.cn}}.}\par}
\vspace{0.55in}
\begin{abstract}
\normalsize
We study randomized mechanisms for scheduling indivisible jobs on $n$ unrelated machines. Each machine is controlled by a selfish agent who privately knows its processing time for each job. A mechanism assigns every job to a machine and makes payments. We require truthfulness in expectation: no agent can increase its expected payment minus its expected processing cost by misreporting, regardless of the other agents' reports. The makespan is the maximum total processing time assigned to a machine; the objective is to minimize its expectation. We prove that the optimal approximation ratio is $(n+1)/2$ for every $n\ge7$, relative to the minimum makespan of an integral assignment. The lower bound holds for every $n$ and allows unrestricted computation and arbitrary correlations among assignments, assuming finite expected payments and a guarantee uniform over all finite job counts. The matching mechanism uses correlated rounding and runs in expected polynomial bit time on rational inputs. These guarantees hold for all finite nonnegative processing times. For two machines, we characterize the optimal approximation ratio exactly as an analytic constant, approximately $1.505995705434$, and give a mechanism attaining it. For three machines the optimal ratio is $2$. For four, five, and six machines we obtain upper bounds of $3$, $105/32$, and $29/8$, respectively, compared with the general lower bound $(n+1)/2$. The proofs for $n\ge7$ do not rely on computer-assisted certificates; the two- and three-machine results use exact rational computer-assisted certificates.
\end{abstract}
\vfill
\end{titlepage}
\pagenumbering{arabic}
\hypersetup{pageanchor=true}
\section{Introduction}\label{sec:intro}
In scheduling on unrelated machines, each of $m$ indivisible jobs can
take a different amount of time on each of $n$ machines. Write $t_{ij}$
for the finite nonnegative processing time of job $j$ on machine $i$.
An assignment $A:[m]\to[n]$ places each job on one machine. A machine's
load is the sum of the processing times of its assigned jobs. The
makespan is the largest of these loads.
\[
 C_{\max}(t,A)=\max_{i\in[n]}\sum_{j:A(j)=i}t_{ij},
 \qquad \OPT(t)=\min_A C_{\max}(t,A).
\]
When the processing times are known, the algorithm of Lenstra, Shmoys,
and Tardos~\cite{LST90} finds a schedule of makespan at most $2\OPT(t)$
in polynomial time.

Nisan and Ronen~\cite{NR01} introduced the strategic version. Each
machine is controlled by a selfish agent, and its processing-time
vector $t_i=(t_{i1},\ldots,t_{im})$ is private information. A
direct-revelation mechanism asks each agent to report its vector, then
uses these reports to assign the jobs and determine payments.
An agent's utility is its payment minus the processing time
of its assigned jobs, evaluated at its true type. A truthful mechanism
makes reporting the entire vector honestly a dominant strategy,
regardless of the other reports. The question is how well such a
mechanism can approximate the minimum makespan.

The Vickrey--Clarke--Groves (VCG) mechanism truthfully minimizes
total processing cost by assigning each job to a fastest machine.
For $n\ge2$, suppose each of $n$ jobs takes one unit of time on machine $1$
and $1+\varepsilon$ elsewhere, where $0<\varepsilon<1$. VCG puts all jobs
on machine $1$, for makespan $n$. Any schedule using another machine
takes at least $1+\varepsilon$, and assigning one job to each machine
attains this bound. The ratio $n/(1+\varepsilon)$ approaches $n$, the
worst-case ratio of VCG~\cite{NR01}. Nisan and Ronen conjectured that
every deterministic truthful mechanism has worst-case ratio at least $n$.
Christodoulou, Koutsoupias, and
Kov\'acs~\cite{CKK23,CKK26} proved the conjecture even for mechanisms
with unrestricted computation.

Randomization permits two notions of truthfulness. A mechanism is
\emph{universally truthful} if it samples a deterministic truthful
mechanism from a distribution independent of the reports. It is
\emph{truthful in expectation} (TIE) if honest reporting maximizes
each agent's expected utility for every fixed report of the other
agents. Universal truthfulness implies TIE, but TIE only requires the
incentive inequalities after averaging over the mechanism's randomness.
We ask how small a TIE mechanism can make the ratio of its expected
makespan to the minimum makespan of an integral assignment.

For general $n$, Mu'alem and Schapira~\cite{MS18} proved the TIE lower
bound $2-1/n$, whereas Lu and Yu~\cite{LY08} gave a TIE mechanism with
ratio $(n+5)/2$. These bounds leave open whether the loss must grow
linearly with the number of machines.
Anastasiadis, Christodoulou, Koutsoupias, Kov\'acs, and
Schecker~\cite{ACKKS26} proved an $\Omega(n)$ lower bound for
universally truthful mechanisms with discrete support, and an upper
bound $n/2+o(n)$. Their lower bound does not apply to general TIE
mechanisms. We prove a lower bound of $(n+1)/2$ for unrestricted TIE
families and attain it for every $n\ge7$.

\subsection{Results}
Fix $n$ machines. A family specifies one mechanism for each finite job
count $m$. Its approximation ratio is at most $R$ if, for every $m$ and
every profile $t$, the assignment $A$ drawn on truthful reports satisfies
$\E[C_{\max}(t,A)]\le R\,\OPT(t)$. Let $\rho(n)$ be the infimum of these
bounds over truthful-in-expectation families. For $n\ge7$, we determine
$\rho(n)$ and give a mechanism that attains it.

\Needspace{5\baselineskip}
\begin{theorem}[Optimal ratio for at least seven machines]\label{thm:main}
The optimal approximation ratio of truthful-in-expectation scheduling
on $n\ge7$ unrelated machines is
\[
                         \rho(n)=\frac{n+1}{2}.
\]
\end{theorem}
The lower bound $\rho(n)\ge(n+1)/2$ holds for every $n$, even with
unrestricted computation and arbitrary correlations among job assignments.
We have $\rho(1)=1$, as the sole machine must process every job. The mechanism
in Section~\ref{overview:upper} attains the lower bound for $n\ge7$ and
gives the upper bounds below for four, five, and six machines.

\begin{theorem}[Smaller machine counts]\label{thm:small}
For two and three machines,
\[
 \rho(2)=R_*\approx1.505995705434,\qquad \rho(3)=2.
\]
Here $R_*$ is the two-machine optimum characterized analytically in
Section~\ref{sec:n2}, with an exact rational enclosure
in~\eqref{n2:decimal-bound}. For four, five, and six machines,
\[
 \frac52\le\rho(4)\le3,\qquad
 3\le\rho(5)\le\frac{105}{32},\qquad
 \frac72\le\rho(6)\le\frac{29}{8}.
\]
\end{theorem}
One load estimate gives the sharp bound for $n\ge8$ and the three
upper bounds for $n=4,5,6$. At $n=7$ the estimate gives $4+1/32$, which
Section~\ref{upper7:appendix-a.-seven-machines-a-plane-slack-certificate}
sharpens to four. The exact ratios for four, five, and six machines
remain open. The two-machine optimum and the three-machine upper bound
use the exact rational certificates in
Sections~\ref{n2:verification-subsection} and~\ref{n3:computational}.

\paragraph{Efficient mechanisms.}
For $n\ge4$ we implement the same mechanism that proves the upper
bound. For two and three machines, we approximate an auxiliary schedule.
For two machines we also replace the analytic optimum by a piecewise
rational rule.

\begin{theorem}[Efficient truthful mechanisms]\label{thm:efficient}
On nonnegative rational reports, the following approximation ratios
are attainable by truthful-in-expectation mechanisms, uniformly over
every finite number of jobs.
\begin{enumerate}[label=\textup{(\roman*)}]
\item For $n\ge4$, the upper bounds in Theorems~\ref{thm:main}
and~\ref{thm:small} are attainable in expected time polynomial in the
binary input length, including when $n$ is part of the input.
\item For $n\in\{2,3\}$ and any rational $0<\varepsilon<1$, the ratios are
\[
 1.505996014(1+\varepsilon)\quad(n=2),\qquad
 2(1+\varepsilon)\quad(n=3).
\]
The expected running time is polynomial in the combined binary encoding
length of the reports and $\varepsilon$, and in $1/\varepsilon$.
\end{enumerate}
\end{theorem}
Truthfulness and the makespan bounds hold for every finite nonnegative
real profile. On rational reports, the mechanisms sample assignments
and rational payments using unbiased bits. Each machine's payment
covers its reported cost in every outcome. The piecewise rational two-machine rule has worst-case ratio
strictly below $1.505996014$, the bound used in the theorem.
Section~\ref{sec:implementation} computes that ratio exactly and proves
the running-time bounds.

\subsection{Relation to prior work}
\paragraph{Deterministic and randomized truthfulness.}
The deterministic resolution of Christodoulou, Koutsoupias, and
Kov\'acs~\cite{CKK23,CKK26} and the discrete-support universal lower
bound in~\cite{ACKKS26} use the structure of
deterministic truthful rules. General TIE mechanisms need not be distributions
over such rules: their incentive constraints hold after averaging.
Our lower bound uses the assignment probabilities and expected payments,
as in the TIE setting of Mu'alem and Schapira~\cite{MS18}, while allowing
the joint assignment distribution to depend on the entire report and
to correlate different jobs arbitrarily.

\paragraph{Task independence and fractional scheduling.}
Allowing each job $j$ to be split into shares $q_{ij}\ge0$ with
$\sum_i q_{ij}=1$ gives the fractional optimum
$\OPT_{\mathrm{frac}}(t)=\min_q\max_i\sum_j t_{ij}q_{ij}\le\OPT(t)$.
Christodoulou, Koutsoupias, and Kov\'acs introduced the truthful
inverse-square rule Square at ICALP 2007~\cite{CKK07}. It achieves
ratio $(n+1)/2$ against this fractional optimum. In a \emph{task-independent}
fractional rule, each job's shares depend only on its own column of reports.
Their matching lower bound holds for every such rule, even without
truthfulness (see the 2010 journal version~\cite[Theorems~4.2, 4.4 and~5.2]{CKK10}).
Lu and Yu~\cite{LY08} proved the same lower bound for randomized
integral TIE rules that assign each job using only its column and make
independent random choices across jobs. On the lower-bound instances of
Christodoulou, Koutsoupias, and Kov\'acs, $\OPT_{\mathrm{frac}}(t)=\OPT(t)=1$.
Their lower bound on a machine's mean load therefore applies to integral
assignments with column-local probabilities even when the assignments
are correlated. Theorem~\ref{thm:locality} derives this locality almost
everywhere for a limiting family with the original approximation guarantee.

\paragraph{From fractional loads to integral schedules.}
Lu and Yu~\cite{LY08} round their Bounded-Square fractional assignment
using the theorem of Kumar, Marathe, Parthasarathy, and Srinivasan~\cite{KMPS05},
obtaining the TIE factor $(n+5)/2$. We preserve the same assignment
probabilities using the slot matching construction of Shmoys and
Tardos~\cite{ST93}. These probabilities fix each machine's expected load.
The matching controls which jobs it can receive together and hence the
range of its load. Section~\ref{overview:upper} combines these ranges with
the mean loads to bound the expected maximum.

\paragraph{Correlated assignments on two machines.}
On two machines, a job can be assigned by comparing the ratio of its
processing times with a random threshold. Lu~\cite{Lu09} drew the
thresholds jointly for two jobs, and Chen, Du, and Zuluaga~\cite{CDZ15}
used copulas to correlate thresholds for any number of jobs.
For each fixed job count, Kuryatnikova and Vera~\cite[Corollary~3 and
Theorem~7]{KV20} characterized the best ratio in this class of universally
truthful, scale-free mechanisms by a minimax formula over joint threshold
distributions. For two jobs their lower and upper bounds are
$1.5059953$ and $1.5059964$, respectively. Their parameter counts jobs.
Here $\rho(2)$ concerns two machines and arbitrarily many jobs. We
identify the exact constant $R_*$ by extending a two-job upper bound
to every job count with an auxiliary assignment, and by reducing an
unrestricted family to the same two-job minimax problem for the lower
bound (Section~\ref{sec:n2}).

\subsection{Organization}
Section~\ref{sec:model} defines the scheduling model.
Section~\ref{sec:overview} derives the lower bound from a perturbed star,
the upper bound from correlated rounding, and the reductions used for
two and three machines. Sections~\ref{sec:lower} and~\ref{sec:generalupper}
prove the locality and load estimates. Sections~\ref{sec:n2} and~\ref{sec:n3}
give the exact results for two and three machines.
Section~\ref{sec:implementation} gives the algorithms on rational
reports. Appendix~\ref{sec:supplement} describes how to reproduce the
finite certificate checks.
\label{page:introduction-end}

\Needspace{9\baselineskip}
\section{Preliminaries}\label{sec:model}
\paragraph{Processing times and assignments.}
Write $[k]=\{1,\ldots,k\}$, with $[0]=\varnothing$. There are $n$ machines and $m$ jobs.
A type profile is a matrix $t=(t_{ij})\in[0,\infty)^{n\times m}$.
Its row $t_i$ is machine $i$'s private type, and $t_{ij}$ is its true
processing time for job $j$. We consider direct-revelation mechanisms.
Write $b_i\in[0,\infty)^m$ for the vector of processing times that machine
$i$ reports and $b=(b_1,\ldots,b_n)$ for the bid profile.
Machine $i$ bids truthfully when $b_i=t_i$.
Write $t_{\cdot j}$ for column $j$, $t_{-i}$ for all other rows, and
$t_{-ij}$ for all entries except $t_{ij}$. Use the same conventions for $b$.
For $J\subseteq[m]$, $t_J$ consists of the columns indexed by $J$ in increasing order.
The assignment set is
$\mathcal A_{m,n}=\{A:[m]\to[n]\}$. Machine $i$ has load
$L_i(t,A)=\sum_{j:A(j)=i}t_{ij}$. The makespan of $A$ and the minimum makespan
over all integral assignments are $C_{\max}(t,A)$ and $\OPT(t)$,
respectively, where
\[
 C_{\max}(t,A)=\max_{i\in[n]}L_i(t,A),\qquad
 \OPT(t)=\min_{A\in\mathcal A_{m,n}}C_{\max}(t,A).
\]
At fixed types $t$ and bids $b$, the bids determine the distribution
of $A$. We write $L_i$ and $C_{\max}$ for the resulting random load and
makespan, evaluated at the true times $t$.

\paragraph{Allocation rule, payment scheme, and truthfulness.}
At bids $b$, the allocation rule draws an assignment from
$y^{m,n}(b)=(y_A^{m,n}(b))_{A\in\mathcal A_{m,n}}$, where
$y_A^{m,n}(b)$ is the probability of the complete assignment $A$.
Write $p_i^{m,n}(b)\in\mathbb R$ for machine $i$'s finite expected net
payment, counting transfers to $i$ positively and charges negatively
as in~\cite[Section~2.1]{CKK26}. The marginal probability of assigning
job $j$ to machine $i$ is
\[
 q_{ij}^{m,n}(b)
 =\Prb_{A\sim y^{m,n}(b)}\bigl(A(j)=i\bigr)
 =\sum_{\substack{A\in\mathcal A_{m,n}\\A(j)=i}}y_A^{m,n}(b).
\]
For fixed $n$, we write $y^m,p_i^m,q_{ij}^m$ and
$q_i^m=(q_{ij}^m)_{j\in[m]}$. We also omit $m$ when it is fixed.
We call a joint assignment distribution with prescribed $q_{ij}$ a
\emph{coupling}. In $q_{ij}(s,b_{-ij})$, only entry $b_{ij}$ is replaced
by $s$. At bids $b=(b_i,b_{-i})$, a machine of type $t_i$ has expected
processing cost $t_i\cdot q_i^m(b)=\sum_j t_{ij}q_{ij}^m(b)$ and expected
utility $p_i^m(b)-t_i\cdot q_i^m(b)$. Dominant-strategy truthfulness
in expectation requires
\begin{equation}\label{eq:tie}
 p_i^m(t_i,b_{-i})-t_i\cdot q_i^m(t_i,b_{-i})
 \ge p_i^m(b_i,b_{-i})-t_i\cdot q_i^m(b_i,b_{-i})
\end{equation}
for every machine $i$, true type $t_i$, alternative bid $b_i$, and
profile $b_{-i}$ of the other machines' bids. A deviation may change
the entire row. Only marginals and expected payments enter~\eqref{eq:tie},
so paying $p_i^m(b)$ deterministically also preserves truthfulness.
Expected makespan depends on the joint assignment law.

Truthfulness alone permits either sign for $p_i^m(b)$. Adding a finite
function of $b_{-i}$ to $p_i^m$ adds the same amount to both sides
of~\eqref{eq:tie}. For the mechanisms in Theorem~\ref{thm:efficient}, $\widehat\pi_{ij}$ is the realized
payment to machine $i$ for job $j$. The total $\widehat P_i=\sum_j\widehat\pi_{ij}$
covers the reported cost of the jobs assigned to $i$ in every outcome and is
therefore nonnegative (Theorem~\ref{thm:efficient}). The per-job expectations are
$\pi_{ij}=\E[\widehat\pi_{ij}]$, so $p_i=\E[\widehat P_i]=\sum_j\pi_{ij}$.

The lower bound allows arbitrary correlations among job assignments and
imposes no continuity, differentiability, joint measurability in the
reports, or computational restriction.

\paragraph{Uniform approximation over job counts.}
We evaluate approximation at truthful bids $b=t$. Below we write $t$
as the input to a rule when a single profile is under consideration.
For fixed $n$, a family consists of one mechanism for each finite $m\ge0$. It is
$R$-approximate, or has approximation ratio at most $R$, if
\begin{equation}\label{eq:ratio}
 \sum_{A\in\mathcal A_{m,n}}y_A^m(t)C_{\max}(t,A)\le R\OPT(t)
\end{equation}
for every $m$ and every profile $t$. When $\OPT(t)=0$, this requires
zero makespan with probability one. The infimum of such finite bounds
over truthful-in-expectation families at fixed $n$ is $\rho(n)$.
For $n\ge2$, assigning each job to a cheapest machine and paying the
winning machine the second-lowest report for that job gives ratio at
most $n$. Its makespan is at most its total processing cost, which is
at most $n\OPT(t)$. For $n=1$, the forced allocation is optimal and
zero payments are truthful. We henceforth take $n\ge2$.

\paragraph{The comparison schedule.}
At profile $t$, let $A$ be the mechanism's random assignment and fix
an optimal assignment $A^*$ with makespan $T=\OPT(t)$. Thus $A^*(j)$
is the machine that processes job $j$ in this schedule, and
$L_i(t,A^*)=\sum_{j:A^*(j)=i}t_{ij}\le T$ for every machine $i$.

The two- and three-machine mechanisms also use an auxiliary assignment
$A^{\mathrm{aux}}$. The machine $A^{\mathrm{aux}}(j)$ is called the
\emph{owner} of job $j$. Its identity can affect which jobs are assigned
together. Write $T_{\mathrm{aux}}=C_{\max}(t,A^{\mathrm{aux}})$ for the
makespan of this schedule. The exact mechanisms choose an optimal
auxiliary assignment from the reported profile. At truthful bids this
is an optimal comparison schedule $A^*$. Section~\ref{sec:implementation}
allows an approximate auxiliary schedule.
In Section~\ref{sec:n2}, $s,t$ denote scalar cost ratios
when they are arguments of the two-job objective. We write $C_f$ for
the optimum expected makespan over couplings with the prescribed
marginals and $R(f)$ for the resulting worst-case ratio.
\label{page:preliminaries-end}

\section{The main argument}\label{sec:overview}
For the lower bound, we show that the root of a star must receive a
large expected load. For the matching upper bound, we control the
expected maximum load by correlating the assignments of different jobs.

\subsection{The unrestricted lower bound}\label{overview:lower}
Consider the star in Figure~\ref{fig:perturbed-star}, with machines as
vertices and jobs as edges~\cite[Section~3]{CKK26}. Job~$1$ is a loop
at the root, machine~$1$. For $2\le j\le n$, job~$j$ is the spoke joining
the root to leaf~$j$. Each job takes one unit of time at an endpoint and
a large finite time elsewhere.

Putting the loop on the root and every spoke on its leaf gives makespan
one, which is optimal. This assignment uses the whole star to tell the
root from a leaf. In a single spoke's column the two endpoints are
equally cheap, so a rule that treats machine labels symmetrically gives
them equal probabilities. As the cost on the other machines grows, a
fixed approximation guarantee forces the probability of using them to zero.
Thus the root's probability approaches one half for each isolated spoke
and one for the isolated loop.

If these probabilities carry over to the full star, the root's expected
load approaches $1+(n-1)/2$. The makespan is at least the root's load
in every outcome. Moreover, the mean load at a fixed profile is
$\sum_j t_{1j}q_{1j}(t)$ regardless of how assignments are correlated.
This same calculation underlies the fractional lower bound of
Christodoulou, Koutsoupias, and Kov\'acs~\cite[Theorem~5.2]{CKK10}.
For an unrestricted mechanism, we must justify carrying the one-job
probabilities over to the star. The next theorem gives a limiting
family in which we can do so. Section~\ref{sec:lower} constructs that
family. Here we apply its one-column probabilities to the star.

\Needspace{9\baselineskip}
\begin{theorem}[Locality of a limiting family]\label{thm:locality}
Fix $n\ge2$ and a truthful-in-expectation family with finite uniform
approximation ratio $R$. There exist measurable functions
$g_i:(0,\infty)^n\to[0,1]$ and, for each $m\ge1$, a measurable assignment
distribution $y^{m,*}$ with marginals $q_{ij}^{m,*}$, defined for almost
every $t\in(0,\infty)^{nm}$, such that for all $i\in[n]$ and $j\in[m]$,
\[
 q_{ij}^{m,*}(t)=g_i(t_{\cdot j}),\qquad
 \sum_A y_A^{m,*}(t)C_{\max}(t,A)\le R\OPT(t).
\]
The same one-column functions $g_i$ apply to every job $j$ at every job count $m$.
The assignment laws and their marginals are equivariant under job and machine
permutations up to null sets.
\end{theorem}

At almost every positive $n$-job profile, \emph{column locality} gives
\[
 q_{1j}^{n,*}(t)=g_1(t_{\cdot j})=q_{11}^{1,*}(t_{\cdot j}).
\]
On the left, all $n$ jobs are present. On the right, column $t_{\cdot j}$
is presented alone and therefore has job index~$1$. The other columns
can affect which jobs the root receives together, but its probability
of receiving this particular job is already fixed by $g_1$. This is
enough to compute the root's expected load. Since the limiting family
still satisfies $\E[C_{\max}]\le R\OPT(t)$, a lower bound on the ratio
$R$ applies to the original family.

\Needspace{6\baselineskip}
\paragraph{Applying locality to the star.}\label{lower:full-dimensional-perturbed-star}

\begin{theorem}[Unrestricted lower bound]\label{thm:lower}
For every $n\ge1$, any truthful-in-expectation family with a finite approximation ratio $R$ uniform over all finite job counts satisfies $R\ge(n+1)/2$.
\end{theorem}
\begin{proof}
For $n=1$, every complete assignment has makespan $\OPT$, so a positive
instance forces $R\ge1$. Assume henceforth that $n\ge2$.

Use the limiting family of Theorem~\ref{thm:locality}. The theorem holds
Lebesgue-almost everywhere, and the unit-cost star could be exceptional.
We therefore perturb the processing times while keeping the same two
cheap endpoints on each spoke.

\Needspace{7\baselineskip}
\paragraph{The profile distribution.}
Fix $\tau>0$ and choose $\overline{t}>1+\tau$ as the left endpoint of the
high-cost interval. There are $m=n$ jobs. Job~$1$ is the loop at machine~$1$,
and job~$j$ joins machines $1$ and $j$ for $2\le j\le n$. The low- and high-cost laws are
$\underline{\nu}=\operatorname{Unif}[1,1+\tau]$ and
$\overline{\nu}=\operatorname{Unif}[\overline{t},\overline{t}+\tau]$. Draw all $n^2$ entries independently,
with the probability distribution of each entry $t_{ij}$ given by
\[
\operatorname{Law}(t_{ij})=
\begin{cases}
\underline{\nu},&i=1\text{ or }i=j,\\
\overline{\nu},&\text{otherwise}.
\end{cases}
\]
Each column and the full profile have Lebesgue densities, so the theorem
applies with probability one at both job counts. We first draw the true
times $t$ and then draw the assignment at the truthful bids $b=t$.
Assigning job $j$ to machine $j$ gives
\[
 \OPT(t)\le\max_{j\in[n]}t_{jj}\le1+\tau.
\]
Figure~\ref{fig:perturbed-star} illustrates the construction.

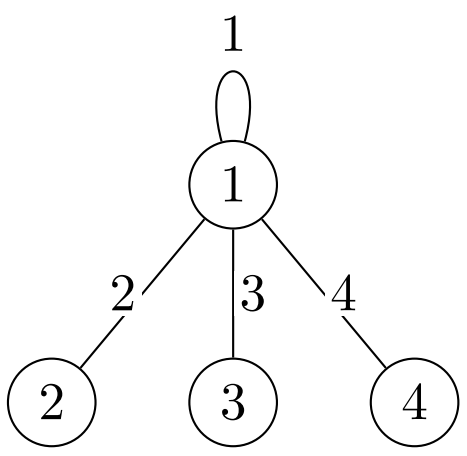
\begin{figure}[ht]
\centering
\begin{minipage}[c]{0.42\textwidth}
\centering
\begin{tikzpicture}[scale=0.8,every node/.style={font=\small}]
\node[draw,circle,inner sep=3pt] (r) at (0,1) {$1$};
\node[draw,circle,inner sep=3pt] (a) at (-1.5,-0.8) {$2$};
\node[draw,circle,inner sep=3pt] (b) at (0,-0.8) {$3$};
\node[draw,circle,inner sep=3pt] (c) at (1.5,-0.8) {$4$};
\draw (r) edge[loop above,min distance=8mm] node[above] {$1$} (r);
\draw (r)--node[midway,left,fill=white,inner sep=1pt] {$2$}(a);
\draw (r)--node[midway,right,fill=white,inner sep=1pt] {$3$}(b);
\draw (r)--node[midway,right,fill=white,inner sep=1pt] {$4$}(c);
\end{tikzpicture}
\end{minipage}\hfill
\begin{minipage}[c]{0.53\textwidth}
\[
 \bigl(\operatorname{Law}(t_{ij})\bigr)_{i,j=1}^4
 =\begin{pmatrix}
 \underline{\nu}&\underline{\nu}&\underline{\nu}&\underline{\nu}\\ \overline{\nu}&\underline{\nu}&\overline{\nu}&\overline{\nu}\\ \overline{\nu}&\overline{\nu}&\underline{\nu}&\overline{\nu}\\ \overline{\nu}&\overline{\nu}&\overline{\nu}&\underline{\nu}
 \end{pmatrix}.
\]
\end{minipage}
\caption{The perturbed star for four machines. In column $j$, the processing
time on machine $i$ has law $\underline{\nu}=\operatorname{Unif}[1,1+\tau]$
if vertex $i$ is an endpoint of edge $j$, and
$\overline{\nu}=\operatorname{Unif}[\overline{t},\overline{t}+\tau]$ otherwise.
Only the root is an endpoint of the loop for job~$1$. All entries are
independent. Assigning the loop to the root and each spoke to its leaf
gives makespan at most $1+\tau$.}
\label{fig:perturbed-star}
\end{figure}

\paragraph{A job in isolation.}
Present column $t_{\cdot j}$ alone. Sending this job outside its endpoints
costs at least $\overline{t}$, whereas one endpoint can process it in
time at most $1+\tau$. The one-job guarantee therefore gives, almost surely,
\[
\overline{t}\sum_{i\notin \{1,j\}}g_i(t_{\cdot j})
\le\sum_i t_{ij} g_i(t_{\cdot j})
\le R\min_i t_{ij}
\le R(1+\tau).
\]
The total probability of sending the job outside $\{1,j\}$ is at most
$R(1+\tau)/\overline{t}$. For the loop, $\{1,j\}=\{1\}$, so
\[
\E[g_1(t_{\cdot1})]\ge1-\frac{R(1+\tau)}{\overline{t}}.
\]
For a spoke $j\ge2$, the two endpoint times can differ in a given draw,
but their joint law is unchanged when machines $1$ and $j$ are swapped.
Machine equivariance exchanges $g_1$ and $g_j$ under this swap. Averaging
over the column therefore assigns half the total endpoint probability
to the root, giving
\[
\E[g_1(t_{\cdot j})]
=\frac12\E[g_1(t_{\cdot j})+g_j(t_{\cdot j})]
\ge\frac12\left(1-\frac{R(1+\tau)}{\overline{t}}\right).
\]

\paragraph{The full star.}
Now present all $n$ jobs together. By locality, the conditional probability
that the root receives job $j$ is exactly the one-job probability
$g_1(t_{\cdot j})$ just bounded. Each job the root receives costs it at
least one. Adding the contributions from the loop and the $n-1$ spokes
therefore gives
\[
 \E[L_1]
 =\sum_{j=1}^n \E[t_{1j}g_1(t_{\cdot j})]
 \ge\sum_{j=1}^n \E[g_1(t_{\cdot j})]
  \ge\left(1+\frac{n-1}{2}\right)
      \left(1-\frac{R(1+\tau)}{\overline{t}}\right).
\]
The full-profile guarantee gives $\E[L_1]\le\E[C_{\max}]\le R(1+\tau)$.
For fixed $\tau>0$, combining this with the lower bound above and letting
$\overline{t}\to\infty$ gives $R(1+\tau)\ge(n+1)/2$. Letting $\tau\downarrow0$
yields $R\ge(n+1)/2$.
\end{proof}

The star argument compares the $n$-job instance with each of its jobs in
isolation. Theorem~\ref{thm:locality} uses the guarantee over all job counts
to obtain the same functions $g_i$ for both counts. With $n\ge2$ machines
and a single job, assigning it to a lowest bidder and paying that machine
the second-lowest bid is truthful and achieves ratio one.

\label{page:lower-overview-end}

\subsection{The correlated upper bound}\label{overview:upper}\label{intro:correlation}
On the star, one machine's mean load gave a lower bound on the expected
makespan. For an upper bound we must control the expected maximum over
all machines. At a fixed profile, the assignment probabilities $q_{ij}$
and expected payments determine each machine's expected utility. The
expected maximum also depends on which jobs go to the same machine.
Consider two jobs that each take one unit
of time on either of two machines. If the jobs choose machines independently
and uniformly, they land on the same machine with probability $1/2$, so
the expected makespan is $3/2$. Choosing uniformly between the two
assignments that split the jobs keeps $q_{ij}=1/2$ for every $i,j$ and
gives makespan one in every outcome.

We couple the probabilities $q_{ij}$ prescribed by Lu and Yu's
Bounded-Square rule~\cite{LY08}, which truncates the inverse-square rule
Square~\cite[Definition~4.1]{CKK10}. Using the slot construction of
Shmoys and Tardos~\cite{ST93}, we order each machine's fractional job
shares by decreasing processing time and draw a single matching for all
jobs. This bounds the range of each machine's load by the processing time
of its longest eligible job.

\Needspace{17\baselineskip}
\begin{samepage}
\begin{theorem}[Bounded-Square allocation with slot rounding]\label{thm:generalupper}
The Bounded-Square allocation rule with slot rounding gives a
truthful-in-expectation mechanism for every finite number of jobs.
For $n\ge7$ its expected makespan is at most $(n+1)\OPT(t)/2$.
For $n=4,5,6$, the respective upper bounds are $3\OPT(t)$,
$(105/32)\OPT(t)$, and $(29/8)\OPT(t)$. Expected payments are finite.
If $\OPT(t)=0$, every supported assignment has zero makespan.
\end{theorem}
\end{samepage}

\paragraph{The mechanism.}
For each job $j$ with positive minimum, let
$S_j=\{i:t_{ij}\le2\min_k t_{kj}\}$ be its eligible machines. Set
\[
 q_{ij}=\frac{\ind\{i\in S_j\}\,t_{ij}^{-2}}
                   {\sum_{k\in S_j}t_{kj}^{-2}}.
\]
For a zero-minimum job, allocate uniformly among its zero-cost machines.
At machine $i$, sort jobs with $q_{ij}>0$ by decreasing $t_{ij}$,
breaking ties by job index. Lay intervals of lengths $q_{ij}$ end to
end in that order and divide the list into slots of capacity one.
Form a matrix with one row per job and one column per slot. Its
job--slot entry is the length of their overlap. Each row sums to one,
and only the last slot of a machine can have a deficit.

There are $K=\sum_i\lceil\sum_jq_{ij}\rceil$ slots. Their deficits
sum to $K-m$, so $K-m$ unit dummy rows suffice to fill them
successively. Decompose the resulting doubly stochastic matrix into perfect matchings
by the procedure in Section~\ref{upper:marginal-preserving-correlated-rounding}.
Draw a matching with its weight in the decomposition. Each real job goes
to the machine that owns its matched slot. The probability that job $j$
goes to machine $i$ is the sum of its overlaps with $i$'s slots, namely
$q_{ij}$. The empty instance has its empty assignment.
The expected per-job payment to machine $i$ is
\[
 \pi_{ij}(t)=t_{ij}q_{ij}(t)
       +\int_{t_{ij}}^\infty q_{ij}(u,t_{-ij})\operatorname{d}u,
\]
where only entry $(i,j)$ varies in the integral. Sum these payments over
jobs. Holding the other reports
fixed, this probability is nonincreasing in $t_{ij}$ and vanishes above
twice the minimum opponent cost. Section~\ref{upper:marginal-preserving-correlated-rounding}
checks this at the cutoff and at zero-cost ties. Lemma~\ref{lem:local-payments}
then gives finite payments and truthfulness in expectation.

\paragraph{Three estimates and their roles.}
Fix a profile with $T=\OPT(t)>0$ and write $\mu_i=\E[L_i]$. In each
outcome choose a maximum-load machine $I$ by fixed tie-breaking, and
let $\theta_i=\Prb(I=i)$. The realized makespan is
$\sum_i\ind\{I=i\}L_i$. Writing the expectation of each product as
$\theta_i\mu_i$ plus the covariance of its two factors gives
\[
 \E[C_{\max}]=\theta\cdot\mu+
       \sum_i\operatorname{Cov}\bigl(\ind\{I=i\},L_i\bigr).
\]
The first term weights each machine's mean load by its chance of being
selected. When $\theta_i>0$, its covariance term is
$\theta_i(\E[L_i\mid I=i]-\mu_i)$. The range of $L_i$ limits how far
this conditional mean can exceed $\mu_i$. Write $\loadwidth(L)$ for
the length of the smallest closed interval containing a bounded load
$L$ almost surely.
\Needspace{14\baselineskip}
\begin{proposition}[Load estimates]\label{prop:upper-estimates}
For this mechanism, let $d_i=\max\{t_{ij}:q_{ij}>0\}$, with $d_i=0$
for empty support. The load ranges satisfy
\begin{equation}\label{eq:upper:width-interface}
 \loadwidth(L_i)\le d_i\le2T.
\end{equation}
The mean-load vector satisfies, for every $\lambda\in[0,\infty)^n$,
\begin{equation}\label{eq:upper:mean-interface}
 \lambda\cdot\mu\le\frac T2\left(\sum_i \lambda_i+n\lVert \lambda\rVert_2\right).
\end{equation}
These means and ranges bound the expected maximum by
\begin{equation}\label{eq:upper:covariance-interface}
 \E[\max_i L_i]\le\theta\cdot\mu+
       \sum_i\loadwidth(L_i)\theta_i(1-\theta_i).
\end{equation}
\end{proposition}
For the first estimate, sorting puts the processing times in
successive slots in intervals with disjoint interiors. All these
intervals lie in $[0,d_i]$. Since a matching selects one entry per
slot, the total range is at most $d_i$ (Lemma~\ref{lem:width}).

For the second, fix a job $j$ with positive minimum processing time and
$A^*(j)=i$. Divide its mean contributions by its processing time on
machine $i$, so $v_k=t_{kj}q_{kj}/t_{ij}$. If $i\in S_j$, the inverse-square
rule gives $\lVert v\rVert_2^2=1/(t_{ij}^2\sum_{k\in S_j}t_{kj}^{-2})=v_i$.
With $e_i$ the $i$th coordinate vector, we obtain the sphere identity
\[
 \lVert v-e_i/2\rVert_2^2=\lVert v\rVert_2^2-v_i+\frac14=\frac14.
\]
Cauchy--Schwarz therefore yields
\[
 \lambda\cdot v=\frac{\lambda_i}2+\lambda\cdot(v-e_i/2)
 \le\frac{\lambda_i+\lVert \lambda\rVert_2}{2}.
\]
If $i\notin S_j$, then $\lVert v\rVert_2<1/2$ gives the same bound.
Zero-minimum jobs contribute zero. Multiply by $t_{ij}$ and sum first
over the jobs with $A^*(j)=i$, whose total cost is at most $T$, then
over machines $i$. This gives~\eqref{eq:upper:mean-interface}
(Lemma~\ref{lem:meanball}).

For the last estimate, subtract the smallest supported load from
$L_i$ in its covariance with $\ind\{I=i\}$. The difference is
nonnegative and at most $\loadwidth(L_i)$. Only the event $I=i$
can contribute positively, giving the factor $\theta_i(1-\theta_i)$
in~\eqref{eq:upper:covariance-interface} (Lemma~\ref{lem:covariance}).

\paragraph{Combining the estimates.}
Because $\theta$ is a probability vector,
$1/\sqrt n\le\tnorm\le1$ and
$\sum_i\theta_i(1-\theta_i)=1-\tnorm^2$.
The mean estimate holds for every nonnegative weight vector, so we may
use $\lambda=\theta$ even though $\theta$ depends on the matching law.
Combining the mean and covariance estimates with $d_i\le2T$ gives
\begin{equation}\label{eq:upper:quadratic}
 \frac{\E[C_{\max}]}{T}
 \le\frac{\theta\cdot\mu}{T}+2\sum_i\theta_i(1-\theta_i)
 \le\frac12+\frac n2\tnorm+2(1-\tnorm^2).
\end{equation}
The two terms depend on the same $\theta$. When a single machine
usually has the largest load, $\tnorm$ is close to one and the range
term is small. A more dispersed $\theta$ gives a smaller bound on
the weighted mean. For $n\ge8$,
the difference between $(n+1)/2$ and this bound is
\[
 (1-\tnorm)\left(\frac n2-2(1+\tnorm)\right)\ge0.
\]
For $n=4,5,6$, completing the square gives
\[
 \frac12+\frac n2\tnorm+2(1-\tnorm^2)
 =\frac52+\frac{n^2}{32}-2(\tnorm-n/8)^2
 \le\frac52+\frac{n^2}{32}.
\]
The resulting ratios are $3$, $105/32$, and $29/8$, respectively.
When $T=0$, every job has a zero-cost machine, and Bounded-Square assigns
each job to such a machine. Thus every outcome has zero makespan.

\paragraph{The additional improvement for seven machines.}
At $n=7$, the bound exceeds four by
$2(1-\tnorm)(\tnorm-3/4)$; only $3/4<\tnorm<1$ needs attention.
Normalize $T=1$ and choose a machine $i$ with
$\theta_i=\max_\ell\theta_\ell$. Recall that $d_i$ is its largest supported
job cost. Keeping this machine's own width bound sharpens the range estimate to
\begin{equation}\label{eq:upper:seven-interface}
 \E[C_{\max}]\le\theta\cdot\mu+2(1-\tnorm^2)
                         -(2-d_i)\theta_i(1-\theta_i).
\end{equation}
If $d_i\le1$, this already covers the excess. Otherwise choose a job
$j$ with $q_{ij}>0$ and $t_{ij}=d_i$, and let $k=A^*(j)$. On machine
$k$ this job costs at most one. Thus $k\ne i$ and $d_i/2\le t_{kj}\le1$.
Both $i$ and $k$ are eligible, since the eligibility cutoff is
at least $d_i>1$. Thus the job's normalized mean vector
$v_\ell=t_{\ell j}q_{\ell j}/t_{kj}$ satisfies
\[
 d_i v_i=t_{kj} v_k,\qquad
 \lVert v-e_k/2\rVert_2=1/2.
\]
The plane equation follows from the inverse-square probabilities.
On the sphere, the loss in $\theta\cdot v$ from its maximum equals
$\tnorm$ times the squared distance to the maximizing point. The plane
equation bounds that distance from below. Lemma~\ref{lem:seven-geometry}
uses this loss to sharpen the mean estimate for the same job that
controls the range term. Lemmas~\ref{lem:seven-elimination}
and~\ref{lem:seven-scalar} show that the two improvements cover the
excess over four. This completes Theorem~\ref{thm:generalupper}.

\paragraph{Exact implementation and payments.}
For rational reports, all slot entries lie on one rational grid of
polynomial bit length. A perfect-matching decomposition uses at most
$(m+n)^2$ matchings, since each subtraction removes a positive entry.
Sampling its rational weights with unbiased bits preserves the allocation
law exactly (Proposition~\ref{prop:slots-efficient}).

The expected-payment integral need not be rational. For one machine--job
pair, write $s=t_{ij}$, $q(u)=q_{ij}(u,t_{-ij})$ and
$B=2\min_{k\ne i}t_{kj}$. Let $X=\ind\{A(j)=i\}$. If $s<B$, draw
$U,V$ independently and uniformly on $[0,1]$, independently of $X$.
The rational payment is
\[
 \widehat\pi=sX+(B-s)\ind\{V\le q(s+(B-s)U)\}.
\]
Its expectation is $sq(s)+\int_s^B q(u)\operatorname{d}u$, and it is at least
$sX$ in every outcome. For $s\ge B$, pay $sX$.
The indicator asks whether $(U,V)$ lies below the graph of a
nonincreasing function. After $k$ bits of each uniform, the dyadic
rectangles where the comparison is still undecided have total area at
most $3\cdot2^{-k}$. Rational endpoint comparisons therefore suffice
in expected polynomial bit time (Lemma~\ref{lem:paymentbits}).

\label{page:upper-overview-end}

\begin{proof}[Proof of Theorem~\ref{thm:main}]
Theorem~\ref{thm:lower} gives $\rho(n)\ge(n+1)/2$ for every $n$.
Theorem~\ref{thm:generalupper} supplies the reverse inequality for every
$n\ge7$, with the same uniform guarantee over all finite job counts.
Their equality proves the claim. Proposition~\ref{prop:slots-efficient}
and Lemma~\ref{lem:paymentbits} give the implementation for $n\ge4$.
\end{proof}

\subsection{Two and three machines: auxiliary schedules and a shared seed}
\label{overview:small}
Theorem~\ref{thm:lower} gives $3/2$ for two machines and $2$ for three.
For two machines a stronger lower bound and a matching coupling give
$R_*$. For three machines a different coupling of the Bounded-Square
marginals attains two.

\paragraph{A common reduction to a few columns.}
Remove zero-minimum jobs, which the mechanisms allocate at zero cost.
If any jobs remain, fix an optimal assignment $A^*$ of makespan
$T=\OPT(t)>0$. Keep these owners fixed and divide each column $j$ by
its processing time $t_{A^*(j),j}$ on its owner. Under both
constructions the same random seed $\omega$ then selects the same
winner for that column. If $v^j(\omega)$ is the normalized job's
load vector, its contribution to the original load vector $L(\omega)$
is $t_{A^*(j),j}v^j(\omega)$.

To bound the makespan, sample one representative $J_i$ per owner,
independently of the seed. Select a job $j$ owned by $i$ with
probability $t_{ij}/T$. These probabilities sum to at most one because
machine $i$ has load at most $T$ in $A^*$. Assign the remaining
probability to a dummy with $v^\varnothing=0$. Writing
$\boldsymbol J=(J_i)_{i\in[n]}$, the choice of weights gives
\[
 \frac{L(\omega)}T
 =\E_{\boldsymbol J}\!\left[\sum_{i} v^{J_{i}}(\omega)\,\middle|\,\omega\right].
\]
Convexity of the maximum, followed by averaging over $\omega$, gives
\[
 \frac{\E[C_{\max}]}T
 \le\E_{\boldsymbol J,\omega}\!\left[\left\|\sum_{i} v^{J_{i}}(\omega)\right\|_\infty\right].
\]
The representative jobs are sampled only for this bound. For every
tuple, the mechanism uses the original owners and the same law for
$\omega$. Hence a bound $R$ on each tuple gives $RT$ on the original
instance (Lemma~\ref{lem:representative}). On two machines, a tuple
with one real job is covered by a one-job bound. On three machines, a
dummy can be filled by a job forced to its owner, which only increases
the makespan. Changing the auxiliary schedule may change which jobs
go together, but each $q_{ij}$, and hence each machine's expected
utility at the prescribed payments, stays the same.

\paragraph{Two machines: an analytic optimum and its active constraints.}
A job with positive costs $(t_{1j},t_{2j})$ goes to machine one with
probability $f(t_{1j}/t_{2j})$, where $f$ is nonincreasing and
$f(x)+f(1/x)=1$. For two representatives with distinct owners, the
normalized columns have the form $(1,s)$ and $(t,1)$. By exchanging
both machine and job labels, we may take $s\le t$. The probabilities
of assigning these jobs to machine one are $1-f(s)$ and $f(t)$.
Using opposite ends of the same uniform interval gives the minimum
expected makespan for this pair. Section~\ref{n2:scalar-subsection}
shows why these pairs give the scalar problem
\[
 R_*=\inf_f\sup_{0<s\le t}
 \bigl[1+s f(s)-(1+s-t)f(t)\bigr].
\]
For the upper bound we construct a rule satisfying every constraint
at $R_*$. For the lower bound we use pairs with $s\le1\le t$ where
the bound is attained. Reciprocity lets us write $f(t)=1-f(1/t)$.
We choose the tight pairs so that $1/t$ in one pair is the first
ratio of the next. A weighted sum then cancels the unknown values of
a competing rule at every interior ratio. At the first ratio, $1$,
symmetry fixes $f(1)=1/2$, and the weight of the last term tends to
zero. Section~\ref{sec:n2} constructs these pairs and the optimal
rule from the same analytic curve.

\paragraph{Three machines: a fixed mixture of two clock maps.}
Keep the Bounded-Square $q_{ij}$, but replace slot rounding by a
fair mixture of two shared exponential-clock constructions. One indexes
clocks by cyclic displacement from the owner; the other uses an owner
clock and clocks indexed by unordered machine pairs. The clocks within
one column are independent, so both maps satisfy $\Prb(A(j)=i)=q_{ij}$.
Sharing them across columns creates different joint assignment distributions.
One global fair coin chooses the map for the whole instance.

The owner reduction leaves three positive columns with diagonal entries
one. For this normalized matrix we prove
\[
 \frac12\bigl(\E[C_{\rm cyc}]+\E[C_{\rm pair}]\bigr)\le2.
\]
A closed support cover has $129$ symmetry orbits. The $54$ smaller cases
have the explicit mathematical arguments in Section~\ref{n3:analytic};
the other $75$ use exact rational certificates on $90{,}791$ terminal
boxes. Each certificate combines uniform probability bounds with a linear
programming dual witness. The closed cover makes the bound uniform over
all real profiles. This proves the approximation ratio of two; the unrestricted
lower theorem proves its optimality.

\paragraph{From existence to efficient exact mechanisms.}
For two and three machines and rational $0<\varepsilon<1$, a
fixed-dimension approximation scheme replaces the optimal auxiliary schedule
by a $(1+\varepsilon)$-approximate schedule.
The common-seed construction preserves the $q_{ij}$ exactly, so this
replacement multiplies the approximation ratio without weakening
truthfulness. The two-machine efficient mechanism uses a piecewise rational
probability function with a separately certified approximation ratio close
to $R_*$. Exact sampling with unbiased bits and randomized rational
payments implement the required allocation and payment expectations.
For the general-machine construction, slot rounding already avoids
computing an auxiliary schedule. These implementation arguments give
Theorem~\ref{thm:efficient}.

\begin{proof}[Proof of Theorem~\ref{thm:small}]
Theorem~\ref{n2:main} identifies $\rho(2)=R_*$.
Theorems~\ref{thm:lower} and~\ref{thm:n3upper} give $\rho(3)=2$.
For $n=4,5,6$, combine Theorems~\ref{thm:lower} and~\ref{thm:generalupper}.
\end{proof}
\label{page:overview-end}

\section{Proof of the locality theorem}\label{sec:lower}
Fix $n\ge2$ and a truthful-in-expectation family with finite uniform
ratio $R$. The star used only the job counts $1$ and $n$. The proof of
Theorem~\ref{thm:locality} uses the freedom to add arbitrarily many jobs
to construct a family with the same bound $R$.

To see why the number of jobs matters, consider three machines and
distinguish job~$1$. An intermediate step of the proof gives, for
almost every positive $m$-job profile,
\[
 q_{i1}^{m,*}(t)=q_{i1}^{1,*}(t_{\cdot1})
       +\sum_{j=2}^m h_{i,1}(t_{\cdot1},t_{\cdot j}).
\]
The first term is machine $i$'s probability of receiving job~$1$ in
isolation. Each summand is the change in that probability when job~$j$
is the only other job. The same function $h_{i,1}$ occurs at every job
count. For almost every fixed first column, draw the other $m-1$ columns
independently and uniformly from the same box. The mean of the sum is
$m-1$ times the mean correction for one added job. The left side stays
in $[0,1]$ for every $m$, so the mean correction must be zero. Applying
this to every box with rational endpoints shows that the correction
vanishes almost everywhere.

For general $n$, we need a bound on how many columns can occur in one
term, and the formula for each term must be the same across job counts.
Section~\ref{lower:free-padding-normalization-in-the-original-all-m-model}
first enforces consistency under deletion of a job with a zero-cost machine, and
Section~\ref{lower:measurable-grid-proxies} takes a common measurable
limit across job counts. The resulting incentive identities and the
condition that each job goes to one machine force interactions involving
at least $n$ jobs to vanish (Section~\ref{lower:muxf6bius-potentials-and-direct-discrete-laplace}).
Each term in a marginal can therefore involve at most $n-2$ other
columns, with the same formula at every job count. After fixing the
first $m$ jobs, adding independent columns up to job count $m'$ gives
an average probability that is a polynomial in $m'-m$ of degree at
most $n-2$. Since this polynomial stays in $[0,1]$, it is constant.
Section~\ref{lower:bounded-polynomial-collapse-without-diagonal-evaluations}
uses its constancy to delete one positive column at a time. Throughout
this argument we retain the joint assignment law to preserve the bound
on expected makespan.

\paragraph{A normalized cost potential.}
Fix the other machines' bids and denote them here by $t_{-i}$.
For machine $i$ of type $t_i$ bidding truthfully, define its normalized
expected payment and cost potential by
\[
 \bar p_i^m(t)=p_i^m(t)-p_i^m(0,t_{-i}),\qquad
 V_i^m(t)=t_i\cdot q_i^m(t)-\bar p_i^m(t).
\]
The potential is the negative of normalized truthful expected utility.
At any fixed report, expected cost minus normalized payment is affine in the
machine's true row. Truthfulness makes this affine function an upper
bound on the potential, with equality when the true row equals the report.
This gives the concave form of Rochet's representation~\cite{Rochet87}.

\Needspace{10\baselineskip}
\begin{lemma}[Normalized potentials]\label{lem:potential}
Any truthful-in-expectation family can be normalized without changing its allocations so that
\[
\bar p_i^m(0,t_{-i})=0,\qquad -\lVert t_i\rVert_1\le \bar p_i^m(t)\le0.
\]
For fixed $t_{-i}$, the potential $V_i^m(t)=t_i\cdot q_i^m(t)-\bar p_i^m(t)$ is concave, coordinatewise nondecreasing, and $1$-Lipschitz in the $\ell_1$ norm as a function of $t_i$. For any types $u,v\in[0,\infty)^m$,
\begin{equation}\label{eq:potential-support}
V_i^m(v,t_{-i})\le V_i^m(u,t_{-i})+(v-u)\cdot q_i^m(u,t_{-i}),
\end{equation}
and $0\le V_i^m(t)\le\sum_j t_{ij}$.
\end{lemma}
\begin{proof}
The subtracted term is independent of machine $i$'s report, so all incentive
comparisons are unchanged. Truthfulness at true zero gives $\bar p_i^m(t)\le0$, and truthfulness at $t_i$ against report zero gives
\[
\bar p_i^m(t)\ge t_i\cdot q_i^m(t)-t_i\cdot q_i^m(0,t_{-i})\ge-\sum_jt_{ij}.
\]
Fix $t_{-i}$ and suppress it and the superscript $m$. Truthfulness for
true type $v$ against report $u$ gives
\[
 V_i(v)=v\cdot q_i(v)-\bar p_i(v)
       \le v\cdot q_i(u)-\bar p_i(u)
       =V_i(u)+(v-u)\cdot q_i(u).
\]
This is~\eqref{eq:potential-support}, with equality when $u=v$. Hence
\[
 V_i(v)=\inf_{u\in[0,\infty)^m}
       \bigl\{V_i(u)+(v-u)\cdot q_i(u)\bigr\},
\]
an infimum of affine functions of $v$, so $V_i$ is concave.
Applying the support inequality in both directions gives
\[
 (v-u)\cdot q_i(v)
 \le V_i(v)-V_i(u)
 \le(v-u)\cdot q_i(u).
\]
In particular, $(v-u)\cdot(q_i(v)-q_i(u))\le0$, the weak-monotonicity
inequality. Since every coordinate of $q_i$ lies in $[0,1]$, the same bounds imply
\[
 |V_i(v)-V_i(u)|\le\lVert v-u\rVert_1,
 \qquad u\le v\ \Longrightarrow\ V_i(u)\le V_i(v).
\]
Finally, $V_i(0)=0$, so monotonicity and the Lipschitz bound give
$0\le V_i(t_i)\le\sum_j t_{ij}$. These assertions hold pointwise for
each fixed $t_{-i}$ and do not presume joint measurability.
\end{proof}

\subsection{Adding jobs with a zero-cost machine}\label{lower:free-padding-normalization-in-the-original-all-m-model}

Call a column \emph{free} if at least one of its entries is zero. Such a
job can always be added to a schedule at no cost. The next lemma makes
this observation hold for the mechanism's assignment distribution and
payments on the retained jobs, as well as for the optimum.

\begin{lemma}[Consistency under adding free jobs]\label{lem:padding}
Every truthful-in-expectation $R$-approximate family admits a
truthful-in-expectation $R$-approximate replacement that is equivariant
under permutations of jobs and machines. In this family, a job whose
column contains a zero is assigned to a zero-cost machine. Deleting
that job preserves the joint assignment distribution on the remaining
jobs and the normalized expected payments.
\end{lemma}
\begin{proof}
We first find a family whose assignment distribution on the original
jobs survives padding by any free column. Replicating a free job then
forces its processing cost to zero, which gives payment consistency.
We average over job permutations at the outset so that padding
operations commute, and over machine permutations at the end.

\subsubsection*{Step 1: A compact space of families}\label{lower:a-compact-convex-space-of-complete-families}

Represent a family $F=(F_m)_{m\ge0}$ by its assignment distributions
and cost potentials, $F_m=(y^m,V^m)$. The distribution $y^m$
determines $q_i^m$, and the potential determines the normalized payment
$\bar p_i^m(t)=t_i\cdot q_i^m(t)-V_i^m(t)$. With this payment,
truthfulness is exactly the supporting inequality~\eqref{eq:potential-support}.
By Lemma~\ref{lem:potential}, at each $m$ and profile $t$ the pair
$(y^m(t),V^m(t))$ belongs to the compact convex set
\[
\left\{(y_A)_{A\in\mathcal A_{m,n}}:y_A\ge0,\ \sum_Ay_A=1\right\}
\times\prod_i[0,\lVert t_i\rVert_1].
\]
For $m=0$ this specifies the empty assignment and zero potentials.
Take the product over all $m$ and $t$. Within it, let $\mathcal K_R$
consist of the job-equivariant families satisfying
\eqref{eq:potential-support}, \eqref{eq:ratio}, and
$V_i^m(0,t_{-i})=0$ at every job count. At fixed reports these are closed
linear equalities and inequalities in the coordinates $y$ and $V$.
Thus $\mathcal K_R$ is a closed convex subset of the compact product.
To obtain a member, normalize the given family and average
over job permutations at each $m$. Each condition survives averaging.

\subsubsection*{Step 2: Padding preserves the feasible family space}\label{lower:the-padding-operator}

Fix a free column $c\in[0,\infty)^n$. At job count $m$, the padding
operator $\mathsf P_c$ runs $F_{m+1}$ on $(t,c)$ and projects its
assignment onto the original jobs. For the potential, hold $c$ fixed
and use machine $i$'s zero report on the original jobs as the
normalization point. This gives
\begin{equation}\label{eq:lower:padding-potential}
(\mathsf P_cV)_i^m(t)
=V_i^{m+1}(t,c)-V_i^{m+1}((0,t_{-i}),c).
\end{equation}
The added coordinate is $c_i$ in both evaluations. Before normalization,
the projected payment is $\bar p_i^{m+1}(t,c)-c_i q_{i,m+1}^{m+1}(t,c)$.
It deducts exactly the expected cost of the job being discarded.
Normalizing this payment at the zero report on the retained jobs gives
\eqref{eq:lower:padding-potential}. Each coordinate of the new family is a finite linear combination of old coordinates,
so $\mathsf P_c$ is continuous and affine in the product topology.

\paragraph{Truthfulness in expectation.}\label{lower:full-row-truthfulness}

Apply~\eqref{eq:potential-support} in $F_{m+1}$ to the own rows
$(u,c_i)$ and $(v,c_i)$. Their difference is $(v-u,0)$, so only the
assignment probabilities of the retained jobs enter the bound.
Subtracting the same constant in~\eqref{eq:lower:padding-potential}
from both sides gives exactly~\eqref{eq:potential-support} for the
projected family. Its potential is zero at own row zero. Applying the
support inequality in both directions between $0$ and $t_i$ gives
$0\le(\mathsf P_cV)_i^m(t)\le\lVert t_i\rVert_1$ as required.

\paragraph{Performance.}\label{lower:performance}

Adding the free job at a zero-cost machine extends an optimal assignment
of $t$ at no extra cost. Conversely, deleting it from any assignment
of $(t,c)$ can only decrease the makespan. Thus $\OPT(t,c)=\OPT(t)$,
and projecting the padded assignment gives expected makespan at most
$R\OPT(t)$ on the retained jobs.

\paragraph{Job symmetry.}\label{lower:job-symmetry}

Relabeling the retained jobs before or after padding gives the same
result by the symmetry of \(F_{m+1}\). Thus \(\mathsf P_c\) maps \(\mathcal K_R\) into
itself.

\subsubsection*{Step 3: Padding operators commute}\label{lower:commutation-including-the-normalization-constants}

For two free columns $c,d$, both orders of padding retain the same
assignment distribution by job symmetry of $F_{m+2}$. For potentials,
apply~\eqref{eq:lower:padding-potential} twice. The inner normalization
sets machine $i$'s costs on the original jobs and on column $c$ to
zero. It has the same value in both terms of the outer difference and
cancels. This leaves
\[
 (\mathsf P_c\mathsf P_dV)_i^m(t)
 =V_i^{m+2}(t,c,d)-V_i^{m+2}((0,t_{-i}),c,d).
\]
Interchanging $c$ and $d$ preserves the right side by job symmetry.
Hence $\mathsf P_c\mathsf P_d=\mathsf P_d\mathsf P_c$ on $\mathcal K_R$.

\subsubsection*{Step 4: A family fixed by every padding operator}\label{lower:simultaneous-invariance-under-every-free-column}

View the assignment probabilities and potentials as coordinates in a
product of copies of $\mathbb R$ with its locally convex product
topology. Step~1 gives a nonempty compact convex subset $\mathcal K_R$.
Steps~2 and~3 show that the maps $\mathsf P_c$,
indexed by all free columns $c$, are commuting continuous affine
self-maps of $\mathcal K_R$. The Markov--Kakutani
theorem~\cite[Theorem~1]{Kakutani38} therefore gives a single family
$F\in\mathcal K_R$ with $\mathsf P_c F=F$ for every free $c$.
Iterating this identity shows that padding by any finite list of free
columns preserves the joint assignment distribution on the original jobs.

\subsubsection*{Step 5: Free jobs have zero cost and consistent payments}\label{lower:replication-forces-zero-cost-assignment-of-every-free-job}

Fix the family obtained in Step 4. The fixed-point identity already
preserves the assignment distributions on the retained jobs. We now show that it also preserves the
normalized payments, by first proving that every free job has zero
processing cost under this family.

Take a profile $t$ with a free job $j$, write $c=t_{\cdot j}$,
and fix a machine $i$ with $c_i>0$. Pad $t$ by $N$ further copies
of this column. Invariance under $\mathsf P_c$ preserves
$q_{ij}^m(t)$, and job symmetry gives this same probability to all
$N+1$ copies. Their contribution to machine $i$'s expected load gives
\[
 (N+1)t_{ij}q_{ij}^m(t)\le R\OPT(t)
 \qquad\text{for every finite }N.
\]
The right side stays fixed as $N$ grows because the optimum is unchanged
and the ratio $R$ is uniform over job counts. Since $t_{ij}>0$, we get
$q_{ij}^m(t)=0$. Thus every free job is assigned only to zero-cost machines.

\paragraph{Payment consistency.}\label{lower:raw-normalized-payments-also-become-deletion-consistent}

It remains to show that the normalization term
$V_i^{m+1}((0,t_{-i}),c)$ in~\eqref{eq:lower:padding-potential} is
zero. If $c_i=0$, this is own-row normalization. If $c_i>0$, another
machine can process the added job for free. Hold the opponents fixed
and let machine $i$ report zero on the original jobs and $s>0$ on the
added job. The probability of assigning that job to $i$ is zero for
every $s>0$, by the replication argument. The supporting inequality in
both directions therefore makes the potential constant for $s>0$.
By continuity in the own row (Lemma~\ref{lem:potential}), the constant
is its value at $s=0$, namely zero.

Equation~\eqref{eq:lower:padding-potential} at the fixed point now gives
$V_i^{m+1}(t,c)=V_i^m(t)$. The expected processing cost is also the
same at both job counts because the retained probabilities agree
and the extra job incurs zero cost. Recovering payments as cost minus potential gives
\[
\bar p_i^m(t)=\bar p_i^{m+1}(t,c).
\]
Thus deleting a free job preserves both the retained joint assignment
distribution and the normalized expected payments.

\subsubsection*{Step 6: Impose machine symmetry}\label{lower:machine-symmetry}

Relabel the input rows, output assignments, and potentials by a
machine permutation. This preserves truthfulness and performance and
takes $\mathsf P_c$ to the operator for the relabeled free column.
Hence the simultaneous fixed points form a convex set invariant under
machine permutations. Averaging a fixed family over those permutations
makes it machine-equivariant while preserving job symmetry and all
the fixed-point identities. This is the family used below.
\end{proof}

\subsection{Measurable limits and incentive derivatives}\label{lower:measurable-grid-proxies}

The family of Lemma~\ref{lem:padding} need not be jointly measurable in
the reports. We evaluate it on countable grids and take limits of the
resulting measurable step functions. We retain the joint assignment laws
to preserve the bound on expected makespan, and the potentials to recover the marginal
probabilities as weak derivatives $\partial_{ij}V_i=q_{ij}$, where
$\partial_{ij}=\partial/\partial t_{ij}$. A single subsequence across
job counts lets us compare these limits when columns are deleted.

\Needspace{9\baselineskip}
\begin{lemma}[A common measurable limit]\label{lem:lower:measurable}
For the normalized family of Lemma~\ref{lem:padding}, let $G_h$ floor
every report coordinate to the nonnegative $h$-grid. There is one
sequence $h_\ell\downarrow0$ along which the grid approximations of all assignment probabilities and potentials, at every job count,
converge weak-star on every bounded profile box.
Their limits are measurable and, almost everywhere, satisfy feasibility,
job and machine equivariance, and the performance bound
\[
 \sum_A y_A^{m,*}(t)C_{\max}(t,A)\le R\OPT(t).
\]
They also satisfy $\partial_{ij}V_i^{m,*}=q_{ij}^{m,*}$ in the
distributional sense. Finite constant-coefficient
linear identities between the grid approximations, including components lifted
from subprofiles by ignoring omitted columns, survive the same extraction.
\end{lemma}
\begin{proof}
We construct the common limit, pass the linear identities, and then recover
the incentive derivatives.

\subsubsection{A common weak-star limit}\label{lower:countable-extraction}

For $h=1/k$, with $k$ a positive integer, write
$\lfloor s\rfloor_h=h\lfloor s/h\rfloor$ and
$G_h(t)=(\lfloor t_{ij}\rfloor_h)_{i,j}$.
Define the grid approximations
\[
y^{m,h}=y^m\circ G_h,\qquad q^{m,h}=q^m\circ G_h,\qquad V_i^{m,h}=V_i^m\circ G_h.
\]
Each is Borel measurable, since it is constant on a countable partition
into grid boxes. No original report-measurability assumption is used.

For fixed $m$, these functions are uniformly bounded on every box
$(0,M)^{nm}$: each coordinate of $y^{m,h}$ and $q^{m,h}$ by $1$, and $V_i^{m,h}$ by $mM$.
There are only countably many choices of $m$, $i$, assignment coordinate,
and positive integer $M$.
Sequential weak-star compactness of bounded subsets of \(L^\infty\) on
each finite box follows from separability of its \(L^1\) predual. A
diagonal subsequence \(h_\ell\to0\) therefore gives simultaneous measurable limits in this
weak-star topology
\[
y^{m,h_\ell}\overset{*}{\rightharpoonup}y^{m,*},\quad q^{m,h_\ell}\overset{*}{\rightharpoonup}q^{m,*},\quad V_i^{m,h_\ell}\overset{*}{\rightharpoonup}V_i^{m,*}.
\]
Weak-star convergence here means that the integral of each displayed
coordinate against any $L^1$ test function on the box converges.
In particular, this determines its average on every subbox of positive
volume. The limits agree on nested boxes, so they define measurable
functions throughout the open orthant, up to Lebesgue-null sets.

\subsubsection{Feasibility, performance, and consistency}\label{lower:properties-surviving-the-common-extraction}

First, nonnegativity, normalization of assignment distributions, and the
identities $q_{ij}=\sum_{A:A(j)=i}y_A$ are weak-star closed. Thus $y^{m,*}$ is a probability distribution over the
finite assignment set, with \(q_{ij}^{m,*}=\sum_{A:A(j)=i}y_A^{m,*}\) almost everywhere.

To pass the performance guarantee to the limit, observe that every
fixed assignment $A$ satisfies
\[
C_{\max}(t,A)\le C_{\max}(G_h(t),A)+mh.
\]
Rounding down can only reduce the makespan of an optimal assignment,
so \(\OPT(G_h(t))\le\OPT(t)\). Applying the guarantee at the grid point
$G_h(t)$ gives
\begin{equation}
\sum_A y_A^{m,h}(t)C_{\max}(t,A)\le R\operatorname{OPT}(t)+mh.\label{eq:lower:3.1}
\end{equation}
The cost coefficients are continuous and bounded on each finite box.
Testing \eqref{eq:lower:3.1} against arbitrary nonnegative \(L^1\) functions and
passing to the limit gives the performance bound for $y^{m,*}$ at
almost every positive profile.

Flooring commutes with job and machine permutations; the family has both
symmetries by Lemma~\ref{lem:padding}. The finite linear
equivariance identities therefore survive weak-star convergence.

Finally, we must compare the common limits at different job counts.
For a job subset $J$, lift a lower-dimensional function $f(t_J)$
by ignoring the other columns. On the corresponding finite boxes,
Fubini gives, for an $L^1$ test function $\varphi$,
$\varphi_J(t_J):=\int\varphi(t)\operatorname{d}t_{[m]\setminus J}\in L^1$ and
\[
 \int f_h(t_J)\varphi(t)\operatorname{d}t
 =\int f_h(t_J)\varphi_J(t_J)\operatorname{d}t_J.
\]
Hence lifting commutes with weak-star convergence. Every finite linear
identity using the common lower-dimensional components remains valid
after extraction, as needed for kernels shared across all job counts.

\Needspace{10\baselineskip}
\subsubsection{Recovering incentive derivatives}\label{lower:recovering-weak-own-derivatives}

Let $e_{ij}$ be the unit vector for entry $t_{ij}$. Applying the
supporting inequality~\eqref{eq:potential-support} in both directions
between adjacent grid reports gives
\begin{equation}\label{eq:lower:3.2}
 q_{ij}^{m,h}(t+he_{ij})\le
 \frac{V_i^{m,h}(t+he_{ij})-V_i^{m,h}(t)}{h}
 \le q_{ij}^{m,h}(t).
\end{equation}
To take the weak limit, move the grid shift onto a test function.
Let $\varphi\ge0$ be smooth and compactly supported in the open
orthant, and extend it by zero. Multiply~\eqref{eq:lower:3.2} by
$\varphi$ and integrate. For sufficiently small $h$, a change of
variables in each shifted term gives
\begin{equation}\label{eq:lower:3.3}
 \int q_{ij}^{m,h}(t)\varphi(t-he_{ij})\,\mathrm{d}t
 \le \int V_i^{m,h}(t)\frac{\varphi(t-he_{ij})-\varphi(t)}{h}\,\mathrm{d}t
 \le \int q_{ij}^{m,h}(t)\varphi(t)\,\mathrm{d}t .
\end{equation}
Along the common sequence $h=h_\ell\to0$, both outer terms converge to
$\int q_{ij}^{m,*}\varphi$. The translated test function converges
uniformly to $\varphi$, and its difference quotient in the middle
integral converges uniformly to $-\partial_{ij}\varphi$. All these
tests are supported in a fixed bounded box, where the grid functions
are uniformly bounded and converge weak-star. Hence
$-\int V_i^{m,*}\partial_{ij}\varphi=\int q_{ij}^{m,*}\varphi$.
By linearity this holds for all smooth compactly supported tests,
which gives
\begin{equation}
\partial_{ij}V_i^{m,*}=q_{ij}^{m,*},\label{eq:lower:3.4}
\end{equation}
in the distributional sense. We apply this identity to alternating
sums of potentials at different job counts.
\end{proof}

\subsection{Interactions of bounded order}\label{lower:muxf6bius-potentials-and-direct-discrete-laplace}

Fix a set $J$ of $s$ jobs. To isolate the contribution of all $s$
columns together, form the alternating sum of the grid potentials
\[
 W_i^{J,h}(t_J)=\sum_{J'\subseteq J}(-1)^{s-|J'|}V_i^{|J'|,h}(t_{J'}).
\]
The empty subprofile has potential zero. For two jobs, this sum is
their joint potential minus the two one-job potentials. Define $W_i^{J,*}$ by the
same sum of limiting potentials. Since each subprofile uses the common
limit for its job count, Section~\ref{lower:properties-surviving-the-common-extraction}
gives $W_i^{J,h_\ell}\overset{*}{\rightharpoonup}W_i^{J,*}$ on bounded
boxes. We will show that $W_i^{J,*}=0$ almost everywhere for $s\ge n$.

\paragraph{Estimate near a zero processing time.}
Fix $j\in J$ and pair each term indexed by $J'\ni j$ with the term
indexed by $J'\setminus\{j\}$. Setting $t_{ij}=0$ makes job $j$ free.
It then costs machine $i$ zero, and deleting it preserves both the
retained assignment probabilities and the normalized payment. Hence
\begin{equation}
V_i^{|J'|}(t_{J'}\vert_{t_{ij}=0})=V_i^{|J'|-1}(t_{J'\setminus\{j\}}).\label{eq:lower:4.1}
\end{equation}
Flooring preserves zeros. By~\eqref{eq:lower:4.1}, monotonicity, and
the Lipschitz bound, each pair therefore satisfies
\[
0\le V_i^{|J'|,h}(t_{J'})
      -V_i^{|J'|-1,h}(t_{J'\setminus\{j\}})
\le\lfloor t_{ij}\rfloor_h.
\]
Summing the $2^{s-1}$ pairs with their respective signs gives
\begin{equation}
|W_i^{J,h}(t_J)|\le2^{s-1}\lfloor t_{ij}\rfloor_h\le2^{s-1}t_{ij},\label{eq:lower:4.2}
\end{equation}
uniformly in the opponents' reports and in $h$.

We used deletion consistency before taking the limit, on profiles with
$t_{ij}=0$. What passes to the limit is~\eqref{eq:lower:4.2}, a bound
throughout the open orthant up to a null set. This interior bound will
control the boundary term in the transform argument.

\paragraph{Derivative and feasibility identities.}
For two jobs the derivative takes a particularly simple form.
\[
 \partial_{i1}W_i^{\{1,2\},*}(t)
 =q_{i1}^{2,*}(t)-q_{i1}^{1,*}(t_{\cdot1}).
\]
It measures how adding job $2$ changes the probability of assigning
job $1$ to machine $i$. Now fix any $j\in J$ and differentiate in
$t_{ij}$. Only subprofiles containing $j$ contribute. In each such
subprofile, move column $j$ to the first position and write
$J'\subseteq J\setminus\{j\}$ for the other jobs. Job equivariance
and the weak derivative identity~\eqref{eq:lower:3.4} give
\begin{equation}\label{eq:lower:interaction-gradient}
 \partial_{ij}W_i^{J,*}(t_J)=
 \sum_{J'\subseteq J\setminus\{j\}}
 (-1)^{s-1-|J'|}q_{i1}^{1+|J'|,*}(t_{\cdot j},t_{J'}).
\end{equation}
There are $2^{s-1}$ terms and every probability lies in $[0,1]$.
Together with~\eqref{eq:lower:4.2} this gives, almost everywhere,
\begin{equation}\label{eq:lower:4.3}
 |W_i^{J,*}(t_J)|\le2^{s-1}t_{ij},\qquad
 |\partial_{ij}W_i^{J,*}(t_J)|\le2^{s-1}.
\end{equation}
For each $J'$, the probabilities of assigning the first job sum to
one over machines. Thus for $s\ge2$,
\[
 \sum_i\partial_{ij}W_i^{J,*}
 =\sum_{J'\subseteq J\setminus\{j\}}(-1)^{s-1-|J'|}
 =(1-1)^{s-1}=0.
\]

\subsubsection{A zero-boundary Laplace lemma}\label{lower:the-discrete-transform-identity-and-feasibility}
\begin{lemma}[Vanishing of high-order interactions]\label{lem:lower:laplace}
Let $s\ge n\ge2$, with coordinates $t_{ij}>0$ for $i\in[n]$ and $j\in[s]$.
Suppose the measurable functions $W_1,\ldots,W_n$ have weak partial
derivatives $\partial_{ij}W_i$. For a finite constant $C\ge0$, assume
\[
 |W_i(t)|\le Ct_{ij},\qquad |\partial_{ij}W_i(t)|\le C
\]
for all $i,j$, and $\sum_i\partial_{ij}W_i(t)=0$ for all $j$, almost everywhere.
Then $W_i=0$ almost everywhere for every $i$.
\end{lemma}
\begin{proof}
\textbf{The transform identity.}
For a strictly positive parameter vector $z=(z_{ij})$, write
\[
\mathcal L f(z)=\int_{(0,\infty)^{ns}}e^{-z\cdot t}f(t)\operatorname{d}t.
\]
The bounds in the statement give absolute convergence for all integrals
below. We first establish the exact identity
\begin{equation}\label{eq:lower:4.4}
\mathcal L(\partial_{ij}W_i)(z)=z_{ij}\mathcal L W_i(z).
\end{equation}
The bound $|W_i(t)|\le Ct_{ij}$ suggests a cutoff in $\log t_{ij}$, since
the factor $t_{ij}$ then cancels the derivative of the logarithm.
Take a smooth compactly supported function $\chi$ on $\mathbb R$ with
$0\le\chi\le1$ and $\chi=1$ on $[-1,1]$. For $M\ge1$ put
\[
 \psi_M(t)=\prod_{k,\ell}\chi\bigl((\log t_{k\ell})/M\bigr).
\]
This cutoff has compact support in the open orthant and converges
pointwise to one as $M\to\infty$. Testing the weak derivative against
$e^{-z\cdot t}\psi_M(t)$ gives
\[
 \int (\partial_{ij}W_i)e^{-z\cdot t}\psi_M\,\mathrm{d}t
 =z_{ij}\int W_i e^{-z\cdot t}\psi_M\,\mathrm{d}t
  -\int W_i e^{-z\cdot t}\partial_{ij}\psi_M\,\mathrm{d}t .
\]
Only one factor of the product is differentiated. Hence
$|\partial_{ij}\psi_M|\le\|\chi'\|_\infty/(Mt_{ij})$, and the last
integral is bounded in absolute value by
\[
 \frac{C\|\chi'\|_\infty}{M}\int e^{-z\cdot t}\,\mathrm{d}t
 =\frac{C\|\chi'\|_\infty}{M}\prod_{k,\ell}z_{k\ell}^{-1}
 \longrightarrow 0 .
\]
The other two integrals converge by dominated convergence, proving
\eqref{eq:lower:4.4}.

\paragraph{The rank argument.}
Summing~\eqref{eq:lower:4.4} over machines and using $\sum_i\partial_{ij}W_i=0$ yields
\begin{equation}\label{eq:lower:4.5}
\sum_i z_{ij}\mathcal L W_i(z)=0,\qquad j\in[s].
\end{equation}
When the $s\times n$ matrix whose $j$th row is $(z_{1j},\ldots,z_{nj})$
has rank $n$, this system forces every $\mathcal L W_i(z)$ to vanish.
The determinant of the first $n$ rows is a nonzero polynomial in $z$.
For example, it equals $n+1$ when the diagonal entries are two and the
other entries are one. Full rank therefore holds on a dense open subset
of the positive orthant. The transforms are continuous on the whole
orthant by dominated convergence, so they vanish for every positive $z$.

\paragraph{Uniqueness.}
To conclude that the functions themselves vanish, we give a compact
moment argument. Fix a strictly positive vector $a$. The density
$e^{-a\cdot t}W_i(t)$ defines a finite signed measure. Push it forward
by $x_{k\ell}=e^{-t_{k\ell}}$ to a measure $\nu$ on $[0,1]^{ns}$. For every multiindex
$\kappa\in\mathbb Z_{\ge0}^{ns}$,
\[
\int x^\kappa\operatorname{d}\nu(x)
=\int e^{-(a+\kappa)\cdot t}W_i(t)\operatorname{d}t
=\mathcal L W_i(a+\kappa)=0.
\]
Polynomials are dense in the continuous functions on the compact cube,
so $\nu=0$. The inverse $t_{k\ell}=-\log x_{k\ell}$ on $(0,1)^{ns}$ gives
$W_i=0$ almost everywhere.
\end{proof}

Apply Lemma~\ref{lem:lower:laplace} to the interactions
$W_i^{J,*}$, using~\eqref{eq:lower:interaction-gradient},
\eqref{eq:lower:4.3}, and the feasibility identity above. We obtain
\begin{equation}\label{eq:lower:4.6}
W_i^{J,*}=0\quad\text{almost everywhere whenever }|J|\ge n.
\end{equation}
The restriction $n\ge2$ ensures that the feasibility cancellation applies
whenever $s\ge n$.

\subsubsection{Common kernels across job counts}\label{lower:finite-order-with-the-same-kernels-for-every-job-count}

Summing the interactions over all job subsets recovers the full
potential. Grouping the terms by the subprofile $J'$ gives
\[
\sum_{J\subseteq[m]}W_i^{J,*}(t_J)
=\sum_{J'\subseteq[m]}V_i^{|J'|,*}(t_{J'})
  \sum_{J'\subseteq J\subseteq[m]}(-1)^{|J|-|J'|}
 =V_i^{m,*}(t).
\]
The inner sum is $(1-1)^{m-|J'|}$ when $J'\ne[m]$ and is one
when $J'=[m]$. By~\eqref{eq:lower:4.6}, only interactions with
$|J|<n$ survive on the left. Differentiating in $t_{i1}$ therefore
retains only sets containing job $1$ and at most $n-2$ other jobs.

For a set of $k$ other jobs, let $h_{i,k}$ be the right side
of~\eqref{eq:lower:interaction-gradient} with job $1$ distinguished
and those $k$ jobs as the remaining arguments. This defines the same
kernel at every ambient job count because the formula uses only
subprofiles of these $k+1$ jobs. It is symmetric in the other
$k$ columns and bounded in absolute value by $2^k$, up to null sets.
In particular, $h_{i,0}(t_{\cdot1})=q_{i1}^{1,*}(t_{\cdot1})$.
The differentiated identity is therefore, for every $m\ge1$,
\begin{equation}\label{eq:lower:4.7}
 q_{i1}^{m,*}(t)
 =\sum_{k=0}^{\min(n-2,m-1)}\ \sum_{\substack{J\subseteq\{2,\ldots,m\}\\|J|=k}}
       h_{i,k}(t_{\cdot1};t_J)
\end{equation}
almost everywhere. In using a lower-dimensional identity here, its
exceptional set still has measure zero after the omitted columns are
restored, by Fubini on bounded boxes.

\subsection{Bounded probabilities force locality}\label{lower:bounded-polynomial-collapse-without-diagonal-evaluations}

We now hold the first $m$ columns fixed and ask what changes when
one more job is drawn from a box. The bound on interaction order
lets us answer this by drawing arbitrarily many jobs from the same box.

\begin{lemma}[Bounded consistent kernels imply locality]\label{lem:lower:collapse}
Let $n\ge2$. Suppose bounded measurable
kernels $h_{i,k}:((0,\infty)^n)^{k+1}\to\mathbb R$, symmetric in their last $k$
arguments almost everywhere, give the expansion
\eqref{eq:lower:4.7} for every $m\ge1$. If the resulting $q_{ij}^{m,*}$ lie
in $[0,1]$ almost everywhere for every $m$, then
$h_{i,k}=0$ almost everywhere for every $1\le k\le n-2$.
\end{lemma}
\begin{proof}
Fix a machine $i$ and a retained job count $m\ge1$. For $m'\ge m$,
keep the first $m$ columns $t$ and draw the $m'-m$ new columns
independently and uniformly from a bounded open box $\mathcal B$ of positive
volume, with rational endpoints and closure in $(0,\infty)^n$.
By Fubini, for almost every retained $t$ we may integrate
both~\eqref{eq:lower:4.7} and the bound $0\le q_{i1}^{m',*}\le1$
over the new columns. There are countably many pairs $(m',\mathcal B)$, so we
can fix $t$ in a single full-measure set for all these integrations,
including the expansion at $m'=m$.

In the expansion at $m'$, group the terms by the set
$J\subseteq\{2,\ldots,m\}$ of old jobs they use and the number
$\ell$ of new jobs they use. When $|J|+\ell\le n-2$, this group's
contribution to the averaged probability is
\[
 \binom{m'-m}{\ell}\frac{1}{|\mathcal B|^\ell}
 \int_{\mathcal B^{\ell}} h_{i,|J|+\ell}(t_{\cdot1};t_J,u_1,\ldots,u_\ell)
                  \,\mathrm{d}u_1\cdots\mathrm{d}u_\ell .
\]
Indeed, each choice of $\ell$ new jobs has the same law and uses the
same kernel. For $\ell=0$ the average is the kernel evaluated on the
retained columns alone. The binomial coefficient is zero if
$\ell>m'-m$. Summing over $J$ and $\ell$ gives a polynomial in $m'-m$
of degree at most $n-2$, since the integrals do not depend on $m'$.
It is an average of assignment probabilities and hence stays in $[0,1]$
for every nonnegative integer $m'-m$. A nonconstant polynomial is
unbounded on these integers. Comparing $m'=m$ with $m'=m+1$ therefore gives
\[
 q_{i1}^{m,*}(t)=\frac{1}{|\mathcal B|}\int_{\mathcal B} q_{i1}^{m+1,*}(t,u)\,\mathrm{d}u .
\]

For the chosen $t$, the difference $q_{i1}^{m+1,*}(t,u)-q_{i1}^{m,*}(t)$
thus integrates to zero over every rational box $\mathcal B$. On each bounded
rational box, its rational subboxes determine the finite signed
measure with this density. Exhausting the open orthant by such boxes
shows that the difference vanishes for almost every $u$, hence for
almost every $(t,u)$ by Fubini. Thus deleting the last column
preserves the probability for job~$1$ almost everywhere. Repeated
deletion, from $m$ columns down to the first, gives $q_{i1}^{m,*}(t)=q_{i1}^{1,*}(t_{\cdot1})$
almost everywhere for every $m$.

It remains to read off the kernels. At $m=1$, \eqref{eq:lower:4.7}
gives $q_{i1}^{1,*}=h_{i,0}$. At $m=k+1$, if the kernels of orders
$1,\ldots,k-1$ vanish almost everywhere, their lifts do as well.
The expansion then consists of $h_{i,0}$ and a single order-$k$
term involving all $k$ other jobs. Deletion gives
$h_{i,0}$ for the whole probability, so that term vanishes almost
everywhere. Induction proves the claim for $1\le k\le n-2$.
\end{proof}

\begin{proof}[Completion of the proof of Theorem~\ref{thm:locality}]
Lemmas~\ref{lem:padding} and~\ref{lem:lower:measurable} give measurable
assignment distributions with the required symmetries and approximation
guarantee. Lemma~\ref{lem:lower:laplace} gives the degree bound in
\eqref{eq:lower:4.7}, and Lemma~\ref{lem:lower:collapse} leaves only
$h_{i,0}=q_{i1}^{1,*}$. Choose a $[0,1]$-valued representative $g_i$ of
this one-job probability. Then $q_{i1}^{m,*}(t)=g_i(t_{\cdot1})$ almost
everywhere for every $m$, and job equivariance gives the identity for
every $j$.
\end{proof}

\section{Proof of the correlated upper bound}\label{sec:generalupper}
We first verify that the mechanism of Section~\ref{overview:upper}
realizes its assignment probabilities and payments on every report.
We then prove the three load estimates in
Proposition~\ref{prop:upper-estimates} and the refinement for seven
machines. Throughout, $\mu_i=\E[L_i]$ and $T=\OPT(t)$. The fixed optimal
assignment is used only in the analysis.

\subsection{The allocation rule and payments}\label{upper:allocation-marginals-and-all-report-truthfulness}

We use the Bounded-Square probabilities, decreasing-cost slots, and
matching distribution specified in Section~\ref{overview:upper}.
The following lemma applies the classical monotone-payment
characterization~\cite[Theorems~4.2 and~4.4]{AT01} job by job, and also
serves the two- and three-machine constructions. Its hypotheses concern
the $q_{ij}$; the joint assignment law may depend on the whole report.

\begin{lemma}[Payments for column-local probabilities]\label{lem:local-payments}
Suppose $q_{ij}(t)\ge0$, $\sum_iq_{ij}(t)=1$, and $q_{ij}$ depends only
on column $t_{\cdot j}$. Fixing the other entries in that column, suppose
$q(s)=q_{ij}(s,t_{-ij})$ is nonincreasing on $[0,\infty)$.
Any assignment distribution realizing these probabilities is truthful
in expectation with per-job expected payment
\[
 \pi(s)=sq(s)-\int_0^s q(u)\operatorname{d}u+C,
\]
where $C$ is finite and independent of machine $i$'s entire report.
In particular, if $\int_0^\infty q(u)\operatorname{d}u<\infty$, one may take
\begin{equation}\label{eq:local-payments}
 \pi(s)=sq(s)+\int_s^\infty q(u)\operatorname{d}u.
\end{equation}
Sum the per-job expectations to obtain each machine's expected payment.
These payments and the expected utilities at every finite true type
and finite deviation are finite.
\end{lemma}
\begin{proof}
Fix the opponents' reports. For a true scalar cost $v$ and report $s$,
the gain from reporting truthfully is
\[
 [\pi(v)-vq(v)]-[\pi(s)-vq(s)]
 =\int_v^s\bigl(q(u)-q(s)\bigr)\operatorname{d}u\ge0.
\]
If $s\ge v$, monotonicity makes the integrand nonnegative on $[v,s]$.
If $s<v$, it makes the integrand nonpositive on $[s,v]$, and reversing
the integration limits gives the same inequality. This also covers jumps.
Since each term depends on the machine's report only through its own
entry for that job, summing proves the incentive inequality for any
simultaneous deviation of the entire row. Correlations do not enter this
sum. Boundedness of $q$ gives finite payments in the first formula;
integrability gives finiteness in the second. The realized processing
cost is bounded by the sum of the finite entries of the true row.
\end{proof}

The lemma specifies expectations. Covering the reported processing cost
in each outcome is a property of the realized payment rules
in Sections~\ref{n3:payments} and~\ref{sec:implementation}.

\subsection{Rounding and truthfulness}\label{upper:marginal-preserving-correlated-rounding}

\paragraph{A finite distribution over matchings.}
Whenever the residual matrix has a common positive row and column sum,
each set of rows has at least as many neighboring columns as rows.
Otherwise those columns could not hold the rows' total weight. Hall's
theorem gives a supported perfect matching. Choose one by a fixed
augmenting-path algorithm with ties broken in row and slot order, and
subtract its smallest edge weight from every matched entry. The row
and column sums decrease by the same amount, and at least one positive
entry disappears. Iterating
gives a convex combination of matchings whose weights sum to one.
Drawing a matching with these weights and deleting the dummies sends
each job $j$ to machine $i$ with probability $q_{ij}$. The load bounds
below hold for every such decomposition.

\paragraph{Monotonicity, including the cutoff points.}
Fix a job $j$, a machine $i$, and the opponents' reports for this job.
Write $q(s)=q_{ij}(s,t_{-ij})$ for its assignment probability when
machine $i$ reports $s$. Its eligibility cutoff is
$B=2\min_{k\ne i}t_{kj}$. Suppose first that $B>0$. For $0<s\le B$,
machine $i$ is eligible and the cutoff for its competitors is
$\min\{2s,B\}$. Thus
\[
 q(s)=\left[1+s^2
       \sum_{\substack{k\ne i\\t_{kj}\le\min\{2s,B\}}}t_{kj}^{-2}\right]^{-1}.
\]
As $s$ increases, the summation set only grows and each term
$s^2t_{kj}^{-2}$ is nondecreasing. The denominator is therefore
nondecreasing, including its jumps when competitors enter, so $q(s)$
is nonincreasing on $(0,B]$. At zero, $q(0)=1$, and for $s>B$,
$q(s)=0$. If $B=0$, then $q(s)=0$ for every $s>0$, while $q(0)$ is the reciprocal
of the number of zero-cost machines. Thus $q$ is nonincreasing in this
case as well.

\paragraph{Apply the payment lemma.}
The Bounded-Square $q_{ij}$ are column-local and nonincreasing by the
preceding calculation. For $0\le s\le B$ the expected payment satisfies
\[
 0\le\pi(s)=sq(s)+\int_s^B q(u)\operatorname{d}u
 \le s+(B-s)=B.
\]
At the cutoff the tail integral is zero, so $\pi(B)=Bq(B)$ even if
the machine receives the job with positive probability. For $s>B$ the
payment is zero. Lemma~\ref{lem:local-payments} gives truthfulness in
expectation and finite payments for every finite whole-row deviation.
For the empty instance, the assignment and all payments are empty or zero.

\subsection{First estimate: the range of each load}\label{upper:exact-load-support-width}

The cost ordering prevents the widths of successive slots from
covering the same interval of processing times twice.

\begin{lemma}[Range of a load]\label{lem:width}
For the distribution produced by slot rounding, machine $i$'s load is supported in an interval of width at most $d_i=\max\{t_{ij}:q_{ij}>0\}$, with $d_i=0$ if $q_{ij}=0$ for every job $j$. If $T=\OPT(t)>0$, then $d_i\le2T$.
\end{lemma}
\begin{proof}
If $q_{ij}=0$ for every $j$, the load is zero. Otherwise, fix machine $i$
and order its fractional job list by nonincreasing processing time.
For each slot $k$, let $\overline t_k$ and $\underline t_k$ be the
largest and smallest processing times among the pieces of positive
length in that slot, including any zero-cost dummy padding. Every
supported matching selects one such piece, so its contribution to the
load lies in $[\underline t_k,\overline t_k]$.

The ordering gives $\overline t_{k+1}\le\underline t_k$ for successive
slots. Their cost intervals therefore have disjoint interiors and all
lie in $[0,d_i]$. Every supported load lies between
$\sum_k\underline t_k$ and $\sum_k\overline t_k$, so
\[
 \loadwidth(L_i)\le\sum_k(\overline t_k-\underline t_k)\le d_i.
\]
Finally, every column minimum is at most $T$, and a
positive $q_{ij}$ requires $t_{ij}$ to be at most twice that minimum.
Thus $d_i\le2T$.
\end{proof}

\subsection{Second estimate: the joint geometry of the mean loads}\label{upper:joint-mean-load-inequality}

To bound $\lambda\cdot\mu$, compare each job's contribution with the
time it takes on its machine in the fixed optimum.

\begin{lemma}[Joint mean-load inequality]\label{lem:meanball}
Let $\mu_i=\sum_jt_{ij}q_{ij}$ and $T=\OPT(t)$. For every $\lambda\in[0,\infty)^n$,
\[
\lambda\cdot\mu\le\frac T2\left(\sum_i \lambda_i+n\lVert \lambda\rVert_2\right).
\]

\end{lemma}
\begin{proof}
Fix a machine $i$ and a job $j$ with $A^*(j)=i$ and positive minimum
processing time. Normalize the job's contributions to expected load by
its processing time on machine $i$, so $v_k=t_{kj}q_{kj}/t_{ij}$.
For each eligible machine $k$, the inverse-square rule gives
$v_k=t_{kj}^{-1}/(t_{ij}\sum_{\ell\in S_j}t_{\ell j}^{-2})$.
Squaring and summing over machines gives
\[
 \lVert v\rVert_2^2
 =\frac1{t_{ij}^2\sum_{k\in S_j}t_{kj}^{-2}}.
\]
If $i\in S_j$, the right side is $v_i$. Completing the square gives
$\lVert v-e_i/2\rVert_2=1/2$, with $e_i$ the $i$th coordinate vector.
Hence
\[
 \lambda\cdot v
 =\frac{\lambda_i}2+\lambda\cdot(v-e_i/2)
 \le\frac{\lambda_i+\lVert\lambda\rVert_2}{2}.
\]
If $i\notin S_j$, then $t_{ij}>2\min_k t_{kj}$ and
$\sum_{k\in S_j}t_{kj}^{-2}\ge(\min_k t_{kj})^{-2}$.
Hence $\lVert v\rVert_2\le(\min_k t_{kj})/t_{ij}<1/2$.
The same bound follows from
$\lambda\cdot v\le\lVert\lambda\rVert_2\lVert v\rVert_2$
and $\lambda_i\ge0$. Multiplying by $t_{ij}$ gives
\[
 \sum_k\lambda_k t_{kj}q_{kj}
 \le\frac{t_{ij}}2(\lambda_i+\lVert\lambda\rVert_2).
\]
A zero-minimum job contributes zero and satisfies this unscaled
inequality as well, even if its cost in the optimum is positive.

Group the jobs by their machines in the fixed optimum. Each group
takes at most $T$ time on its machine, so
\[
\lambda\cdot\mu
 \le\frac12\sum_i
       \left(\sum_{j:A^*(j)=i}t_{ij}\right)
       (\lambda_i+\lVert\lambda\rVert_2) 
 \le\frac T2\left(\sum_i\lambda_i+n\lVert\lambda\rVert_2\right).
\]
\end{proof}

\subsection{Third estimate: from load geometry to expected makespan}\label{upper:expected-actual-maximum}

The mean estimate bounds $\theta\cdot\mu$. To bound the expected makespan,
we also need to account for how much higher a machine's load can be when
it is selected as a maximizer. The next lemma bounds this contribution
using only the machine's load range and its probability of being selected.

\begin{lemma}[Covariance bound for a maximum]\label{lem:covariance}
Let $L_1,\ldots,L_n$ be jointly distributed bounded loads and put $\mu_i=\E[L_i]$.
Choose a maximizing machine $I$ using a fixed tie-breaking rule and
write $\theta_i=\Prb(I=i)$. Then
\[
\E[\max_i L_i]\le \theta\cdot\mu+\sum_i\loadwidth(L_i)\theta_i(1-\theta_i).
\]

\end{lemma}
\begin{proof}
Let $\underline L_i$ be the lower endpoint of the smallest closed
interval supporting $L_i$. Since $\E[\ind\{I=i\}-\theta_i]=0$,
subtracting this endpoint gives
\[
 \E[\ind\{I=i\}L_i]-\theta_i\mu_i
 =\E\bigl[(\ind\{I=i\}-\theta_i)(L_i-\underline L_i)\bigr]
 \le \theta_i(1-\theta_i)\loadwidth(L_i).
\]
Indeed, the product is nonpositive on $\{I\ne i\}$. On $\{I=i\}$ it
is at most $(1-\theta_i)\loadwidth(L_i)$, and this event has probability $\theta_i$.
Summing over $i$ proves the claim because
$\max_i L_i=\sum_i\ind\{I=i\}L_i$ in every outcome.
\end{proof}
This proves the last of the three estimates in
Proposition~\ref{prop:upper-estimates}. Use the maximizer probabilities
$\theta_i$ as weights in the mean bound and apply
$\loadwidth(L_i)\le 2T$ in the covariance bound. The result
is~\eqref{eq:upper:quadratic}, which Section~\ref{overview:upper}
maximizes over $\tnorm$ for $n\ge8$ and $n=4,5,6$. If $T=0$, every job has a
zero-cost machine and is assigned to one with probability one.

\subsection{A sharp instance for the construction}\label{upper:sharpness-and-small-n-scope}

Return to the star of Section~\ref{overview:lower}. For this explicit
rule the assignment probabilities are defined at ties, so we can use
the unperturbed star. Set all cheap processing times to one and all
others to three. The root is cheap for every job, and each leaf is
cheap only for its own spoke. The diagonal schedule has makespan one
and is optimal. Bounded-Square gives the root the loop with
certainty and each of the $n-1$ spokes with probability one-half.
In every outcome the root's load is at least one and each leaf's load
is at most one. The expected makespan therefore equals the root's
expected load, $(n+1)/2$, under every coupling. This proves sharpness
for the construction wherever its upper bound is $(n+1)/2$.

\subsection{The seven-machine refinement}\label{upper7:appendix-a.-seven-machines-a-plane-slack-certificate}
At seven machines the right side of~\eqref{eq:upper:quadratic} equals
$4+1/32-2(\tnorm-7/8)^2$. To recover the possible $1/32$ excess, choose
a machine $i$ most likely to be selected as the maximizer. If its longest
eligible job takes at most $T=\OPT(t)$ there, the narrower load range
suffices. Otherwise that job must run on a different machine in the
optimal assignment. Its times on the two machines then restrict how much
it can contribute to the mean term $\theta\cdot\mu$.

\begin{proposition}[Seven machines]\label{prop:seven}
The Bounded-Square slot mechanism has expected makespan at most
$4\OPT$ on seven machines, for every finite job count and all finite
nonnegative processing times.
\end{proposition}
\begin{proof}
Divide all times by $T>0$. This preserves the Bounded-Square probabilities,
the slot ordering, and the matching law, and divides every load by $T$.
With the optimum now equal to one, the three load estimates give
\[
 \E[C_{\max}]\le\frac12+\frac72\tnorm+2(1-\tnorm^2).
\]
The excess over four is $2(1-\tnorm)(\tnorm-3/4)$, so only
$3/4<\tnorm<1$ needs attention. Choose a machine $i$ with maximal
selection probability $\theta_i$. Since $\sum_k\theta_k=1$,
$\tnorm^2\le\theta_i$ and $\theta_i>1/2$. Moreover,
\begin{equation}\label{eq:seven:1}
 \tnorm^2\le\theta_i^2+(1-\theta_i)^2,
 \qquad
 \frac{1+\sqrt{2\tnorm^2-1}}2\le\theta_i\le\tnorm.
\end{equation}
The first inequality bounds the sum of squares of the other
coordinates by $(\sum_{k\ne i}\theta_k)^2=(1-\theta_i)^2$.
Solving this quadratic on $\theta_i>1/2$ gives the lower endpoint;
the upper endpoint follows from the norm.

Recall that $d_i=\max\{t_{ij}:q_{ij}>0\}$, with $d_i=0$ for empty
support. Lemma~\ref{lem:seven-geometry} gives the desired bound when
$d_i\le1$. Otherwise, we combine the saving in the range bound for
machine $i$ with the saving in the mean bound for a job attaining $d_i$.
Lemma~\ref{lem:seven-elimination} reduces the required
improvement to the explicit inequality~\eqref{eq:seven:5}, and
Lemma~\ref{lem:seven-scalar} proves that inequality. Thus the expected
makespan is at most four. Rescale to arbitrary positive optimum; when
the optimum is zero, every supported edge has zero cost. Together with
Theorem~\ref{thm:lower}, this proves $\rho(7)=4$.
\end{proof}

\Needspace{10\baselineskip}
\subsubsection{A supported job improves both estimates}
\label{upper7:geometry}
\label{upper7:a.1-a-supported-job-gives-a-weighted-mean-deficit}
\begin{lemma}[Combined improvement]\label{lem:seven-geometry}
Let $n=7$. Normalize $T=1$, assume $3/4<\tnorm<1$, and choose a machine $i$
with $\theta_i=\max_\ell\theta_\ell$. If $d_i\le1$, then
$\E[C_{\max}]\le4$. Otherwise, choose a job $j$ with
$q_{ij}>0$ and $t_{ij}=d_i$, and let $k=A^*(j)$.
Then $k\ne i$, $d_i/2\le t_{kj}\le1$, and
\[
\begin{aligned}
 \E[C_{\max}]\le{}&\frac12+\frac72\tnorm+2(1-\tnorm^2)
                    -(2-d_i)\theta_i(1-\theta_i)\\
 &-\frac{t_{kj}\bigl(d_i\theta_i-t_{kj}(\tnorm+\theta_k)\bigr)_+^2}
 {4\bigl((d_i^2+t_{kj}^2)\tnorm-d_i t_{kj}\theta_i
                          +t_{kj}^2\theta_k\bigr)}.
\end{aligned}
\]
Here $(\cdot)_+=\max\{\cdot,0\}$, and the denominator is positive.
\end{lemma}
\begin{proof}
\textbf{The improvement in the range estimate.}
The width of $L_i$ is at most $d_i$, and the other machines' load
widths are at most two. Hence the sum of the covariance bounds is at most
\[
 d_i\theta_i(1-\theta_i)+2\sum_{\ell\ne i}\theta_\ell(1-\theta_\ell)
 =2(1-\tnorm^2)-(2-d_i)\theta_i(1-\theta_i).
\]
If $d_i\le1$, the improvement is at least
\[
 \theta_i(1-\theta_i)\ge\tnorm(1-\tnorm)
 \ge2(1-\tnorm)(\tnorm-3/4),
\]
because $1/2<\theta_i\le\tnorm\le1$ and
$2(\tnorm-3/4)\le\tnorm$. Thus assume $d_i>1$ and fix the job $j$
in the statement. The column minimum is at least $d_i/2$, whereas its
machine in the optimum processes it in time at most one. Therefore
$d_i/2\le t_{kj}\le1<d_i$, so $k\ne i$. Both machines are
eligible under Bounded-Square, since twice the column minimum is
at least $d_i>1$.

\paragraph{The restriction on this job's expected loads.}
Let $v_\ell=t_{\ell j}q_{\ell j}/t_{kj}$ be this job's normalized
mean contribution on machine $\ell$, as in Lemma~\ref{lem:meanball}.
The inverse-square probabilities give
\begin{equation}\label{eq:seven:2}
 \lVert v\rVert_2^2=v_k,\qquad
 \frac{v_i}{v_k}
 =\frac{d_iq_{ij}}{t_{kj} q_{kj}}
 =\frac{t_{kj}}{d_i}.
\end{equation}
Thus $v$ lies on the sphere of radius $1/2$ centered at $e_k/2$
and in the plane $d_i v_i=t_{kj}v_k$. Without the plane
constraint, the best point in direction $\theta$ lies one radius from
the center in that direction. Call it $v^{\mathrm{ball}}$. The loss at
the actual vector $v$ is exactly proportional to squared distance.
\begin{equation}\label{eq:seven:3}
 v^{\mathrm{ball}}=\frac12\left(e_k+\frac{\theta}{\tnorm}\right),
 \qquad
 \theta\cdot(v^{\mathrm{ball}}-v)=\tnorm\,\lVert v^{\mathrm{ball}}-v\rVert_2^2 .
\end{equation}
Indeed, $\lVert v\rVert_2^2=v_k$ and
$\lVert v^{\mathrm{ball}}\rVert_2^2=v_k^{\mathrm{ball}}$, so
expanding the squared distance gives the second identity. The value
$\theta\cdot v^{\mathrm{ball}}=(\theta_k+\tnorm)/2$ is the
normalized bound for this job in Lemma~\ref{lem:meanball}.
Figure~\ref{fig:seven-section} shows how the plane can force a loss.

\begin{figure}[!htbp]
\centering
\begin{tikzpicture}[x=1cm,y=1cm,>=stealth,font=\small]
\fill[black!4] (0,0) circle (1.3);
\draw (0,0) circle (1.3);
\draw[very thick] (-1.252,.35)--(1.252,.35);
\fill (0,0) circle (1.5pt) node[below] {$e_k/2$};
\draw[->,thick] (0,0)--(.78,1.04) node[pos=.55,left] {$\theta$};
\fill (.78,1.04) circle (1.7pt);
\node[anchor=south,align=center] at (.78,1.25) {ball maximizer\\$v^{\mathrm{ball}}$};
\fill (1.252,.35) circle (1.7pt);
\draw[->] (1.55,-1.20)--(1.30,.28);
\node[anchor=north,align=center] at (1.55,-1.25) {best point\\in the section};
\end{tikzpicture}
\caption{For job $j$, let $k=A^*(j)$ be its machine in the fixed optimum.
The disk depicts its load ball and the chord its section by $d_i v_i=t_{kj}v_k$. In this schematic the
ball maximizer in direction $\theta$ lies off the chord. For the actual
point $v$ on the sphere, \eqref{eq:seven:3} gives the loss in $\theta\cdot v$
in terms of its squared distance from $v^{\mathrm{ball}}$.}
\label{fig:seven-section}
\end{figure}
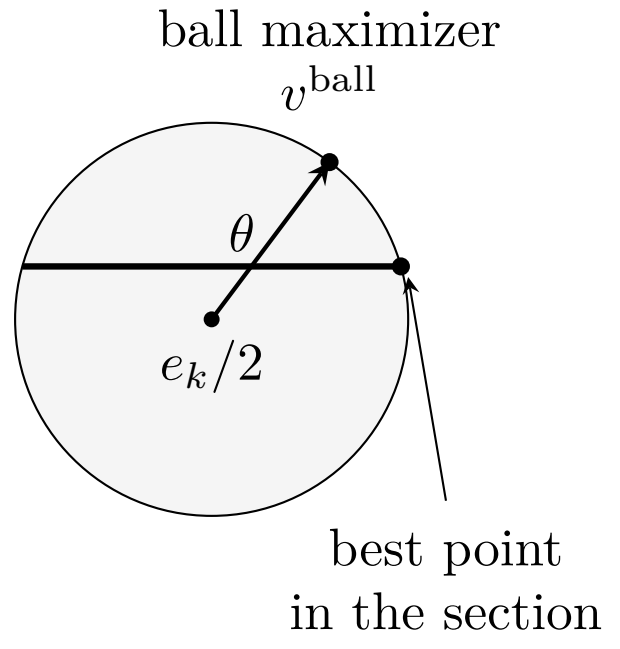

\paragraph{Quantifying the improvement in the mean estimate.}
To bound the squared distance in~\eqref{eq:seven:3}, evaluate the
plane equation at $v^{\mathrm{ball}}$ and subtract its value at $v$,
where it vanishes. At both points the $k$th coordinate equals the
squared norm. Thus $v_k^{\mathrm{ball}}-v_k=(v^{\mathrm{ball}}+v)
\cdot(v^{\mathrm{ball}}-v)$, giving
\[
 d_i v_i^{\mathrm{ball}}-t_{kj}v_k^{\mathrm{ball}}
 =\bigl(d_i e_i-t_{kj}(v^{\mathrm{ball}}+v)\bigr)
       \cdot(v^{\mathrm{ball}}-v).
\]
The squared norm of the first factor is bounded by
\[
 \bigl\lVert d_i e_i-t_{kj}(v^{\mathrm{ball}}+v)\bigr\rVert_2^2
 \le d_i^2-2t_{kj}
       \bigl(d_i v_i^{\mathrm{ball}}-t_{kj}v_k^{\mathrm{ball}}\bigr).
\]
To see this, expand the square, use
$\lVert v^{\mathrm{ball}}+v\rVert_2^2\le
2(v_k^{\mathrm{ball}}+v_k)$, and cancel the terms
involving $v$ by the plane equation. Now apply Cauchy--Schwarz to the
preceding inner product. Substituting
\[
 d_i v_i^{\mathrm{ball}}-t_{kj}v_k^{\mathrm{ball}}
 =\frac{d_i\theta_i-t_{kj}(\tnorm+\theta_k)}{2\tnorm}
\]
and multiplying by $t_{kj}\tnorm$ bounds the saving
$\Delta_{\mathrm{mean}}=t_{kj}\theta\cdot(v^{\mathrm{ball}}-v)$
in the selected job's contribution as follows.
\begin{equation}\label{eq:seven:4}
 \Delta_{\mathrm{mean}}\ge
 \frac{t_{kj}\bigl(d_i\theta_i-t_{kj}(\tnorm+\theta_k)\bigr)^2}
 {4\bigl((d_i^2+t_{kj}^2)\tnorm-d_i t_{kj}\theta_i
                         +t_{kj}^2\theta_k\bigr)}.
\end{equation}
Division is valid because the denominator, apart from the factor of
four, is at least $(d_i^2-d_i t_{kj}+t_{kj}^2)\tnorm>0$.
Replacing the square by the square of the positive part gives the
bound in the statement and permits the monotone comparison in the
next lemma.

Every other job retains its original mean bound, so
$\theta\cdot\mu\le(1+7\tnorm)/2-\Delta_{\mathrm{mean}}$.
Adding the range estimate proves the lemma. Both improvements use
the same probabilities $\theta$ and the same selected job $j$.
\end{proof}

\subsubsection{Eliminating the selected job's costs}
\label{upper7:elimination}
\label{upper7:a.2-eliminating-the-remaining-job-parameter}
\begin{lemma}[Endpoint reduction]\label{lem:seven-elimination}
Under the assumptions of Lemma~\ref{lem:seven-geometry}, it suffices
to prove, for every $1\le d_i\le2$,
\begin{equation}\label{eq:seven:5}
 (2-d_i)\theta_i(1-\theta_i)
 +\frac{\bigl((d_i+1)\theta_i-\tnorm-1\bigr)_+^2}
        {4\bigl((d_i^2+1)\tnorm+1-(d_i+1)\theta_i\bigr)}
 \ge2(1-\tnorm)(\tnorm-3/4).
\end{equation}
\end{lemma}
\begin{proof}
Keep $\theta$ fixed and write $k=A^*(j)$ for the selected job's machine
in the optimum. We minimize the lower bound from Lemma~\ref{lem:seven-geometry} over the cost region
$d_i\ge1$, $d_i/2\le t_{kj}\le1$. Along any ray in this region
the two costs scale by the same positive factor. The numerator of the
fraction scales cubically and its denominator quadratically.
Thus this lower bound on the mean saving scales linearly, and the range saving
$(2-d_i)\theta_i(1-\theta_i)$ is affine in $d_i$. Their sum takes its
minimum at one of the two ends of the ray segment. At the lower end
$d_i=1$, so the range term alone covers
$2(1-\tnorm)(\tnorm-3/4)$ by the proof of Lemma~\ref{lem:seven-geometry}.

At the upper end $t_{kj}=1$, and the mean saving is at least
\[
 \frac{\bigl(d_i\theta_i-\tnorm-\theta_k\bigr)_+^2}
        {4\bigl((d_i^2+1)\tnorm-d_i\theta_i+\theta_k\bigr)}.
\]
Since $k\ne i$, we have $\theta_k\le1-\theta_i$.
Increasing $\theta_k$ to $1-\theta_i$ decreases the numerator
and increases the positive denominator. This gives the fractional
term on the left of~\eqref{eq:seven:5}. Thus that inequality covers
the remaining endpoint and hence the whole segment.
\end{proof}

\subsubsection{The remaining scalar inequality}
\label{upper7:scalar}
\begin{lemma}[Scalar improvement]\label{lem:seven-scalar}
Inequality~\eqref{eq:seven:5} holds for every probability vector
$\theta$ with $3/4\le\tnorm\le1$, every machine $i$ with
$\theta_i=\max_k\theta_k$, and every $1\le d_i\le2$.
\end{lemma}
\begin{proof}
We seek a single lower bound valid for all $1\le d_i\le2$.
Apart from its factor of four, the denominator in~\eqref{eq:seven:5}
is at least one at $d_i=1$ since $\theta_i\le\tnorm$. Its derivative
$2d_i\tnorm-\theta_i$ is positive. Replacing it by its value
$5\tnorm+1-3\theta_i$ at $d_i=2$ therefore gives a lower bound on
the fraction.

For any slope $\alpha\ge0$, completing the square gives the affine bound
\[
 \frac{\bigl((d_i+1)\theta_i-\tnorm-1\bigr)_+^2}{4(5\tnorm+1-3\theta_i)}
 \ge \alpha\bigl((d_i+1)\theta_i-\tnorm-1\bigr)-\alpha^2(5\tnorm+1-3\theta_i).
\]
Indeed, the positive part in the numerator is at least its argument,
and the denominator is positive.
If $0\le\alpha\le1-\theta_i$, the terms involving $2-d_i$ combine
to give $(2-d_i)\theta_i(1-\theta_i-\alpha)\ge0$.
Thus the left-hand side of~\eqref{eq:seven:5} is at least
\begin{equation}\label{eq:seven:6}
 \alpha(3\theta_i-\tnorm-1)-\alpha^2(5\tnorm+1-3\theta_i).
\end{equation}
Take $\alpha=\min\{1/10,1-\tnorm\}$, which is admissible since
$\theta_i\le\tnorm$. Up to $\tnorm=9/10$ the slope is constant.
Above that threshold it supplies the factor $1-\tnorm$ in the required
improvement.

\Needspace{12\baselineskip}
\paragraph{When $\tnorm\le9/10$.}
\label{upper7:a.3-first-branch-a-positive-bernstein-certificate}
With $\alpha=1/10$, multiplying the desired bound by $100$ reduces it to
\[
 33\theta_i-15\tnorm-11\ge200(1-\tnorm)(\tnorm-3/4).
\]
Substitute the lower bound on $\theta_i$ from~\eqref{eq:seven:1}.
It suffices to prove
\[
 \frac{33}{2}\bigl(1+\sqrt{2\tnorm^2-1}\bigr)-15\tnorm-11-200(1-\tnorm)(\tnorm-3/4)\ge0 .
\]
At $\tnorm=3/4$ the left side is $(33/\sqrt2-23)/4>0$. Its derivative with respect to
$\tnorm$ is $\tnorm(400+33/\sqrt{2\tnorm^2-1})-365$.
Cauchy--Schwarz applied to $(2,1)$ and $(1,\sqrt{2\tnorm^2-1})$ gives
$\sqrt{10}\,\tnorm\ge 2+\sqrt{2\tnorm^2-1}$. Multiplication and the
arithmetic--geometric mean inequality give
\[
 \sqrt{10}\,\tnorm\left(400+\frac{33}{\sqrt{2\tnorm^2-1}}\right)
 \ge 833+400\sqrt{2\tnorm^2-1}+\frac{66}{\sqrt{2\tnorm^2-1}}
 \ge 833+40\sqrt{66}>365\sqrt{10}.
\]
The last comparison follows from $40\sqrt{66}>324$ and
$365\sqrt{10}<1157$. Thus the derivative is positive throughout the
interval.

\paragraph{When $\tnorm\ge9/10$.}
\label{upper7:a.4-second-branch-concavity-and-endpoint-bounds}
Here $\alpha=1-\tnorm$. To compare the largest coordinate with the
norm, use~\eqref{eq:seven:1} to obtain $2\theta_i(1-\theta_i)\le1-\tnorm^2$.
Since $\theta_i\ge\tnorm^2\ge81/100>4/5$ and $1+\tnorm\le2$, this
gives $1-\theta_i\le5(1-\tnorm)/4$, or
$\theta_i\ge(5\tnorm-1)/4$.

With this choice of $\alpha$, subtracting the right side of
\eqref{eq:seven:5} from~\eqref{eq:seven:6} leaves
\[
 (1-\tnorm)\bigl[3(2-\tnorm)\theta_i+5\tnorm^2-7\tnorm-1/2\bigr].
\]
Substituting the lower bound on $\theta_i$ bounds the bracket below
by $(5\tnorm^2+5\tnorm-8)/4$. This quadratic increases on $[9/10,1]$
and equals $11/80$ at $9/10$. The surplus is therefore nonnegative,
including at $\tnorm=1$. This proves~\eqref{eq:seven:5}.
\end{proof}


\section{The exact optimum for two machines}
\label{n2:section}\label{sec:n2}

For two machines, choosing the $q_{ij}$ and
choosing the joint distribution of assignments are separate steps.  The general two-job minimax
formulation of Kuryatnikova and Vera~\cite[Theorem~7, equation~(26)]{KV20}
provides the antecedent for the first problem.  Their two-job scalar problem concerns
universally truthful, monotone, task-independent, scale-free rules.
Here we optimize the coupling at each reported profile and then extend
the same probability function to every finite job count.  This extension
preserves truthfulness in expectation; its correlations may depend on
the reports.

\begin{theorem}[Exact two-machine optimum]\label{n2:main}
There is a uniquely specified analytic constant $R_*$ such that
\begin{equation}\label{n2:decimal-bound}
  1.5059957054321575104<R_*<1.5059957054363210457.
\end{equation}
It is defined by the rational function $\Psi$ and the analytic boundary root
$r_*$ in \eqref{n2:parameters}--\eqref{n2:boundary-equation}, through
$R_*=\Psi(r_*)$.  There is a truthful-in-expectation mechanism attaining
approximation ratio $R_*$ on two machines for every finite number of jobs and
every finite nonnegative processing-time matrix.  Its expected payments are finite,
and every assignment in its support has zero makespan whenever
$\OPT=0$. The optimal approximation ratio is
\[
                         \rho(2)=R_*.
\]
\end{theorem}

The proof combines the analytic construction in
Lemma~\ref{n2:curve-certificate} with the exact rational verification
in Theorem~\ref{n2:verification}. Theorem~\ref{n2:owner-transfer}
extends the upper bound to every finite job count, and
Lemma~\ref{n2:locality-bridge} obtains the matching lower bound from
Theorem~\ref{thm:locality}. Section~\ref{sec:implementation} gives an
efficient implementation based on a piecewise rational probability function.

\subsection{Optimal coupling and the scalar objective}
\label{n2:scalar-subsection}

Throughout this section $n=2$. We first analyze two jobs and then
extend the construction to every finite $m$. The functional $R(f)$
below measures the performance of a fixed probability function $f$;
$\rho(2)$ is the optimum over all truthful families on two machines.
We start with strictly positive processing times.
Zero entries are handled in Theorem~\ref{n2:owner-transfer}.

For the rules studied in this subsection, the cost ratio $t_{1j}/t_{2j}$ determines the
probability that job $j$ goes to machine one. Let $\mathcal F$
consist of the nonincreasing functions
$f:(0,\infty)\to[0,1]$ satisfying
\begin{equation}\label{n2:reciprocity}
                         f(x)+f(1/x)=1.
\end{equation}
An increase in this ratio can only decrease the probability
$f(t_{1j}/t_{2j})$. Reciprocity expresses symmetry under exchange of
the machines and gives $f(1)=1/2$. The function may have jumps.

For a positive two-job profile $P=(t_{ij})\in(0,\infty)^{2\times2}$,
the rule specifies $q_{1j}=f(t_{1j}/t_{2j})$ for $j=1,2$.
Write $C_f(P)$ for the smallest expected makespan among all joint
assignment laws $y$ with these two probabilities. Thus we minimize over
probability vectors $(y_A)_{A\in\mathcal A_{2,2}}$ subject to the two
marginal constraints:
\[
 C_f(P)=\min_{y}\sum_{A\in\mathcal A_{2,2}}y_A C_{\max}(P,A),
 \qquad \sum_{A:A(j)=1}y_A=q_{1j}\quad(j=1,2).
\]
Its worst-case ratio on two-job profiles is
\begin{equation}\label{n2:R2-definition}
 R(f)=\sup_{P\in(0,\infty)^{2\times2}}\frac{C_f(P)}{\OPT(P)} .
\end{equation}
The next lemma expresses this ratio in terms of two scalar cost ratios
$s,t$. We later use it to bound the expected makespan for any finite
job count.

\begin{lemma}[Two-ratio reduction]\label{n2:scalar-reduction}
For $f\in\mathcal F$, define $\Phi_f(s,t)=1+s f(s)-(1+s-t)f(t)$.
Then the two-job approximation ratio is
\begin{equation}\label{n2:scalar-objective}
 R(f)=\sup_{0<s\le t<\infty}\Phi_f(s,t).
\end{equation}
\end{lemma}

The scalar objective is affine in the values of $f$.  Its solution has
two parts: a \emph{primal} rule for which every $\Phi_f(s,t)$ is at most
$R_*$, and a \emph{dual} argument showing that every competing rule has
a scalar supremum of at least $R_*$.  Theorem~\ref{n2:owner-transfer}
extends the upper bound to every finite job count.
Lemma~\ref{n2:locality-bridge} derives the matching lower bound from
unrestricted truthful-in-expectation families.  We first prove the
scalar reduction, including the optimal coupling on which it rests.

\paragraph{Optimize the joint assignment at a fixed profile.}
Once $q_{11}$ and $q_{12}$ are fixed, only one joint probability remains to choose.

\begin{lemma}[Optimal two-job coupling]\label{n2:frechet}
Fix a positive processing-time matrix $t\in(0,\infty)^{2\times2}$
and probabilities $q_{11},q_{12}\in[0,1]$ for assigning jobs $1,2$
to machine $1$. Among all joint assignments with these probabilities,
expected makespan is minimized when
\[
 \Prb(A(1)=A(2)=1)=\max\{0,q_{11}+q_{12}-1\}.
\]
One minimizing assignment uses a single uniform $U\in[0,1]$:
job $1$ goes to machine $1$ when $U\le q_{11}$, and job $2$ goes to
machine $1$ when $U\ge1-q_{12}$. Each job goes to machine $2$ otherwise.
\end{lemma}
\begin{proof}
Write $y_{(1,1)}=\Prb(A(1)=A(2)=1)$, using the assignment-probability
notation $y_A$ from Section~\ref{sec:model}.
The prescribed $q_{11}$ and $q_{12}$ then determine the other three
probabilities. The following table lists all outcomes; in the first
column, $(i,k)$ means that job $1$ goes to machine $i$ and job $2$
goes to machine $k$.
\begin{center}
\begin{tabular}{ccc}
\toprule
Assignment $(A(1),A(2))$ & Probability & Makespan\\
\midrule
$(1,1)$ & $y_{(1,1)}$ & $t_{11}+t_{12}$\\[2pt]
$(1,2)$ & $q_{11}-y_{(1,1)}$ & $\max\{t_{11},t_{22}\}$\\[2pt]
$(2,1)$ & $q_{12}-y_{(1,1)}$ & $\max\{t_{21},t_{12}\}$\\[2pt]
$(2,2)$ & $1-q_{11}-q_{12}+y_{(1,1)}$ & $t_{21}+t_{22}$\\
\bottomrule
\end{tabular}
\end{center}
The second row, for example, has probability $q_{11}-y_{(1,1)}$ because its
probability plus the first row's equals $\Prb(A(1)=1)=q_{11}$.
All four probabilities are nonnegative exactly when
\[
 \max\{0,q_{11}+q_{12}-1\}\le y_{(1,1)}\le\min\{q_{11},q_{12}\}.
\]
Multiplying each makespan by its probability gives
\[
 \begin{aligned}
 \E[C_{\max}]
 &=y_{(1,1)}(t_{11}+t_{12})+(q_{11}-y_{(1,1)})\max\{t_{11},t_{22}\}\\
 &\quad +(q_{12}-y_{(1,1)})\max\{t_{21},t_{12}\}
       +(1-q_{11}-q_{12}+y_{(1,1)})(t_{21}+t_{22}).
 \end{aligned}
\]
This is affine in $y_{(1,1)}$. Its coefficient of $y_{(1,1)}$ is strictly positive:
\[
 \sum_{i,j}t_{ij}-\max\{t_{11},t_{22}\}-\max\{t_{21},t_{12}\}
 =\min\{t_{11},t_{22}\}+\min\{t_{21},t_{12}\}>0.
\]
The minimum therefore occurs at the smallest feasible value of $y_{(1,1)}$.
For the stated threshold rule, the two events have probabilities
$q_{11}$ and $q_{12}$, and both occur precisely on
$[0,q_{11}]\cap[1-q_{12},1]$. This interval has length
$\max\{0,q_{11}+q_{12}-1\}$, as required.
\end{proof}

\paragraph{Normalize the worst-case profile.}
We next remove the two column scales and use machine symmetry to order
the remaining ratios.

\begin{proof}[Proof of Lemma~\ref{n2:scalar-reduction}]
For scalar ratios $0<s\le t$, consider the processing-time matrix
\[
 P=\begin{pmatrix}1&t\\s&1\end{pmatrix}.
\]
Here $q_{11}=f(1/s)=1-f(s)$ and $q_{12}=f(t)$. Since $f$ is
nonincreasing, $q_{11}+q_{12}\le1$, so Lemma~\ref{n2:frechet} gives
zero probability to $(1,1)$. The other assignments $(1,2)$, $(2,1)$
and $(2,2)$ have probabilities $1-f(s)$, $f(t)$ and $f(s)-f(t)$;
their makespans are $1$, $t$ and $1+s$, respectively. Therefore
\[
 C_f(P)=1-f(s)+t f(t)+(1+s)(f(s)-f(t))=\Phi_f(s,t).
\]
The assignment $(1,2)$ has makespan one, so $\OPT(P)\le1$ and
\[
 \Phi_f(s,t)=C_f(P)\le\frac{C_f(P)}{\OPT(P)}\le R(f).
\]

Conversely, fix a positive two-job profile $P=(t_{ij})$ and an optimal
assignment of makespan $T$. If the optimum splits the jobs, relabel the
columns so that job $j$ goes to machine $j$. Multiply column $j$ by
$T/t_{jj}\ge1$. Both cost ratios and hence both assignment probabilities
stay fixed, while the makespan of every assignment can only increase.
The diagonal assignment still has makespan $T$, so the optimum remains
$T$. Dividing all entries by $T$ now gives the matrix displayed above,
with $s=t_{21}/t_{11}$ and $t=t_{12}/t_{22}$. If $s>t$, exchanging
both machines and both jobs interchanges the ratios and preserves the
coupling value by reciprocity. We may therefore take $s\le t$.
The normalized matrix has coupling value $\Phi_f(s,t)$, which gives
$C_f(P)/T\le\Phi_f(s,t)$.

Suppose instead that the selected optimum puts both jobs on one
machine. For a job whose slower machine takes $r\ge1$ times as long as its faster machine, let $s\downarrow0$ in $\Phi_f(s,r)$.
Since $0\le s f(s)\le s$, this gives
\begin{equation}\label{n2:one-job-ratio}
 1+(r-1)f(r)\le\sup_{0<s\le t}\Phi_f(s,t).
\end{equation}
The left side is the one-job expected-cost ratio when machine one is
slower. Reciprocity gives the same bound when machine two is slower.
Thus each job's expected assigned cost is at most this scalar supremum
times its cost on the selected owner. The two owner costs sum to $T$,
and the total assigned cost bounds the makespan in every outcome.
Consequently
\[
 \frac{C_f(P)}T\le\sup_{0<s\le t}\Phi_f(s,t).
\]
This proves the reverse inequality in both cases.
\end{proof}

The piecewise family in \cite[Definition~2]{KV20} restricts the general
optimization: Proposition~4 and equations~(29)--(30) give an upper bound,
not a new characterization over all cumulative distribution functions.
To compare the integrands, take a continuous $f\in\mathcal F$ with
$R(f)<\infty$ and put $F=1-f$. Then $F$ is a nondecreasing threshold
cumulative distribution function, whereas $f$ is nonincreasing.
Under $(x,y)=(1/s,t)$, the integrand in their equation~(30) becomes
\[
 t-s+(1+s-t)F(t)+sF(1/s)
 =1+s f(s)-(1+s-t)f(t)=\Phi_f(s,t),
\]
and $xy\ge1$ becomes $s\le t$. This identity uses reciprocity of $f$,
not a piecewise rational representation. Our definition also permits jumps;
the midpoint and smoothing arguments below handle them explicitly.
Their $n$ counts jobs; throughout this paper, $n$ counts machines and
$m$ counts jobs.

\subsection{A coupling for every finite job count}
\label{n2:owner-subsection}

The reduction below applies to any number of machines. It isolates
the convexity argument used here and again for three machines.
The owner labels are fixed before the allocation seed is drawn.

\begin{lemma}[Reduction to one representative per owner]\label{lem:representative}
Fix a positive processing-time matrix and an auxiliary assignment $A^{\mathrm{aux}}$ of
makespan $T_{\mathrm{aux}}>0$, with owner $A^{\mathrm{aux}}(j)$ for job $j$. Suppose one random seed
$\omega$ specifies every job's winner from its column and owner, and
scaling a column by a positive factor preserves its winner when the
seed and owner are held fixed. Write $L(\omega)=(L_i(\omega))_{i\in[n]}$ for the original
load vector, and let $v^j(\omega)$ be the load vector of column
$t_{\cdot j}/t_{A^{\mathrm{aux}}(j),j}$. Thus its $k$th coordinate is
$t_{kj}/t_{A^{\mathrm{aux}}(j),j}$ if that job is assigned to $k$, and zero otherwise.
For the proof, draw one representative $J_i$ per owner. These draws are
independent of each other and of $\omega$, with
\[
 \Prb(J_{i}=j)=\frac{t_{i j}}{T_{\mathrm{aux}}}\quad(A^{\mathrm{aux}}(j)=i),\qquad
 \Prb(J_{i}=\varnothing)=1-\frac{1}{T_{\mathrm{aux}}}\sum_{j:A^{\mathrm{aux}}(j)=i}t_{i j}.
\]
The dummy probability is nonnegative because each owner's load is at most $T_{\mathrm{aux}}$.
Write $\boldsymbol J=(J_i)_{i\in[n]}$ and set $v^{\varnothing}=0$. Then
\begin{equation}\label{eq:representative-bound}
 \frac{\E[C_{\max}]}{T_{\mathrm{aux}}}
 \le\E_{\boldsymbol J,\omega}\!\left[\left\|\sum_{i} v^{J_{i}}(\omega)\right\|_\infty\right].
\end{equation}
In particular, a uniform bound $R$ for every such representative tuple
gives $\E[C_{\max}]\le R T_{\mathrm{aux}}$.
\end{lemma}
\begin{proof}
Fix $\omega$. The normalized column has the same winner as job $j$
because its owner is unchanged. Multiplying $v^j(\omega)$ by
$t_{A^{\mathrm{aux}}(j),j}$ therefore recovers that job's contribution
to $L(\omega)$. Summing over jobs gives
\begin{equation}\label{eq:representative-identity}
 \frac{L(\omega)}{T_{\mathrm{aux}}}
 =\sum_{i}\sum_{j:A^{\mathrm{aux}}(j)=i}\frac{t_{i j}}{T_{\mathrm{aux}}} v^j(\omega)
 =\E_{\boldsymbol J}\left[\sum_{i} v^{J_{i}}(\omega)\mid\omega\right].
\end{equation}
Taking the maximum coordinate and applying Jensen's inequality gives
\[
 \frac{C_{\max}(\omega)}{T_{\mathrm{aux}}}
 \le\E_{\boldsymbol J}\!\left[\left\|\sum_{i}v^{J_{i}}(\omega)\right\|_\infty
       \,\middle|\,\omega\right].
\]
Averaging over $\omega$ proves \eqref{eq:representative-bound}.
Independence ensures that conditioning on any representative tuple leaves
the original seed law unchanged. All sums are finite. The construction
does not sample representatives; they are used only in the proof.
\end{proof}

The lemma requires the same seed for every representative tuple.
Choosing a separate favorable coupling for each tuple would not give
\eqref{eq:representative-identity}. Independence of the $q_{ij}$ from
the owner is a separate condition, used for incentives rather than
for this load identity.

\begin{theorem}[Two-job extremality]
\label{n2:owner-transfer}
For each $f\in\mathcal F$ with $R(f)<\infty$, there is a truthful-in-expectation mechanism for every finite number of jobs whose
approximation ratio is at most $R(f)$.  It assigns each positive-cost job $j$ to machine one with probability $f(t_{1j}/t_{2j})$, has finite expected payments, and has zero makespan in every supported outcome when $\OPT=0$.
\end{theorem}
\begin{proof}
First consider positive costs $t_{ij}$. Choose an optimal auxiliary
assignment $A^*$ at the reported profile, using fixed
deterministic tie-breaking. Its makespan is
\[
 T=\max\left\{\sum_{j:A^*(j)=1}t_{1j},\ \sum_{j:A^*(j)=2}t_{2j}\right\}=\OPT.
\]
Put $q_{1j}=f(t_{1j}/t_{2j})$.  For one uniform $U\in[0,1]$, assign
\begin{equation}\label{n2:owner-rule}
 A(j)=1\quad\Longleftrightarrow\quad
 \begin{cases}
 U\le q_{1j},&A^*(j)=1,\\
 U\ge1-q_{1j},&A^*(j)=2.
 \end{cases}
\end{equation}
The orientation depends on the selected optimal assignment, but $\Prb(A(j)=1)=q_{1j}$ for every job $j$.

For any representative pair $(j,k)$ with $A^*(j)=1$ and $A^*(k)=2$, the
normalized profile in Lemma~\ref{lem:representative} is
\begin{equation}\label{n2:synthetic-pair}
 P_{jk}=\begin{pmatrix}
 1&t_{1k}/t_{2k}\\ t_{2j}/t_{1j}&1
 \end{pmatrix}.
\end{equation}
Assigning each representative to its nominated owner has makespan one, so
$\OPT(P_{jk})\le1$. Scaling preserves the ratios and the winners in
\eqref{n2:owner-rule}. The two owner classes use opposite orientations
of the same $U$, hence Lemma~\ref{n2:frechet} gives
\[
 \E_U[C_{\max}(P_{jk})]=C_f(P_{jk})
 \le R(f)\OPT(P_{jk})\le R(f).
\]
A representative tuple may contain a dummy, even if both owner classes
are nonempty. For a single real representative with costs $(a,b)$,
\eqref{n2:one-job-ratio} and Lemma~\ref{n2:scalar-reduction} bound its
expected cost directly:
\begin{equation}\label{n2:one-job-bound}
 af(a/b)+b(1-f(a/b))\le R(f)\min\{a,b\}\le R(f),
\end{equation}
since its owner entry is one. Two dummies contribute zero. Thus every
representative tuple, including those for an empty owner class, obeys
the same bound. Lemma~\ref{lem:representative} now yields
\begin{equation}\label{n2:owner-bound}
 \E[C_{\max}]\le R(f)T=R(f)\OPT.
\end{equation}

For a job $j$ with a zero entry, assign it uniformly among the
machines reporting zero. Its assignment probabilities are
\begin{equation}\label{n2:zero-extension}
 q_{ij}(t)=\frac{\ind\{t_{ij}=0\}}
                 {\ind\{t_{1j}=0\}+\ind\{t_{2j}=0\}}\qquad(i=1,2).
\end{equation}
Remove these jobs before applying the positive-cost construction. Their removal leaves the integral optimum
unchanged: deleting them cannot increase it, and any assignment of the
remaining jobs extends at no cost.  They contribute no load in any
supported outcome.  In particular, $\OPT=0$ implies that all jobs have a
zero entry, and the mechanism then has makespan zero surely.  The empty
instance is treated in the same way.

It remains to verify truthfulness in expectation.  For fixed opponent reports,
each $q_{ij}$ depends only on column $t_{\cdot j}$ and is nonincreasing
in $t_{ij}$, including at the boundary in~\eqref{n2:zero-extension}.
For machine two on a positive column, reciprocity gives
$q_{2j}(t)=f(t_{2j}/t_{1j})$.
Give machine $i$ the deterministic payment
\begin{equation}\label{n2:payments}
 p_i(t)=\sum_{j=1}^{m}\left(
 t_{ij}q_{ij}(t)
 -\int_0^{t_{ij}}q_{ij}(s,t_{-ij})\operatorname{d}s\right).
\end{equation}
Use the first formula in Lemma~\ref{lem:local-payments}, with $C=0$.
It applies to every $f\in\mathcal F$ here, without an assumption on
the tail integral. It proves finite expected payments and truthfulness
in expectation against simultaneous deviations of the whole row.
The report-dependent orientation changes only correlations.
\end{proof}

\subsection{The reverse reduction from unrestricted mechanisms}
\label{n2:reverse-subsection}

Theorem~\ref{thm:locality} supplies local assignment probabilities
almost everywhere, whereas $R(f)$ takes a supremum over every positive
profile. The next lemma bridges these two domains. Scaling and symmetry
reduce the one-job probability to a function of the cost ratio; averaging
the limits from either side of each ratio then gives the required
probabilities at every profile, including discontinuities.

\begin{lemma}[Homogeneity and midpoint values]
\label{n2:homogeneity-midpoint}
Every unrestricted two-machine truthful-in-expectation family with a
finite uniform approximation ratio $S$ yields a function $f_{\mathrm{mid}}\in\mathcal F$
such that every positive two-job profile has a coupling with
$\Prb(A(j)=1)=f_{\mathrm{mid}}(t_{1j}/t_{2j})$ for $j=1,2$
and expected makespan at most $S\OPT$.
\end{lemma}
\begin{proof}
We first impose global scale invariance before taking the measurable
limit in Theorem~\ref{thm:locality}. Work in the common fixed-point set
of the free-padding operators on $\mathcal K_S$ from
Lemma~\ref{lem:padding}. It is nonempty, compact, and convex. Its
coordinates are the assignment laws $y^m$ and cost potentials $V_i^m$.
For $\beta>0$, scale these coordinates by
\[
 (\mathsf S_\beta y)^m(t)=y^m(\beta t),\qquad
 (\mathsf S_\beta V)_i^m(t)=\frac{V_i^m(\beta t)}{\beta}.
\]
This map is continuous and affine in the pointwise product topology.
Applying~\eqref{eq:potential-support} at the scaled reports and dividing
by $\beta$ gives the support inequality for the new potentials. The
corresponding normalized payment is $\bar p_i^m(\beta t)/\beta$.
Both the makespan and the optimum scale by $\beta$, so the approximation
guarantee is preserved. For a free column $c$, substituting in the
padding formula~\eqref{eq:lower:padding-potential} gives
\[
                         \mathsf P_c\mathsf S_\beta=\mathsf S_\beta\mathsf P_{\beta c}.
\]
Since $\beta c$ is free, $\mathsf S_\beta$ preserves the common
fixed-point set. The scaling operators commute, so the Markov--Kakutani
theorem gives a family fixed by all of them. Averaging this
family over the two machine permutations preserves all padding and
scaling identities and makes the family machine-equivariant.

For this pointwise family, put $f(r)=q_{11}^1(r,1)$.
Global homogeneity gives
\[
                           q_{11}^1(a,b)=f(a/b)\qquad(a,b>0).
\]
One-job incentive compatibility makes $f$ nonincreasing; feasibility
and machine equivariance imply $f(r)+f(1/r)=1$.  In particular, $f$
is measurable and has only countably many discontinuities.

Apply the grid construction and common limit from the proof of
Theorem~\ref{thm:locality} to this chosen family.  At any positive
$(a,b)$ for which $f$ is continuous at $a/b$, the value $q_{11}^1(G_h(a,b))$ converges to $f(a/b)$, because the ratio of the grid-rounded
coordinates converges to $a/b$.  The exceptional profiles lie on
countably many ratio rays and have two-dimensional measure zero.
Bounded convergence on finite boxes identifies the extracted one-job
limit $g_1(a,b)$ with $f(a/b)$ almost everywhere. The same one-column
function $g_1$ occurs at every job count by Theorem~\ref{thm:locality}.
Thus at almost every positive
two-job matrix $t$,
\[
 q_{1j}^{2,*}(t)=f(t_{1j}/t_{2j})\qquad(j=1,2).
\]
The same limiting assignment distribution satisfies the approximation
inequality almost everywhere. We next extend these two conclusions
together to every positive profile.

Because $f$ is nonincreasing and bounded, its left and right limits
$f(r-)$ and $f(r+)$ exist at every $r>0$. Define
\[
 f_{\mathrm{mid}}(r)=\frac{f(r-)+f(r+)}2.
\]
This function is nonincreasing. It is also reciprocal: inversion
reverses the side of approach, so reciprocity of $f$ gives
\[
 f(r-)+f((1/r)+)=1,\qquad f(r+)+f((1/r)-)=1.
\]
Fix a positive two-job profile $P=(t_{ij})$. Approach it once with
both ratios below their target values and once with both above.
Either approach is possible within the full-measure set just obtained:
scaling the first row of $P$ by a factor just below or above one makes
both comparisons strict, and they remain strict on an open ball about
the scaled profile. Such balls can be taken arbitrarily close to $P$.

On each side, choose a sequence of profiles from these balls converging
to $P$. The joint assignment laws have convergent subsequences in the
simplex on the four assignments. Denote the two limits by $y^-$ and
$y^+$. Every assignment's makespan, and hence the optimum, is
continuous in the profile. Thus both laws have expected makespan at
most $S\OPT(P)$ at $P$. For job $j$, their machine-one probabilities
are the left and right limits of $f$ at $t_{1j}/t_{2j}$. Their
half-and-half mixture therefore has marginal
\[
 \sum_{A:A(j)=1}\frac{y_A^-+y_A^+}{2}
 =f_{\mathrm{mid}}(t_{1j}/t_{2j}) \qquad (j=1,2)
\]
and still satisfies the same makespan bound. This argument applies
also when the two target ratios coincide.
\end{proof}

\begin{lemma}[Reduction to the scalar problem]\label{n2:locality-bridge}
Theorem~\ref{thm:locality} and Theorem~\ref{n2:owner-transfer} imply
\begin{equation}\label{n2:rho-scalar}
 \rho(2)=\inf_{f\in\mathcal F}R(f)
 =\inf_{f\in\mathcal F}\sup_{0<s\le t}\Phi_f(s,t).
\end{equation}
\end{lemma}
\begin{proof}
Given a truthful family of ratio $S$, Lemma~\ref{n2:homogeneity-midpoint}
gives a rule with $R(f_{\mathrm{mid}})\le S$. Taking the infimum over
families gives $\inf_{f\in\mathcal F}R(f)\le\rho(2)$.
Conversely, Theorem~\ref{n2:owner-transfer} constructs a truthful family
of ratio at most $R(f)$ for every $f$ with finite $R(f)$. This gives
$\rho(2)\le\inf_{f\in\mathcal F}R(f)$. Lemma~\ref{n2:scalar-reduction}
identifies $R(f)$ with the supremum of $\Phi_f$.
\end{proof}

\paragraph{Attainment and continuous approximation.}
For completeness, the scalar infimum is attained even without the
explicit optimizer constructed below.  The class $\mathcal F$ is closed
in the compact product $[0,1]^{(0,\infty)}$, and each inequality
$\Phi_f(s,t)\le S$ is pointwise closed.  A finite sublevel is nonempty:
the rule that selects the strictly faster machine and splits ties has
factor at most two, since the sum of minimum job costs is at most the sum
of the two optimal loads.  Compactness therefore gives attainment.
There is also no infimum gap from imposing continuity. For $\sigma>0$,
define the logarithmic average
\[
 f_\sigma(r)=\frac1{2\sigma}\int_{-\sigma}^{\sigma}f(re^u)\operatorname{d}u
 =\frac1{2\sigma}\int_{\log r-\sigma}^{\log r+\sigma}f(e^v)\operatorname{d}v.
\]
The second formula is a moving average of a bounded measurable function,
so it is continuous in $\log r$. Monotonicity is inherited from $f$;
the substitution $u\mapsto-u$ and reciprocity of $f$ give
$f_\sigma(r)+f_\sigma(1/r)=1$. Hence $f_\sigma\in\mathcal F$.

Fix a positive two-job matrix $P=(t_{ij})$. For $u\in[-\sigma,\sigma]$, let
$P^{(u)}$ be obtained by multiplying its first row by $e^u$. Use the
optimal $f$-coupling from Lemma~\ref{n2:frechet} at $P^{(u)}$, but
evaluate its makespan at $P$. For every assignment $A$,
\[
 C_{\max}(P,A)\le\max\{1,e^{-u}\}C_{\max}(P^{(u)},A),\qquad
 \OPT(P^{(u)})\le\max\{1,e^u\}\OPT(P).
\]
Therefore its expected makespan on the original profile is at most
\[
 \max\{1,e^{-u}\}C_f(P^{(u)})
 \le e^{|u|}R(f)\OPT(P)\le e^\sigma R(f)\OPT(P).
\]
Mixing these couplings uniformly over $u\in[-\sigma,\sigma]$ gives
$\Prb(A(j)=1)=f_\sigma(t_{1j}/t_{2j})$, so
\begin{equation}\label{n2:smoothing}
 R(f_\sigma)\le e^\sigma R(f).
\end{equation}
Letting $\sigma\downarrow0$ proves equality of the infima over continuous
rules and all rules in $\mathcal F$.

\subsection{A lower bound from tight constraints}
\label{n2:dual-subsection}

\begin{figure}[hbp]
\centering
\begin{tikzpicture}[x=1.45cm,y=1.45cm,>=stealth]
\fill[black!5] (0,0)--(1,1)--(0,1)--cycle;
\fill[black!11] (0,1)--(1,1)--(1,2.8)--(0,2.8)--cycle;
\fill[black!3] (1,1)--(2.8,2.8)--(1,2.8)--cycle;
\draw[->] (0,0)--(3,0) node[right] {$s$};
\draw[->] (0,0)--(0,3) node[above] {$t$};
\draw (0,0)--(2.8,2.8) node[right] {$s=t$};
\draw[dashed] (0,1)--(1,1)--(1,2.8);
\node[below] at (1,0) {$1$};
\node[left] at (0,1) {$1$};
\node at (.33,.70) {\textup{(i)}};
\node at (.5,1.95) {\textup{(ii)}};
\node at (1.54,2.35) {\textup{(iii)}};
\end{tikzpicture}
\caption{The scalar domain $0<s\le t$. Regions \textup{(i)},
\textup{(ii)}, and \textup{(iii)} correspond to $t\le1$, $s\le1\le t$,
and $s\ge1$, respectively. The unbounded regions are truncated.}
\label{fig:ratio-domains}
\end{figure}
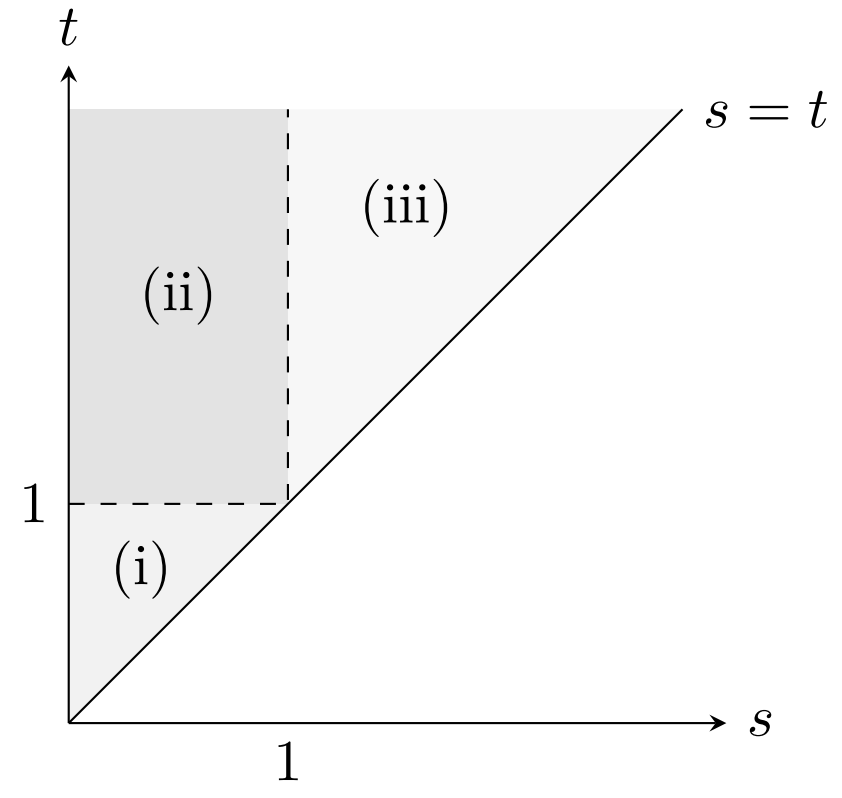

To prove that a candidate $f_0$ attains a value $\Gamma$, we must
bound $\Phi_{f_0}(s,t)$ by $\Gamma$ in all three regions of
Figure~\ref{fig:ratio-domains}. The matching lower bound needs only
a sequence of pairs in the middle region. There, writing the pair
as $(s,1/c)$ expresses both job probabilities through the restriction
of the rule to $(0,1]$. If this pair is tight and the next pair
starts at $c$, reciprocity makes the two constraints involve the same
unknown value $f(c)$ of any competing rule. The weights in the next
lemma are chosen to cancel that value.

A \emph{cross pair} has $s\le1\le t$. A \emph{contact} is a pair at
which $\Phi_{f_0}(s,t)=\Gamma$.

\begin{lemma}[A telescoping lower bound]\label{n2:contact-chain}
Let $f_0\in\mathcal F$ and $\Gamma\in\mathbb R$. Suppose a sequence
of first ratios $s_0=1$, $s_k\in(0,1]$, gives contacts
\begin{equation}\label{n2:dual-hypotheses}
 \Phi_{f_0}(s_k,1/s_{k+1})=\Gamma \qquad(k\ge0).
\end{equation}
Starting at $w_0=1$, choose successive weights by
\[
 w_{k+1}s_{k+1}=w_k(1/s_{k+1}-1-s_k).
\]
If the weights are positive and $w_ks_k\to0$, then
$R(f)\ge\Gamma$ for every $f\in\mathcal F$. If also $R(f_0)\le\Gamma$,
then $f_0$ attains the scalar minimum $\Gamma$.
\end{lemma}
\begin{proof}
Fix $f\in\mathcal F$ with $R(f)<\infty$. Write
$\Delta_k=f(s_k)-f_0(s_k)$. At the second ratio $1/s_{k+1}$,
reciprocity gives the difference $-\Delta_{k+1}$. Subtracting the
candidate's equality from the competitor's objective therefore gives
\begin{equation}\label{n2:dual-row}
 R(f)-\Gamma\ge
 \Phi_f(s_k,1/s_{k+1})-\Phi_{f_0}(s_k,1/s_{k+1})
 =s_k\Delta_k-(1/s_{k+1}-1-s_k)\Delta_{k+1}.
\end{equation}
After multiplication by $w_k$, the right side is
$w_ks_k\Delta_k-w_{k+1}s_{k+1}\Delta_{k+1}$. Summing from $0$ to
$N$ cancels each interior term. The first term vanishes as well,
since $f(1)=f_0(1)=1/2$. This leaves
\begin{equation}\label{n2:dual-telescope}
 (R(f)-\Gamma)\sum_{k=0}^{N}w_k
 \ge -w_{N+1}s_{N+1}\Delta_{N+1}.
\end{equation}
On $(0,1]$ both rules take values in $[1/2,1]$, so $|\Delta_{N+1}|\le1/2$.
Since $\sum_{k=0}^N w_k\ge1$, we obtain
\begin{equation}\label{n2:dual-value}
 R(f)\ge \Gamma-
 \frac{w_{N+1}s_{N+1}}{2\sum_{k=0}^N w_k}.
\end{equation}
The last fraction tends to zero. This proves the lower bound, which
also holds when $R(f)=\infty$. Applying it to $f_0$ proves the
attainment claim.
\end{proof}

\subsection{Analytic construction and certified shape}
\label{n2:curve-subsection}

We seek a rule and a family of cross pairs at which the same bound is
attained. Write a pair as $(a,1/c)$ with $0<a,c\le1$. For a job of
cost ratio $a$, let $h(a)=1-f(a)$ be the probability of assignment to
machine two, and put $g(a)=a h(a)$. Reciprocity gives $f(1/c)=h(c)$,
so the objective is
\[
 \Phi_f(a,1/c)=1+a-g(a)+(1/c-1-a)h(c).
\]
Fix $a$. At an interior maximizing partner $c$, the derivative with
respect to $c$ vanishes:
\[
 \left(\frac1c-1-a\right)h'(c)-\frac{h(c)}{c^2}=0 .
\]
Along a differentiable family of such pairs with constant value
$1+\alpha$, the derivative with respect to $a$ also vanishes. The
term containing the partner's derivative drops out by stationarity,
leaving $g'(a)=1-h(c)$. The contact value itself gives
\[
 g(a)=a(1-h(c))-\alpha+(1/c-1)h(c).
\]
These three equations determine the graph at $a$ from its value and
slope at $c$. We use them to construct a candidate. In
Section~\ref{n2:primal-dual-subsection} we prove that the resulting
partners are global maximizers.

At a fixed contact the partner ratio equals the first ratio, say
$a=c=r$. The actual pair in the scalar objective is $(r,1/r)$.
Stationarity and the slope equation become
\[
 h(r)=(1-r-r^2)\,r h'(r),\qquad 2h(r)+r h'(r)=1 .
\]
Write $d=1-r-r^2$. Solving gives the prescribed value
$h_*=d/(1+2d)$ and scaled slope $\beta=1/(1+2d)$ at $r$:
$h(r)=h_*$ and $r h'(r)=\beta$. We will find the fixed ratio in
\begin{equation}\label{n2:parameter-interval}
 [r_-,r_+]
 =\left[\frac{52943222426}{10^{11}},\frac{52943222427}{10^{11}}\right].
\end{equation}
At the fixed pair, $\Phi_f(r,1/r)=1+r(1-h_*)+(1/r-1-r)h_*$.
Denote this value by $\Psi(r)$. To find the derivative $\lambda_r$
of the partner at $a=c=r$, differentiate $g'(a)=1-h(c)$ along
the contacts. Since $g(a)=a h(a)$, this gives
$r h''(r)=-(2+\lambda_r)h'(r)$. Differentiating the stationarity
equation and using $h(r)=rdh'(r)$ gives
\[
 r^2=\lambda_r\bigl(-2+2d+d\,r h''(r)/h'(r)\bigr)
     =-\lambda_r(2+d\lambda_r).
\]
We take the root with $|\lambda_r|<1$, so iteration brings the
partners toward $r$. The resulting expressions are
\begin{equation}\label{n2:parameters}
 \Psi(r)=\frac{(r-1)(r+1)^2}{r(2r^2+2r-3)},\qquad
 \lambda_r=-\frac{r^2}{1+\sqrt{1-dr^2}}.
\end{equation}
Set $\alpha(r)=\Psi(r)-1$. With $G=g(c)$ and $V=g'(c)$, we have
$cV-G=c^2h'(c)$. The stationarity equation thus gives the first
ratio $a=1/c-1-G/[c(cV-G)]$. The value and slope equations give the
other two coordinates of the map
\begin{equation}\label{n2:contact-map}
 B_r(c,G,V)=\left(a,\ a(1-G/c)-\alpha(r)+(1/c-1)G/c,\ 1-G/c\right).
\end{equation}
Thus the input is the graph triple at the partner $c$ and the output
is the graph triple at $a$.

Write $z$ for the curve parameter and
$\gamma_r(z)=(a_r(z),G_r(z),v_r(z))$ for the ratio, graph value, and
candidate slope at that parameter. Moving from a first ratio to its
partner corresponds to moving from $z$ to $\lambda_r z$. The map $B_r$
reconstructs the first triple from this partner triple, so we seek
\begin{equation}\label{n2:invariance}
 \gamma_r(z)=B_r(\gamma_r(\lambda_r z))\qquad (|z|\le1/2).
\end{equation}
We place the fixed contact at $z=0$ and normalize $a_r'(0)=1$.
The prescribed value and derivative are
\begin{equation}\label{n2:jet}
 \gamma_r(0)=(r,\ rh_*,\ h_*+\beta),\qquad
 \gamma_r'(0)=\left(1,\ h_*+\beta,\ -\frac{\lambda_r\beta}{r}\right).
\end{equation}
Here primes on curve coordinates denote derivatives in $z$. On the
real interval where $a_r$ is increasing, the graph in the ratio
variable is $g(a)=G_r(a_r^{-1}(a))$. Lemma~\ref{n2:contact} verifies
that $v_r(z)=g'(a_r(z))$. Since $-143/1000<\lambda_r<-142/1000$,
successive partners approach the fixed ratio from alternating sides.

The remaining condition comes from the job with equal machine costs.
The curve reaches ratio one at the unique $z_1(r)\in(0.42190,0.42192)$
with $a_r(z_1(r))=1$. Symmetry requires $h(1)=1/2$, or equivalently
\begin{equation}\label{n2:boundary-equation}
 G_r(z_1(r))=\frac12.
\end{equation}
The next lemma constructs the curves and brackets a root of this
equation. Section~\ref{n2:primal-dual-subsection} proves that every
root gives the scalar optimum $\Psi(r)$, and that the root $r_*$
is unique. All terminating decimals below are exact rationals. The
degree-$24$ polynomials used in the finite verification enclose the
analytic curves and their derivatives. Appendix~\ref{supp:analytic}
describes these inputs and the checks on them.

\begin{lemma}[Analytic and shape certificate]\label{n2:curve-certificate}
\textup{(i) Analytic existence and parameter dependence.}
For every $r\in [r_-,r_+]$, equations \eqref{n2:invariance}--\eqref{n2:jet} have
a unique analytic solution on $|z|\le1/2$, and this solution depends
continuously on $r$.

\textup{(ii) Graph and shape.} On $I_z=[-437/5000,211/500]$,
the coordinate $a_r$ is positive and has strictly positive derivative.  Its image
contains $[0.44397,1]$ in its interior.  Write
$g(a)=G_r(a_r^{-1}(a))$ and $h(a)=g(a)/a$ on this image.  There is a
unique zero $b=b(r)\in(0.44397,0.44398)$ of $h$, and on the image of $I_z$
one has
\begin{equation}\label{n2:shape-signs}
 h'>0,\qquad h''<0.
\end{equation}
\textup{(iii) Bounds used beyond the cross region.}
The contact parameter $\eta=a_r(\lambda_r z_1(r))$ satisfies
\begin{equation}\label{n2:chain-enclosures}
 0.47036<\eta<0.47038.
\end{equation}
On the low interval $b\le a\le\eta$, the value and slope satisfy
\begin{equation}\label{n2:low-enclosures}
 0\le h(a)\le\frac1{20},\qquad \frac{17}{10}\le h'(a)\le2.
\end{equation}
On the high interval, the curvature and endpoint slope satisfy
\begin{equation}\label{n2:high-enclosures}
 h''(a)\le-\frac{13}{20}\quad(9/10\le a\le1),\qquad
 \frac9{10}h'(9/10)-h(9/10)>0.
\end{equation}
\textup{(iv) Boundary signs.}
The boundary residual $G_r(z_1(r))-1/2$ is positive at $r_-$
and negative at $r_+$, with the enclosures
\begin{equation}\label{n2:boundary-signs}
 \begin{aligned}
 3.9921524
 &<10^{12}\!\left(G_{r_-}(z_1(r_-))-\frac12\right)<3.9921585,\\
 -3.7791184
 &<10^{12}\!\left(G_{r_+}(z_1(r_+))-\frac12\right)<-3.7791124.
 \end{aligned}
\end{equation}
\end{lemma}
\begin{proof}
We prove (i) by contraction, (ii)--(iii) by derivative estimates and
interval coverage, and (iv) by endpoint enclosures for the same analytic
family. The finite rational inputs must pass the acceptance conditions
in Theorem~\ref{n2:verification}; these numerical premises are distinct
from the analytic implications proved here.

\paragraph{An invariant analytic ball.}
Let $Q_r$ be the polynomial whose constant and
linear coefficients are \eqref{n2:jet} and whose coefficients of orders
$2$ through $24$ are the uniform rational coefficient inputs to the
finite verification in Theorem~\ref{n2:verification}.  In the Banach space of triples holomorphic in
$|z|<1/2$ and continuous on its closure, use the maximum of the three
supremum norms.  Consider the closed ball of radius $10^{-7}$ about
$Q_r$, restricted to functions satisfying \eqref{n2:jet}.

The operator $\mathcal T_r\gamma=B_r(\gamma(\lambda_r\,\cdot))$ preserves these
value and derivative conditions.  At the fixed point, the parameter definitions give
\[
 h_*=d\beta,\qquad 1-h_*=h_*+\beta,\qquad
 \alpha=r\beta+\frac{1-r}{r}h_*.
\]
At $(c,G,V)=(r,rh_*,h_*+\beta)$, the denominator factor $cV-G$
equals $r\beta$. Substitution and $d\lambda_r^2+2\lambda_r+r^2=0$ give
\[
 B_r(\gamma_r(0))=\gamma_r(0),\qquad
 \lambda_r DB_r(\gamma_r(0))\gamma_r'(0)=\gamma_r'(0),
\]
so the constant and linear coefficients are preserved.  Two functions in this ball differ
by a function vanishing to order at least two, so Schwarz's lemma gives
\begin{equation}\label{n2:schwarz}
 \|\gamma(\lambda_r\,\cdot)-\widetilde \gamma(\lambda_r\,\cdot)\|_\infty
 \le |\lambda_r|^2\|\gamma-\widetilde \gamma\|_\infty.
\end{equation}
For inputs $\gamma(\lambda_r z)$ with $\gamma$ in this ball and $|z|\le1/2$,
rational bounds give $|c|\ge0.4571275$ and $|cV-G|\ge0.2498736$,
keeping the rational map away from its poles. The Jacobian satisfies
$\|DB_r\|_\infty<34.874$. The contraction and residual estimates are
\begin{equation}\label{n2:banach-data}
 |\lambda_r|^2\|DB_r\|_\infty<0.703885<\frac45,\qquad
 \|\mathcal T_rQ_r-Q_r\|_\infty<7.756\cdot10^{-10}.
\end{equation}
For each fixed $z$, the set of allowed inputs contains the line segment between $\gamma(\lambda_r z)$ and $\widetilde \gamma(\lambda_r z)$. Thus the Jacobian bound and \eqref{n2:schwarz} give a contraction.  Its residual
plus its Lipschitz constant times $10^{-7}$ is less than
$7.117\cdot10^{-8}<10^{-7}$, proving that the ball is invariant.
The fixed point is holomorphic on the open disk and continuous on its
closure. It also extends across the boundary. The bound $|\lambda_r|<1$ keeps
$\lambda_r z$ inside the disk on a neighborhood of $|z|\le1/2$, and
the strict lower bounds on the two denominators persist on a sufficiently
small such neighborhood. Equation~\eqref{n2:invariance} gives the
extension there. For continuous dependence on $r$, write $\gamma=Q_r+e$, where
$e(0)=e'(0)=0$.  The induced map
\[
 e\longmapsto B_r\bigl((Q_r+e)(\lambda_r\,\cdot)\bigr)-Q_r
\]
is a uniform contraction on the same radius-$10^{-7}$ ball and depends
continuously on $r$.  Its fixed point therefore varies continuously.
Uniqueness also makes the solution real on the real axis.

To verify the residual without discarding any polynomial terms, write
$(c,G,V)=Q_r(\lambda_r z)$. For $j=1,2,3$, the $j$th
coordinate of $B_r(c,G,V)$ is $N_j/D_j$, where
\begin{equation}\label{n2:residual-numerators}
 \begin{alignedat}{2}
 N_1&=(1-c)(cV-G)-G, &\qquad D_1&=c(cV-G),\\
 N_2&=N_1(c-G)+\bigl((1-c)G-\alpha c^2\bigr)(cV-G), & D_2&=c^2(cV-G),\\
 N_3&=c-G, & D_3&=c.
 \end{alignedat}
\end{equation}
For each coordinate, subtract its output polynomial $Q_{r,j}(z)$ times
the corresponding denominator from the numerator.  The weighted
coefficient norm $\sum_k|u_k|2^{-k}$ bounds the resulting polynomial
on the whole disk.  Division by the denominator-modulus lower bounds established above
yields the residual in \eqref{n2:banach-data}.  Interval
coefficients enclose all $r\in [r_-,r_+]$; the only radical, in $\lambda_r$, is
enclosed by rational square-root bounds.  Thus no floating-point
comparison is needed.

\paragraph{Uniqueness with the prescribed value and derivative at zero.}
The contraction proves uniqueness within the ball about $Q_r$. Suppose
there were another analytic solution with the value and derivative
in~\eqref{n2:jet}. Let $u$ be the first nonzero Taylor coefficient
of its difference from $\gamma_r$, of order $k\ge2$. Both solutions
have value $\gamma_r(0)$ at zero. Comparing terms of order $k$ in
\eqref{n2:invariance} gives
\[
 u=\lambda_r^k\,DB_r(\gamma_r(0))\,u.
\]
Indeed, all terms involving only lower coefficients cancel. At $z=0$
the Jacobian bound in~\eqref{n2:banach-data} applies, so
$|\lambda_r|^k\|DB_r(\gamma_r(0))\|_\infty<1$. This contradicts
$u\ne0$. The Taylor series agree, hence the solutions agree on the
disk. Iteration of the contraction gives
\begin{equation}\label{n2:curve-iteration}
 \gamma_r(z)=\lim_{N\to\infty}B_r^{\circ N}
                      \bigl(Q_r(\lambda_r^N z)\bigr).
\end{equation}
By~\eqref{n2:banach-data}, the initial error is at most
$\|\mathcal T_rQ_r-Q_r\|_\infty/(1-4/5)<5\cdot10^{-9}$.
After $N$ iterations it is at most $5\cdot10^{-9}(4/5)^N$.

\paragraph{From disk error to real graph bounds.}
The analytic error must control derivatives on an interval, not just
values at sample points.  Enclose the constant and linear coefficients
over $[r_-,r_+]$ and retain the exact higher coefficients.  Differentiate these
polynomials algebraically.  If $e$ is the analytic error in a coordinate,
Cauchy's estimate gives, for $z\in I_z$,
\begin{equation}\label{n2:cauchy-error}
 |e^{(j)}(z)|\le\frac{j!10^{-7}}{(1/2-|z|)^j},
                    \qquad j=0,1,2.
\end{equation}
On each rational subinterval of $I_z$, Horner evaluation and these error
bounds enclose $a_r,G_r$ and their first two derivatives in $z$.
At a given $z$, let $a=a_r(z)$. Differentiating $g(a_r(z))=G_r(z)$
with respect to $z$ gives the graph derivatives in the first row
below. Differentiating $a h(a)=g(a)$ with respect to $a$ gives the
second row. Here $a_r'(z)>0$ and $a>0$ are checked before division.
\begin{equation}\label{n2:coordinate-derivatives}
 \begin{aligned}
 g'(a)&=\frac{G_r'(z)}{a_r'(z)},&
 g''(a)&=\frac{G_r''(z)a_r'(z)-G_r'(z)a_r''(z)}{[a_r'(z)]^3},\\
 h'(a)&=\frac{g'(a)-h(a)}{a},&
 h''(a)&=\frac{g''(a)-2h'(a)}{a}.
 \end{aligned}
\end{equation}
The rational interval tests in Theorem~\ref{n2:verification} give all
signs in \eqref{n2:shape-signs}, the endpoint coverage, and
\eqref{n2:low-enclosures}--\eqref{n2:high-enclosures} on a finite
cover of $I_z$.  In the low interval the lower bound $h\ge0$ follows
from the zero $b$ and $h'>0$; it is not imposed to the left of $b$.

For the remaining special points, agreement of the constant and linear coefficients improves the error
bound to $|e(z)|\le10^{-7}(2|z|)^2$.  Monotonicity of $a_r$ permits rational bracketing of the parameter
values $a_r^{-1}(0.44397)$, $a_r^{-1}(0.44398)$, and $a_r^{-1}(1)$.  Opposite signs of $h$ at the first two points and
$h'>0$ isolate $b$.  Evaluation at the last bracket gives $z_1$,
and evaluation at $\lambda_r z_1$ gives
\eqref{n2:chain-enclosures}.  The invariant interval and convergence of
all subsequent contacts will follow from monotonicity of $a_r$, these
rational bounds, and $-1<\lambda_r<0$.

\paragraph{Sharper bounds at the endpoints.}
The uniform error bound suffices for the shape inequalities but not for the small boundary
residuals.  The two supplied endpoint coefficient arrays give invariant
balls of analytic functions of radius $10^{-20}$, by the same contraction argument.  The distance from each endpoint polynomial to the uniform polynomial, plus
$10^{-20}$, is less than $10^{-7}$.  The endpoint fixed points therefore coincide with the
uniform family by uniqueness. Rational brackets for $a=1$, together with
these sharper bounds, give \eqref{n2:boundary-signs}.
\end{proof}

It remains to identify the candidate slope $v_r(z)$. We show that
the difference $G_r'(z)-v_r(z)a_r'(z)$ vanishes along the curve.

\begin{lemma}[Contact identity and graph slope]\label{n2:contact}
Along the analytic curve, $G_r'(z)=v_r(z)a_r'(z)$.  In particular,
$v_r(z)=g'(a_r(z))$ on $I_z$.
\end{lemma}
\begin{proof}
Fix $r$ and write $(a,G^+,V^+)=B_r(c,G,V)$. The contact-value
identity gives $G^+=aV^+-\alpha+(1-c)G/c^2$, with $V^+=1-G/c$.
Differentiating and cancelling the term $V^+\operatorname{d}a$ gives
\begin{equation}\label{n2:contact-identity}
\operatorname{d}G^+-V^+\operatorname{d}a
 =\left(-\frac ac+\frac{1-c}{c^2}\right)\operatorname{d}G
   +\frac G{c^2}\left(a+1-\frac2c\right)\operatorname{d}c 
 =\frac{G}{c^2(cV-G)}(\operatorname{d}G-V\operatorname{d}c).
\end{equation}
The second equality substitutes the first coordinate of
\eqref{n2:contact-map} for $a$. Thus the output differential is a scalar
multiple of the input differential.

\Needspace{6\baselineskip}
Set $D(z)=G_r'(z)-v_r(z)a_r'(z)$. Evaluating the differential identity
along the invariant curve, equations \eqref{n2:invariance} and
\eqref{n2:contact-identity} give
\[
 D(z)=\frac{\lambda_r G_r(\lambda_r z)}
 {a_r(\lambda_r z)^2\bigl(a_r(\lambda_r z)v_r(\lambda_r z)-G_r(\lambda_r z)\bigr)}
 D(\lambda_r z).
\]
At $z=0$, the coefficient multiplying $D(\lambda_r z)$ is
$\lambda_r d/r^2$. A first nonzero Taylor coefficient of $D$ of order
$k\ge0$ would therefore satisfy $1=\lambda_r^{k+1}d/r^2$.
This is impossible because $|\lambda_r|<1$ and $0<d/r^2<1$.  Hence $D$ vanishes identically.  Positivity of $a_r'$
on $I_z$ proves the graph-slope assertion.
\end{proof}

\begin{corollary}[Convexity of the contact graph]\label{n2:graph-convexity}
For every $r\in [r_-,r_+]$, the graph function $g$ is strictly convex on $a_r(I_z)$.
\end{corollary}
\begin{proof}
For this fixed $r$, define the partner ratio $\tau(a_r(z))=a_r(\lambda_r z)$.
The bounds on $\lambda_r$ give $\lambda_r I_z\subset I_z$. Since $a_r'>0$,
\[
 \tau'(a_r(z))=\lambda_r\,\frac{a_r'(\lambda_r z)}{a_r'(z)}<0.
\]
By Lemma~\ref{n2:contact}, the third coordinate of~\eqref{n2:invariance}
is $g'(a)=1-h(\tau(a))$. Thus $g''(a)=-h'(\tau(a))\tau'(a)>0$.
\end{proof}

\subsection{Pairs with both ratios at least one}
\label{n2:highhigh-subsection}

For $1\le s\le t$, the reciprocal ratios $x=1/t$ and $y=1/s$ satisfy
$0<x\le y\le1$. Write the objective in these coordinates as
\[
 F(x,y)=1+\frac{h(y)}{y}+\left(\frac1x-1-\frac1y\right)h(x).
\]
Here $F(x,y)=\Phi_f(1/y,1/x)$ when $h(a)=f(1/a)$. Its upper edge
$y=1$ is also a cross pair, $(s,t)=(1,1/x)$. Section~\ref{n2:primal-dual-subsection}
will bound this edge using the contacts. The following lemma extends
that bound to the whole region.
\begin{lemma}[Bound after taking reciprocal ratios]\label{n2:highhigh}
Let $\alpha\ge1/2$, $11/25\le b\le9/20$, and
$47/100\le\eta\le471/1000$, with $b<\eta$. Suppose $h$ is continuous
on $[0,1]$, zero on $[0,b]$, and $C^2$ on $[b,1]$ with one-sided
derivatives at the endpoints. Assume $h(1)=1/2$, $h'\ge0$, and
$h''\le0$ on $[b,1]$. Suppose the upper edge satisfies
\begin{equation}\label{n2:tie-hypothesis}
 F(x,1)=\frac32+(1/x-2)h(x)\le F(\eta,1)=1+\alpha
 \qquad(0<x\le1).
\end{equation}
At $1$ and $9/10$, assume the slope conditions
\begin{equation}\label{n2:contact-hypothesis}
 h'(1)=\frac12-h(\eta),\qquad \frac9{10}h'(9/10)-h(9/10)\ge0.
\end{equation}
On the low interval $b\le x\le\eta$, assume
\begin{equation}\label{n2:low-hypothesis}
 0\le h(x)\le\frac1{20},\qquad \frac{17}{10}\le h'(x)\le2.
\end{equation}
On the high interval, assume the curvature bound
\begin{equation}\label{n2:high-hypothesis}
 h''(y)\le-\frac{13}{20}\quad(9/10\le y\le1).
\end{equation}
Then
\begin{equation}\label{n2:highhigh-bound}
 F(x,y)\le1+\alpha\qquad(0<x\le y\le1).
\end{equation}

\end{lemma}
\begin{proof}
We first use monotonicity in $y$ to reduce the estimate to the upper
edge and a rectangle adjoining it. On the rectangle, concavity will
make the edge maximizer $(\eta,1)$ a maximizer in both variables.
Concavity implies $(yh'(y)-h(y))'=yh''(y)\le0$ on $[b,1]$,
and the endpoint contact gives
\begin{equation}\label{n2:p-minus-h}
                 yh'(y)-h(y)\ge h'(1)-h(1)=-h(\eta).
\end{equation}
Moreover,
\begin{equation}\label{n2:Fy}
                 F_y(x,y)=\frac{yh'(y)-h(y)+h(x)}{y^2}.
\end{equation}
For $x\ge\eta$, monotonicity of $h$ and
\eqref{n2:p-minus-h} imply $F_y\ge0$.  Therefore
$F(x,y)\le F(x,1)\le1+\alpha$.

Now let $x\le\eta$.  For $b\le y\le9/10$, the intermediate slope
test and monotonicity of $yh'(y)-h(y)$ imply $yh'(y)-h(y)\ge0$, hence again
$F_y\ge0$.  If $y\le b$, both values of $h$ vanish and $F=1$.
It is consequently enough to consider $y\ge9/10$.  When $x<b$,
replacing $x$ by $b$ leaves $F$ unchanged.  Thus only the rectangle
$[b,\eta]\times[9/10,1]$ remains.

On $[b,\eta]\times[9/10,1]$ we have
$1/x-1-1/y\ge1000/471-1-10/9=17/1413>0$.
The mixed derivative is $F_{xy}=h'(x)/y^2$. Differentiating in each
variable twice gives
\begin{equation}\label{n2:hessian}
 \begin{aligned}
 F_{xx}&=\frac{2(h(x)-xh'(x))}{x^3}+
           \left(\frac1x-1-\frac1y\right)h''(x),\\
 F_{yy}&=\frac{h''(y)}y-\frac{2(yh'(y)-h(y)+h(x))}{y^3}.
 \end{aligned}
\end{equation}
The low-interval bounds and $h''(x)\le0$ give
\[
 -F_{xx}\ge\frac{2(xh'(x)-h(x))}{x^3}
 \ge2\left(\frac{11}{25}\cdot\frac{17}{10}-\frac1{20}\right)
            \left(\frac{1000}{471}\right)^3>13 .
\]
For the other diagonal entry, \eqref{n2:p-minus-h} gives
$yh'(y)-h(y)+h(x)\ge-h(\eta)\ge-1/20$. Using the curvature bound
at $y$ and then the slope bound at $x$ yields
\[
 -F_{yy}\ge\frac{13}{20}-\frac1{10}\left(\frac{10}{9}\right)^3>\frac12,
 \qquad |F_{xy}|\le2\left(\frac{10}{9}\right)^2<\frac52 .
\]
The Hessian has negative diagonal entries and determinant
\begin{equation}\label{n2:hessian-margin}
 \det\nabla^2 F=(-F_{xx})(-F_{yy})-F_{xy}^2
 >13\cdot\frac12-\left(\frac52\right)^2=\frac14.
\end{equation}
Thus $F$ is concave on the rectangle. The horizontal derivative at $(\eta,1)$ vanishes because
$\eta\in(b,1)$ maximizes $x\mapsto F(x,1)$. The prescribed slope
at one gives the vertical derivative there:
\[
 F_y(\eta,1)=h'(1)-h(1)+h(\eta)=0.
\]
The value there is $F(\eta,1)=1+\alpha$.
The tangent-plane inequality for a concave
function therefore bounds the entire rectangle by $1+\alpha$.
\end{proof}

\subsection{Global feasibility and a matching lower bound}
\label{n2:primal-dual-subsection}

Each root of~\eqref{n2:boundary-equation} gives a candidate rule.
We first bound its objective in the three regions of
Figure~\ref{fig:ratio-domains}. We then use its tight cross pairs in
Lemma~\ref{n2:contact-chain} to prove that every competing rule has
at least the same worst-case ratio.

\begin{theorem}[Exact scalar value]\label{n2:scalar-value}
Equation~\eqref{n2:boundary-equation} has a unique root $r_*$ in $[r_-,r_+]$.
For this root there is a continuous $f_*\in\mathcal F$ such that
\begin{equation}\label{n2:scalar-value-equation}
          \min_{f\in\mathcal F}R(f)=R(f_*)=\Psi(r_*).
\end{equation}
\end{theorem}
\begin{proof}
The boundary residual is continuous in $r$: the curve is continuous in
$r$, and its strictly increasing coordinate $a_r$ has a continuously
varying inverse at $a=1$.  The opposite signs in
\eqref{n2:boundary-signs} therefore give at least one root in the
interior of $[r_-,r_+]$.  Fix any such root $r$. Put $\alpha=\Psi(r)-1$ and let $b$ be the zero of $h$ in
Lemma~\ref{n2:curve-certificate}.  Retain $h(a)=g(a)/a$ on $[b,1]$
and set $h(a)=0$ on $[0,b]$.  Define
\begin{equation}\label{n2:primal-definition}
 f(a)=
 \begin{cases}
  1-h(a),&0<a\le1,\\
  h(1/a),&a\ge1.
 \end{cases}
\end{equation}
The boundary condition gives $h(1)=1/2$; monotonicity and $h(b)=0$
give $0\le h\le1/2$.  Thus $f$ is continuous, nonincreasing,
$[0,1]$-valued, and reciprocal.

\paragraph{Pairs with one ratio on each side of one.}
Fix a first ratio $a$ in the analytic interval $[b,1]$. We will show
that its proposed partner gives the largest objective among all $c$.
For $a,c\in(0,1]$, the scalar objective at $(s,t)=(a,1/c)$ is
\begin{equation}\label{n2:cross-objective}
 \Phi_f(a,1/c)=1+a(1-h(a))+(1/c-1-a)h(c).
\end{equation}
Recall the partner ratio $\tau(a_r(z))=a_r(\lambda_r z)$ from
Corollary~\ref{n2:graph-convexity}. To apply the graph identities on
$[b,1]$, we need $\tau([b,1])\subset(b,1)$. Write $z_b=a_r^{-1}(b)$
and $z_1=z_1(r)$. Since $b<r<1$, we have $z_b<0<z_1$.
At the right endpoint $\tau(1)=a_r(\lambda_r z_1)=\eta>b$, so
$\lambda_r z_1>z_b$. At the other endpoint,
\[
 0<\lambda_r z_b
 <\frac{143}{1000}\frac{437}{5000}<0.42190<z_1.
\]
We used $z_b\in I_z$, $|\lambda_r|<143/1000$, and the enclosure
for $z_1$. Thus multiplication by $\lambda_r$ sends $[z_b,z_1]$ into
$(z_b,z_1)$, as required.

For $a\in[b,1]$ let $c=\tau(a)$. By Lemma~\ref{n2:contact}, the
input triple of $B_r$ is $(c,g(c),g'(c))$, so $cV-G=c^2h'(c)$.
The first coordinate of~\eqref{n2:contact-map} gives
\begin{equation}\label{n2:A-contact}
 \frac1c-1-\frac{h(c)}{c^2h'(c)}=a\qquad(c=\tau(a)).
\end{equation}
Its third coordinate and the contact identity yield
\begin{equation}\label{n2:slope-contact}
                             g'(a)=1-h(\tau(a)).
\end{equation}
Finally, the second coordinate says, with $c=\tau(a)$,
\[
              g(a)=a(1-h(c))-\alpha+(1/c-1)h(c).
\]
Substituting this into \eqref{n2:cross-objective} gives the active
equalities
\begin{equation}\label{n2:active-cross}
                         \Phi_f(a,1/\tau(a))=1+\alpha.
\end{equation}
It remains to compare the contact with the other possible partners.
For fixed $a\in[b,1]$, only $(1/c-1-a)h(c)$ in
\eqref{n2:cross-objective} depends on $c$. This term is positive
precisely when $b<c<1/(1+a)$. The contact lies in this interval:
by~\eqref{n2:A-contact}, at $c=\tau(a)$ we have
$1/c-1-a=h(c)/(c^2h'(c))>0$.
In the positive interval the derivative has the sign of
\[
 c^2\partial_c\Phi_f(a,1/c)=c\bigl(1-(1+a)c\bigr)h'(c)-h(c).
\]
This expression vanishes at $\tau(a)$ by the same contact equation
and is strictly decreasing throughout the interval, since
\begin{equation}\label{n2:cross-derivative}
 \partial_c\bigl(c^2\partial_c\Phi_f(a,1/c)\bigr)
 =-2(1+a)c h'(c)+c\bigl(1-(1+a)c\bigr)h''(c)<0.
\end{equation}
The objective therefore increases up to $\tau(a)$ and decreases
after it within this interval. Outside it the $c$-dependent term is
nonpositive. This proves that $\tau(a)$ is the unique global maximizer
and~\eqref{n2:active-cross} bounds every partner by $1+\alpha$.
For $a<b$, we have $g(a)=0$ and the cross objective is affine in $a$
with slope $1-h(c)\ge0$. Its value at $b$ bounds this last case.

\paragraph{Pairs with both ratios at most one.}
For $0<s\le t\le1$, we have $1/2\le f(s),f(t)\le1$ and
$1+s-t\ge0$.  Therefore
\begin{equation}\label{n2:lowlow}
 \Phi_f(s,t)\le1+s-\frac{1+s-t}{2}
             =\frac{1+s+t}{2}\le\frac32< \Psi(r).
\end{equation}
The final strict inequality follows directly from
\eqref{n2:parameters} on $[r_-,r_+]$.

\paragraph{Pairs with both ratios at least one.}
For $1\le s\le t$, put $x=1/t$, $y=1/s$.  Reciprocity turns the
objective into $F(x,y)$ from \eqref{n2:highhigh-bound}.
The upper edge is $F(x,1)=\Phi_f(1,1/x)$, so the cross bound with
$a=1$ gives $F(x,1)\le1+\alpha$, with equality at $\eta=\tau(1)$.  Equation~\eqref{n2:slope-contact}
at $a=1$, together with $g'(1)=h(1)+h'(1)$, gives
$h'(1)=1/2-h(\eta)$.  All remaining hypotheses of
Lemma~\ref{n2:highhigh} follow from
Lemma~\ref{n2:curve-certificate}, and $\alpha>1/2$.
Thus $\Phi_f(s,t)\le1+\alpha$ also when $1\le s\le t$. All three regions are covered, so
\begin{equation}\label{n2:primal-upper}
                              R(f)\le \Psi(r).
\end{equation}

\paragraph{Optimality from the contact chain.}
Start with first ratio $s_0=1$ and take each tight partner as the next
first ratio. In the curve parameter this is iteration of multiplication
by $\lambda_r$, so
\begin{equation}\label{n2:dual-definition}
 s_k=a_r(\lambda_r^k z_1),\qquad s_{k+1}=\tau(s_k)\qquad(k\ge0).
\end{equation}
The first two tight pairs are $(1,1/\eta)$ and $(\eta,1/\tau(\eta))$.
Since $\lambda_r^k z_1\in[\lambda_r z_1,z_1]$, every $s_k$ lies in
$[\eta,1]$. All constraints $\Phi_f(s_k,1/s_{k+1})=\Psi(r)$ are tight
by~\eqref{n2:active-cross}.

It remains to check the weights in Lemma~\ref{n2:contact-chain}. At
$c=s_{k+1}$, the contact equation~\eqref{n2:A-contact} gives
\[
 1/s_{k+1}-1-s_k
 =\frac{h(s_{k+1})}{s_{k+1}^2 h'(s_{k+1})}>0.
\]
Thus the weights are positive. Since $s_k\to a_r(0)=r$, their
successive ratios satisfy
\begin{equation}\label{n2:dual-decay}
 \frac{w_{k+1}}{w_k}
 =\frac{1/s_{k+1}-1-s_k}{s_{k+1}}
 \longrightarrow\frac{1-r-r^2}{r^2}=\frac d{r^2}\in(0,1).
\end{equation}
Indeed $d>0$ and $d-r^2=(1+r)(1-2r)<0$ on the parameter interval.
Hence $w_k$ decays geometrically for all sufficiently large $k$.
As $s_k\le1$, the endpoint coefficient $w_ks_k$ tends to zero.
Lemma~\ref{n2:contact-chain} now gives $R(\widetilde f)\ge\Psi(r)$
for every $\widetilde f\in\mathcal F$. The upper bound~\eqref{n2:primal-upper}
therefore gives $\min_{f\in\mathcal F}R(f)=\Psi(r)$.

\paragraph{Uniqueness and the numerical enclosure.}
The preceding construction works for every boundary root in $[r_-,r_+]$.
Every such root must consequently have the same value of $\Psi(r)$.
Direct differentiation gives
\begin{equation}\label{n2:R-derivative}
 \Psi'(r)=\frac{-(r-3)(r+1)(2r-1)}
              {r^2(2r^2+2r-3)^2}>0\qquad(r\in [r_-,r_+]).
\end{equation}
Hence the boundary root is unique.  Evaluating the rational function
at the two rational endpoints gives
\begin{equation}\label{n2:rational-value-bound}
 \frac{45863945683627831142259104319801}{30454234044755662284518208639602}
 <\Psi(r_*)<
 \frac{1100734696398070348526212946031517}{730901617066140697052425892063034}.
\end{equation}
Exact comparison with the two terminating decimals in
\eqref{n2:decimal-bound} proves that enclosure.  This completes
the proof of \eqref{n2:scalar-value-equation}.
\end{proof}

Theorem~\ref{n2:main} now follows from
Theorems~\ref{n2:scalar-value} and \ref{n2:owner-transfer}, together
with Lemma~\ref{n2:locality-bridge}.  The first ratios $s_k$ alternate about $r_*$ and converge to it.
The actual tight pairs $(s_k,1/s_{k+1})$ converge to $(r_*,1/r_*)$.

The optimal rule also admits nonnegative payments. Since $h=0$ on
$[0,b]$, $f(r)=0$ for $r\ge1/b$. If the other machine $k$ reports
$t_{kj}>0$, the probability that machine $i$ receives job $j$ vanishes
once its own reported time reaches $t_{kj}/b$. If $t_{kj}=0$, it
vanishes at every positive own time by~\eqref{n2:zero-extension}.
Thus the per-job payment
\begin{equation}\label{n2:nonnegative-payments}
 \pi_{ij}(t)=t_{ij}q_{ij}(t)
             +\int_{t_{ij}}^\infty q_{ij}(u,t_{-ij})\,\mathrm{d}u
\end{equation}
is finite. Relative to job $j$'s term in~\eqref{n2:payments}, this
adds $\int_0^\infty q_{ij}(u,t_{-ij})\,\mathrm{d}u$. Column locality
makes the added amount independent of machine $i$'s entire report,
so truthfulness is preserved. At a truthful report the expected
utility from job $j$ is the nonnegative tail integral
in~\eqref{n2:nonnegative-payments}.

\subsection{Finite verification of the analytic certificate}
\label{n2:verification-subsection}

\begin{theorem}[Rational verifier specification]\label{n2:verification}
The analytic existence, shape, and boundary assertions used in
Lemma~\ref{n2:curve-certificate}, as well as the enclosure
\eqref{n2:decimal-bound}, admit the following finite verification using
exact rational arithmetic and integer square roots. The required finite
inputs are uniform degree-$24$ rational coefficient arrays, two
endpoint-specific arrays and rational brackets, and parameter-enclosing
shape arrays.  Acceptance consists of the checks
listed below.  Together with the displayed algebraic identities, their
acceptance implies all numerical premises of the scalar proof.
\end{theorem}
\begin{proof}
For a rational interval $[l,u]$, use the usual inclusion-preserving
arithmetic, permitting division only when zero is excluded.  For
$x\ge0$, a square-root enclosure at denominator $D$ is obtained by
integer square root: if $k=\lfloor D\sqrt{x}\rfloor$, then
$k/D\le\sqrt{x}\le(k+1)/D$, with equality detected by squaring.
No approximate comparison enters acceptance.

\begin{enumerate}
\item \emph{Uniform analytic bounds.}
Use the uniform rational coefficient arrays, but replace
orders zero and one by the exact parameter-dependent value and derivative at zero
\eqref{n2:jet}.  Enclose $r\in [r_-,r_+]$, the radical in $\lambda_r$, and
all resulting coefficients.  Bound polynomial disks by weighted
coefficient sums.  Form the full numerators
\eqref{n2:residual-numerators}, subtract output times denominator,
and check the nonzero denominator bounds, the Jacobian bound, the
residual bound, and the strict ball-invariance inequality in
\eqref{n2:banach-data}.  The weaker acceptance thresholds
\[
 |c|>0.45,\quad |cV-G|>0.24,\quad
 |\lambda_r|^2\|DB_r\|_\infty<4/5,\quad
 \|\mathcal T_rQ_r-Q_r\|_\infty<10^{-9}
\]
suffice because $10^{-9}+(4/5)10^{-7}<10^{-7}$. The more precise
bounds in \eqref{n2:banach-data} are outward rational summaries.

\item \emph{Endpoint identification and signs.}
For each endpoint-specific coefficient array and rational bracket,
perform the same invariant-ball calculation with radius $10^{-20}$.  Check that the
weighted coefficient distance to the uniform polynomial plus
$10^{-20}$ is less than $10^{-7}$.  If $[z_-,z_+]$ is the corresponding
bracket, require the upper enclosure of $a_r(z_-)$ to be below one,
the lower enclosure of $a_r(z_+)$ to be above one, and the enclosure
of $G_r([z_-,z_+])-1/2$ to have the sign stated in
\eqref{n2:boundary-signs}.  The endpoint residual bound is
$4.695\cdot10^{-22}$.  Nesting identifies these solutions with the
uniform family, so the signs concern the same boundary function.

\item \emph{Shape arrays and coverage.}
Reconstruct the constant and linear coefficient intervals from
\eqref{n2:parameters} and \eqref{n2:jet}; require the reconstructed shape arrays
to enclose them, with unchanged higher rational coefficients.
Starting with $I_z$, bisect any interval on which a required strict
inequality is not yet certified.  On each accepted leaf, use
Horner evaluation, \eqref{n2:cauchy-error}, and
\eqref{n2:coordinate-derivatives} to require $a_r(z)>0$, $a_r'(z)>0$,
$h'>0$, and $h''<0$.
On a leaf whose $a$-enclosure intersects $[0.44397,0.47038]$,
also require $h\le1/20$ and $17/10\le h'\le2$.
On a leaf intersecting $[9/10,1]$, require $h''\le-13/20$;
on any leaf containing $a=9/10$, require $a h'-h>0$.
Finally require $a_r(-437/5000)<0.44397$ and
$a_r(211/500)>1$.  The frozen data admit a covering partition with
$19$ leaves and maximum bisection depth $8$; acceptance requires
complete interval coverage, not just evaluations at sample points.

\item \emph{Zeros and the first contact.}
Use the sharper error $10^{-7}(2|z|)^2$ and the proved
monotonicity of $a_r$ to bracket the preimages of
$0.44397$, $0.44398$, and $1$.  Require negative and positive
$h$-enclosures at the first two preimages, and the enclosure
$0.42190<z_1<0.42192$ at the third.  Check
$0.47036<a_r(\lambda_r z_1)<0.47038$.  These brackets imply $\eta>b$;
the invariant interval for $\tau$ and the location of every subsequent
contact then follow from the monotonicity argument in the scalar proof.
The positive weights and their decay follow from the contact identity
and \eqref{n2:dual-decay}.

\item \emph{Algebraic identities and final values.}
The fixed-point and derivative identities follow by clearing
denominators and using $d\lambda_r^2+2\lambda_r+r^2=0$, as in the
proof of Lemma~\ref{n2:curve-certificate}. The contact identity
\eqref{n2:contact-identity} and the sign of $\Psi'$ in
\eqref{n2:R-derivative} are algebraic steps in the proof.
The verifier checks the three derivative bounds and the determinant
bound in~\eqref{n2:hessian-margin}. It evaluates $\Psi(r_-)$ and
$\Psi(r_+)$ exactly, checks the fractions printed
in~\eqref{n2:rational-value-bound}, and compares them with the decimal
endpoints in~\eqref{n2:decimal-bound}.
\end{enumerate}

The contraction argument, Cauchy estimates, monotone inversion, and
interval inclusion justify each implication from these finite checks to
the analytic assertions.  Convexity of $g$, the invariant interval of
$\tau$, and the decay of the contact-chain weights follow from the
mathematical arguments above.  The numerical premises require the finite
coefficient arrays and brackets to pass all the stated checks; those
inputs and their verification code are described in
Appendix~\ref{sec:supplement}.
The certificate uses neither an optimization routine nor unvalidated
Taylor tails.  In particular, finite numerical sampling is not a
substitute for any of the interval coverage requirements.
\end{proof}

\section{Three machines: a shared-clock coupling}\label{sec:n3}
On three machines we keep the Bounded-Square probability $q_{ij}$ that
machine $i$ receives job $j$. These probabilities determine each
machine's expected load. The chance that a machine receives several jobs together
depends on the joint assignment, and so does the expected maximum
load. We choose this joint assignment by drawing one set of clocks for
all jobs. Each job compares independent clocks, but a clock used for
one job can also decide the winner of another.
\begin{samepage}
\begin{theorem}[Three-machine upper bound]\label{thm:n3upper}
There is a truthful-in-expectation mechanism with expected makespan at most $2\OPT$ on every finite nonnegative three-machine instance. Its expected payments and utilities are finite on every finite report profile and deviation, and every supported outcome has zero makespan at zero optimum.
\end{theorem}
\end{samepage}
We first show on one profile why the clock sharing matters.
Section~\ref{n3:transfer} reduces the general makespan bound to three
jobs, each of cost one on its owner. Sections~\ref{n3:cover}--\ref{n3:reproducibility}
prove that normalized bound for the fixed mixture.
\subsection{The shared-clock mechanism}\label{n3:allocation}
Label the three machines $0,1,2$. The jobs still have labels
$1,\ldots,m$, except on a normalized three-job profile, where job $j$
has owner $j\in\{0,1,2\}$. For a job with $\min_k t_{kj}>0$, the
eligible machines are $S_j=\{i:t_{ij}\le2\min_k t_{kj}\}$. The
Bounded-Square rule prescribes
\begin{equation}
 q_{ij}(t)=\frac{\ind\{i\in S_j\}\,t_{ij}^{-2}}{\sum_{k\in S_j}t_{kj}^{-2}}.
 \label{eq:n3:1}
\end{equation}
Assign zero-minimum jobs uniformly among their zero-cost machines.
For the other jobs choose an optimal assignment $A^*$, breaking ties
deterministically, and call $k=A^*(j)$ the owner of job $j$.

Draw one fair coin to choose a map $\kappa$ from Figure~\ref{fig:clock-maps},
then draw its clocks once for the whole instance. Assign job $j$ to
the eligible machine $i$ minimizing $t_{ij}^2E_{\kappa(i,j)}$, breaking
ties by a fixed machine order.
\begin{itemize}
\item \textbf{Cyclic map.} Draw independent rate-one exponential clocks
$E_0,E_1,E_2$ and set $\kappa(i,j)=(i-k)\bmod3$.
\item \textbf{Pair map.} Draw independent rate-one exponential clocks
$E_*,E_{01},E_{02},E_{12}$. Set $\kappa(k,j)=*$ and
$\kappa(i,j)=\{i,k\}$ for $i\ne k$.
\end{itemize}
\begin{figure}[htbp]
\centering
\begin{minipage}{.46\linewidth}
\centering\textbf{Cyclic map}\par\smallskip
$\begin{array}{c|ccc}
 k\backslash i&0&1&2\\\hline
 0&E_0&E_1&E_2\\
 1&E_2&E_0&E_1\\
 2&E_1&E_2&E_0
\end{array}$
\end{minipage}\hfill
\begin{minipage}{.46\linewidth}
\centering\textbf{Pair map}\par\smallskip
$\begin{array}{c|ccc}
 k\backslash i&0&1&2\\\hline
 0&E_*&E_{01}&E_{02}\\
 1&E_{01}&E_*&E_{12}\\
 2&E_{02}&E_{12}&E_*
\end{array}$
\end{minipage}
\caption{Read across the row for job $j$'s owner $k$. Each candidate
$i$ multiplies its clock by $t_{ij}^2$, and the smallest eligible
score wins. Within a row the clocks are independent. Wherever a clock
symbol repeats, even in different rows, it denotes the same draw.
The fair coin selects one table for all jobs.}
\label{fig:clock-maps}
\end{figure}

Write $\omega$ for the coin and clocks used on the positive-minimum
jobs. For a fixed map $\kappa$, write $C_\kappa$ for its makespan and
$y_A^\kappa(t)$ for the probability that it produces assignment $A$.
We suppress $t$ at a fixed profile. The fair coin gives
$y_A=(y_A^{\rm cyc}+y_A^{\rm pair})/2$. A tuple lists the winners in
job order.

\paragraph{How sharing a clock changes the makespan.}
Consider the following matrix, with jobs as columns.
\[
 t=\begin{pmatrix}1&3&2\\1&1&1\\3&3&1\end{pmatrix}.
\]
The diagonal assignment is optimal with makespan one. Job $1$ stays
on machine $1$. Job $0$ moves there with probability $q_{10}=1/2$,
and job $2$ goes to machines $0,1,2$ with probabilities $1/9,4/9,4/9$.
Relative to the diagonal makespan of one, moving job $0$ to machine
$1$ adds one, and sending job $2$ to machine $0$ or $1$ adds two or
one. The increments add except in assignment $(1,1,0)$. Its loads
are $(2,2,0)$, so adding both increments overcounts by two.
For any coupling of these jobs,
\[
 \E[C_{\max}]=1+q_{10}+2q_{02}+q_{12}-2y_{(1,1,0)}
             =\frac{13}{6}-2y_{(1,1,0)}.
\]
To get a bound of two we need $y_{(1,1,0)}\ge1/12$. Independent job
assignments give only $q_{10}q_{02}=1/18$ and hence expected makespan
$37/18$.

In the cyclic map, if job $2$ goes to machine $0$, then $4E_1<E_0$.
This also gives $E_1<E_0$, so job $0$ goes to machine $1$.
Consequently $y_{(1,1,0)}^{\rm cyc}=q_{02}=1/9$.
In the pair map the two moves require the owner clock $E_*$ to exceed
both $E_{01}$ and $4E_{02}$, and job $2$ also requires $4E_{02}<E_{12}$.
Given the alternative clocks, the probability that $E_*$ exceeds both
thresholds is $e^{-\max\{E_{01},4E_{02}\}}\ge e^{-E_{01}}e^{-4E_{02}}$.
The product is what two independent owner clocks would give. With
separate owner clocks the two moves use disjoint draws, so
$y_{(1,1,0)}^{\rm pair}\ge q_{10}q_{02}=1/18$. The fair mixture
therefore gives $y_{(1,1,0)}\ge(1/9+1/18)/2=1/12$.
Lemma~\ref{lem:n3:star} analyzes this pattern for all eligible costs.

\subsection{Truthfulness in expectation}\label{n3:payments}
\paragraph{The probability for one job.}
Fix a positive-minimum job $j$ with any owner. In either clock map
the machines in $S_j$ use distinct clocks. Their scores are independent
exponentials with rates $t_{ij}^{-2}$. Thus for $i\in S_j$,
\[
 \Prb(A(j)=i)
 =\int_0^\infty t_{ij}^{-2}
       \exp\!\left(-u\sum_{k\in S_j}t_{kj}^{-2}\right)\,\mathrm{d}u
 =\frac{t_{ij}^{-2}}{\sum_{k\in S_j}t_{kj}^{-2}} .
\]
This is~\eqref{eq:n3:1}. Changing the owner can change which other
jobs use these clocks, but each job keeps the same $q_{ij}(t)$.
Moreover, for fixed $\omega$ and fixed owner, multiplying the column
by a positive scalar preserves $S_j$ and multiplies all its scores by
the same factor. It preserves that job's winner in every realization.

\paragraph{Payments.}
For each machine and job the expected payment is
\begin{equation}
 \pi_{ij}(t)=t_{ij}q_{ij}(t)
      +\int_{t_{ij}}^\infty q_{ij}(u,t_{-ij})\,\mathrm{d}u.
 \label{eq:n3:5}
\end{equation}
Section~\ref{upper:marginal-preserving-correlated-rounding} proves
that the column rule is nonincreasing in $t_{ij}$, including across
the zero-cost and cutoff boundaries. It vanishes for
$t_{ij}>2\min_{k\ne i}t_{kj}$, so the integral is finite. By Lemma~\ref{lem:local-payments},
summing~\eqref{eq:n3:5} over jobs gives truthfulness in expectation
against deviations of the entire row. The payment and expected
processing cost depend only on the $q_{ij}$, even when a deviation
changes the owners and hence the joint assignment law.

If machine $i$ receives job $j$ and $q_{ij}>0$, pay it
$\widehat\pi_{ij}=\pi_{ij}/q_{ij}$ for that job. Otherwise pay zero.
When $q_{ij}=0$ the tail integral vanishes by monotonicity. This gives
$\E[\widehat\pi_{ij}]=\pi_{ij}$ in both cases and covers the reported
cost in every outcome. Finite expected payments and bounded realized
costs on every finite type give finite expected utilities.
Section~\ref{sec:implementation} implements the payments on rational
inputs. The exact factor-two rule here uses an optimal auxiliary
assignment, which may require exhaustive search.

\subsection{From three columns to every finite job set}\label{n3:transfer}
\begin{samepage}
\begin{theorem}[Normalized three-column inequality]\label{thm:n3normalized}
Let $t$ be a positive finite $3\times3$ processing-time matrix with
$t_{jj}=1$ for $j=0,1,2$. Use machine $j$ as job $j$'s owner and apply
the factor-two cutoff. If $C_{\rm cyc}$ and $C_{\rm pair}$ denote the
makespans under the respective clock maps, then
\begin{equation}
 \frac{\E[C_{\rm cyc}]+\E[C_{\rm pair}]}2\le2. \label{eq:n3:2}
\end{equation}
\end{theorem}
\end{samepage}
The left side is the expected makespan of the global fair mixture.
The diagonal schedule has makespan one and can be suboptimal.
Sections~\ref{n3:cover}--\ref{n3:reproducibility} prove this absolute
bound on every such profile. We now explain why it gives $2\OPT(t)$
on the original jobs.

\paragraph{Write the original loads as an average.}
Set aside the zero-minimum jobs, which contribute no load and can
be restored at zero cost in an optimum. If no jobs remain, every
outcome has makespan zero. Otherwise, write $T=\OPT(t)>0$ and retain
the owners from $A^*$. Fix the seed $\omega$. For a job $j$ owned by
$i$, divide its column by $t_{ij}$. The normalized job has the same
winner for this seed. Write $v^j(\omega)$ for its load vector. Its
contribution to the original vector $L(\omega)$ is exactly
$t_{ij}v^j(\omega)$.

For the proof, draw one representative $J_i$ for each owner,
independently across owners and independently of $\omega$. Choose
an owned job $j$ with probability
\begin{equation}\label{eq:n3:3}
 \Prb(J_i=j)=\frac{t_{ij}}T\qquad (A^*(j)=i).
\end{equation}
These weights sum to at most one because $L_i(t,A^*)\le T$.
Give the remaining probability to a dummy with zero load vector.
Thus $L(\omega)/T$ is the conditional mean of the sum of the three
representative load vectors. Convexity of the maximum gives
\[
 \frac{\E_\omega[C_{\max}]}T
 \le \E_{\boldsymbol J}\E_\omega
       \left[\left\|\sum_{i=0}^2 v^{J_i}(\omega)\right\|_\infty\right],
\]
as in Lemma~\ref{lem:representative}. The draws of representatives occur only in this load calculation.

\paragraph{Bound each representative triple.}
Fix the representatives, keeping their original owners. Index the
representative for machine $i$ as job $i$. It has time one there.
For a dummy, insert a job with time one on machine $i$ and three on
each other machine. The cutoff forces this job to $i$, so the insertion
adds one unit to its load. The other winners are unchanged for every
seed. The completed triple therefore dominates the representative
loads coordinatewise. Since $\boldsymbol J$ was sampled independently
of $\omega$, the completed triple uses the original fair coin and
clock laws. Theorem~\ref{thm:n3normalized} bounds its expected
makespan by two for every fixed triple. Averaging gives
$\E[C_{\max}]\le2T$.

\subsection{A complete closed support cover}\label{n3:cover}
A \emph{support mask} specifies the eligible machines $S_j$ for each
job $j$. Once these sets are fixed, only the supported times enter
the clock comparisons or contribute to the loads. The proof groups
normalized profiles by their support masks and, in a full column, by which alternative has the smaller time. A column with one, two, or
three eligible machines is called a \emph{singleton}, \emph{two-edge},
or \emph{full} column. A singleton job is forced to its sole eligible
machine. An \emph{owner singleton} is forced to its owner.

Every owner entry is one, so each column minimum is at most one.
Ineligible nonowner entries may be set to three. If the owner is
eligible, the possibilities are
\begin{enumerate}
\item the owner alone.
\item the owner and one alternative of cost $x\in[1/2,2]$, with two
choices of alternative.
\item all three machines, with alternative costs
$1/2\le x\le y\le2$ and $y\le2x$, and two choices of the cheaper
alternative.
\end{enumerate}
If the owner is excluded, the minimum is strictly below $1/2$.
Hold $S_j$ fixed and multiply all its supported times by the same
factor until the minimum is $1/2$. This leaves every clock comparison
unchanged and can only increase the realized makespan. We bound this
limiting column with its original eligible set. The possibilities are
\begin{enumerate}[resume]
\item one forced alternative of cost $1/2$, with two choices of machine.
\item both alternatives, with costs $1/2$ and $b\in[1/2,1]$, with two
choices of the minimum.
\end{enumerate}
This gives $1+2+2+2+2=9$ patterns per owner. At an actual input whose
minimum is $1/2$, the cutoff includes the owner. Such inputs occur
in the first three cases. These closed comparison domains may overlap,
which does not affect an upper bound.

Simultaneously relabeling machines and their owned jobs preserves
both clock laws. Indeed, every permutation of the three labels is
$i\mapsto ai+b\pmod3$, where $a\in\{1,-1\}$. This permutes the cyclic
clock labels, and also permutes the unordered-pair labels.
We count the resulting equivalence classes of patterns. A three-cycle
fixes $9$ patterns, since the choice at one owner determines the other
two. For a transposition, the owner it fixes must have an owner
singleton because every other pattern distinguishes one of the two
exchanged alternatives. The choice at either exchanged owner then determines
the other. Thus each nonidentity permutation fixes $9$ patterns, and
Burnside's lemma gives
\[
 \frac{9^3+3\cdot9+2\cdot9}{6}=129.
\]

\begin{table}[tbp]
\centering
\begin{tabular}{p{0.73\linewidth}r}
\toprule
Proof family (applied in order) & Orbits\\\midrule
At most one nonsingleton column (Lemma~\ref{lem:n3:one-random}) &34\\
All owners excluded, not already counted (Lemma~\ref{lem:n3:excluded}) &6\\
One owner singleton and two excluded two-edge columns (Lemma~\ref{lem:n3:fixed-excluded}) &3\\
One forced job with excluded owner and two two-edge columns with eligible owners (Lemma~\ref{lem:n3:forced-binary}) &4\\
All owners eligible, no full column, at most two two-edge columns (Lemma~\ref{lem:n3:owned-binary}) &3\\
One full column, one two-edge column, one owner singleton (Lemmas~\ref{lem:n3:chain}--\ref{lem:n3:star}) &4\\\midrule
Direct bounds &54\\
All three columns full &2\\
Remaining mixed patterns &73\\\midrule
Total &129\\\bottomrule
\end{tabular}
\caption{Coverage of the normalized support patterns. Apply the rows in
order, excluding patterns counted in earlier rows. A boundary profile
may occur in more than one pattern.}\label{tab:n3-cover}
\end{table}
Table~\ref{tab:n3-cover} assigns each pattern to a proof. The direct
bounds cover $54$ symmetry classes. The remaining $75$, including
three two-edge columns with eligible owners, are treated in
Section~\ref{n3:computational}.

\subsection{Direct bounds for the smaller support patterns}\label{n3:analytic}
Some masks require only the marginal probabilities $q_{ij}$. For
example, if two jobs are forced, the third job is the sole source
of randomness in the makespan. When several jobs can move, the proofs
below single out the complete assignment whose probability matters.
Its coefficient in the expected makespan tells us whether we need a
lower or an upper bound on that probability.

\subsubsection{At most one nonsingleton column}
\begin{lemma}[One nonsingleton column]\label{lem:n3:one-random}
In the normalized closed-mask cover, if at most one column is
nonsingleton, then $\E[C_{\max}]\le2$ under either clock map.
\end{lemma}
\begin{proof}
Let $j$ be the possibly nonsingleton column, or any column if all are
singletons. Its diagonal owner is machine $j$. Only this job can be random,
so coupling is immaterial. Write $L_i^{\mathrm{fix}}$ for the load on
machine $i$ from the other two jobs. A fixed job costs $1$ on its owner;
if it is forced to another machine, increase its cost to $1/2$.
These increases can only increase the makespan.

First suppose $j\in S_j$. Then $t_{jj}=1$ and the other supported costs
lie in $[1/2,2]$. Conditional on assigning job $j$ to machine $i$, the
makespan is
\[
 C_{\max}^{(i)}=\max\{L_i^{\mathrm{fix}}+t_{ij},\max_{k\ne i}L_k^{\mathrm{fix}}\}.
\]
Thus the inverse-square rule gives
\begin{equation}\label{eq:n3:6}
 \E[C_{\max}]-2
 =\frac{\sum_{i\in S_j}(C_{\max}^{(i)}-2)t_{ij}^{-2}}
             {\sum_{i\in S_j}t_{ij}^{-2}}.
\end{equation}
The denominator is positive. The following cases bound its numerator.
\begin{itemize}
\item \textbf{Both fixed jobs stay.} Then $C_{\max}^{(j)}=1$ and
$C_{\max}^{(i)}=1+t_{ij}$ for $i\ne j$. With two alternatives $i,k$,
\[
 \E[C_{\max}]-1
 =\frac{t_{ij}^{-1}+t_{kj}^{-1}}{1+t_{ij}^{-2}+t_{kj}^{-2}}
 \le\frac1{\sqrt2}<1.
\]
The inequality follows from Cauchy--Schwarz and
$2(t_{ij}^{-2}+t_{kj}^{-2})\le(1+t_{ij}^{-2}+t_{kj}^{-2})^2/2$.
With one alternative, its expected work is $t_{ij}/(1+t_{ij}^2)\le1/2$.
\item \textbf{One stays and the other is forced to the same machine.}
The sole fixed load is $3/2$ on a machine $i\ne j$. The owner term in
\eqref{eq:n3:6} is $-1/2$, and the only potentially positive term is
$(t_{ij}-1/2)/t_{ij}^2\le1/2$, since $(t_{ij}-1)^2\ge0$.
An absent edge removes this term.
\item \textbf{One stays and the other is forced to machine $j$.}
Here $L_j^{\mathrm{fix}}=1/2$ and $L_i^{\mathrm{fix}}=1$ for one $i\ne j$.
The owner term is $-1/2$, and the only potentially positive alternative
term is $(t_{ij}-1)/t_{ij}^2\le1/4$.
\item \textbf{Both are forced.} Their total fixed load is $1$, and every
machine other than $j$ has fixed load at most $1/2$. If $L_j^{\mathrm{fix}}=1$,
every completion has makespan at most $2$. Otherwise the owner term is
at most $-1/2$, while each of at most two positive alternative terms is
at most $(t_{ij}-3/2)_+/t_{ij}^2\le1/8$ for $t_{ij}\le2$.
\end{itemize}

Now suppose $j\notin S_j$. Every supported cost is at most $1$.
The makespan is pointwise at most $2$ unless one machine $i$ has fixed
load $3/2$. In that case,
$C_{\max}\le3/2+t_{ij}\ind\{A(j)=i\}$. A singleton contributes at
most $1/2$. With two supported machines $i,k$, their minimum cost is
at most $1/2$, and the expected work on $i$ is
\[
 t_{ij}q_{ij}=\frac{t_{ij}t_{kj}^2}{t_{ij}^2+t_{kj}^2}\le\frac12.
\]
Indeed, if $t_{ij}\le1/2$, this is at most $t_{ij}$; otherwise
$t_{kj}\le1/2$, and $t_{ij}^2+t_{kj}^2\ge2t_{ij}t_{kj}$ bounds it by
$t_{kj}/2$. This completes the $34$-orbit family.
\end{proof}

\subsubsection{Excluded owners and short two-edge families}
\begin{lemma}[All owners excluded]\label{lem:n3:excluded}
If all owners are excluded in a normalized closed mask, then
$\E[C_{\max}]\le9/5$ under any coupling with the prescribed $q_{ij}$.
\end{lemma}
\begin{proof}
A two-edge column at the upper-limit costs $(1/2,b)$ has expected processing work
\[
 \frac{2b^2+b}{4b^2+1}\le\frac35,\quad \frac12\le b\le1,
\]
since the inequality is equivalent to $(2b-3)(b-1)\ge0$. A singleton costs at most $1/2$. Makespan is at most total work, giving $9/5$ under any coupling.
\end{proof}

\begin{lemma}[One owner singleton and two excluded columns]\label{lem:n3:fixed-excluded}
In a normalized closed mask with one owner singleton and two excluded
owners whose columns each have two supported edges, $\E[C_{\max}]\le9/5$
under any coupling with the prescribed $q_{ij}$.
\end{lemma}
\begin{proof}
The machine receiving the owner singleton has load $1$ plus incoming work. Either other row receives at most one of the excluded jobs, costing at most $1$, so that machine attains the maximum. Each incoming expectation is either
\[
 \frac{2b^2}{4b^2+1}\le\frac25
 \quad\text{or}\quad \frac b{4b^2+1}\le\frac14.
\]
Thus $\E[C_{\max}]\le9/5$. The two-edge assumption matters; a singleton instead falls in Lemma~\ref{lem:n3:one-random}.
\end{proof}

\begin{lemma}[At most two two-edge columns with eligible owners]\label{lem:n3:owned-binary}
In a normalized closed mask with all owners eligible, no full column,
and at most two two-edge columns, $\E[C_{\max}]\le2$ under any coupling with
the prescribed $q_{ij}$.
\end{lemma}
\begin{proof}
A machine receives at most one unit of work from its own job, so
$C_{\max}\le1+\sum_{j:A(j)\ne j}t_{A(j),j}$ in every outcome.
If machine $i$ is the unique alternative for job $j$, that job's
expected work away from its owner is
$t_{ij}q_{ij}=t_{ij}/(1+t_{ij}^2)\le1/2$. Since at most two jobs
can move, $\E[C_{\max}]\le2$.
\end{proof}

\subsubsection{A forced column and two columns with eligible owners}
\begin{lemma}[A forced column and two two-edge columns]\label{lem:n3:forced-binary}
In a normalized closed mask with one forced excluded-owner column and
two two-edge columns with eligible owners, the fixed fair mixture satisfies $\E[C_{\max}]\le2$
for every pair of supported alternative costs $x,y\in[1/2,2]$.
\end{lemma}
\begin{proof}
\textbf{Normalize the pattern and enumerate assignments.}
The forced job contributes only to its destination's load, so
increasing its cost to $1/2$ can only increase the makespan. The two
clock laws are invariant under simultaneous relabeling of machines
and owners. Label this job's owner $2$ and its destination $0$.
Let $i_0\in\{1,2\}$ and $i_1\in\{0,2\}$ be their jobs' alternative
destinations. Write $x=t_{i_0,0}$ and $y=t_{i_1,1}$ for the two
variable costs, both in $[1/2,2]$. The probabilities of taking the
alternatives are $q_{i_0,0}=1/(1+x^2)$ and $q_{i_1,1}=1/(1+y^2)$.
The probability that both jobs move is the complete-assignment
probability $y_{(i_0,i_1,0)}$, since job~$2$ always goes to machine~$0$.
The table gives the makespan for each choice of destinations.
\begin{center}
\begin{tabular}{ccccc}
\toprule
$(i_0,i_1)$ & Both stay & Only $0$ moves & Only $1$ moves & Both move\\\midrule
$(1,0)$ & $3/2$ & $1+x$ & $3/2+y$ & $\max(x,y+1/2)$\\
$(1,2)$ & $3/2$ & $1+x$ & $\max(3/2,y)$ & $\max(x,y)$\\
$(2,0)$ & $3/2$ & $\max(1,x)$ & $3/2+y$ & $\max(x,y+1/2)$\\
$(2,2)$ & $3/2$ & $\max(1,x)$ & $\max(3/2,y)$ & $x+y$\\\bottomrule
\end{tabular}
\end{center}
\textbf{The probability that both jobs move.}
All owner comparisons use the same owner clock. If the alternative
clock is shared as well, the two moves compare the same pair of clocks
with different thresholds. The events are nested and their joint
probability is $\min(q_{i_0,0},q_{i_1,1})$.
When their alternative clocks are distinct, condition on the shared
owner clock. Its density and the two independent move probabilities give
\begin{equation}
\begin{aligned}
 \int_0^\infty e^{-z}(1-e^{-z/x^2})(1-e^{-z/y^2})\operatorname{d}z
 &=q_{i_0,0}+q_{i_1,1}-1+\frac1{1+x^{-2}+y^{-2}}\\
 &=q_{i_0,0}q_{i_1,1}\left(1+\frac1{1+x^{-2}+y^{-2}}\right).
\end{aligned}                                                    \label{eq:n3:7}
\end{equation}
For $(i_0,i_1)=(1,0)$ the pair map shares the alternative clock. For
$(1,2)$ and $(2,0)$ the cyclic map shares it. In all three rows
the fair coin therefore selects the nested events half the time and
the distinct alternatives half the time. The marginals and the
probability of both moves determine all four move patterns, so their
distribution is the same in these three rows. Since
$x,y\ge1/2$, averaging the shared- and distinct-clock probabilities gives
\[
 y_{(i_0,i_1,0)}\ge
 \frac{\min(q_{i_0,0},q_{i_1,1})+(10/9)q_{i_0,0}q_{i_1,1}}2 .
\]
For every choice of moves the first row of the table has makespan at
least as large as the next two. It remains to bound the first and last rows.

\textbf{The reverse pair $(1,0)$.}
A single move puts both nonforced jobs on one machine. When both move,
they exchange machines. The first row therefore gives
\[
 \E[C_{\max}]=\frac32+(x-1/2)q_{10}+yq_{01}
                  -\min\{x+1/2,y+1\}\,y_{(1,0,0)}.
\]
Substituting the lower bound on $y_{(1,0,0)}$ and clearing the
positive denominators gives
\begin{equation}\label{n3:forced-remainder}
\begin{aligned}
18(1+x^2)(1+y^2)\bigl(2-\E[C_{\max}]\bigr)
 &\ge 9\bigl((x-1)^2+1\bigr)(1+y^2)-18y(1+x^2)\\
 &\quad+\min\{x+1/2,y+1\}\bigl(19+9\min\{x^2,y^2\}\bigr).
\end{aligned}
\end{equation}
It suffices to show that the right side is positive in each of the
three regions determined by the minima.
\begin{enumerate}
\item If $x\le y$, the right side is quadratic in $y$, with leading
coefficient $9((x-1)^2+1)>0$. Completing the square gives the lower bound
\[
 9x^3+\frac{27}{2}x^2+x+\frac{55}{2}
       -\frac{9(1+x^2)^2}{(x-1)^2+1}.
\]
Since $8((x-1)^2+1)-3(1+x^2)=5(x-8/5)^2+1/5>0$, the subtracted
fraction is less than $24(1+x^2)$. The remainder exceeds
$9(x-1/2)^3+3(x-1)^2+(x-1/2)/4+7/4>0$.
\item If $y\le x\le y+1/2$, the right side expands to
\[
 \tfrac92(y-2)^2+\tfrac{23}{2}+9(y-1)^2x^2+(1-9y^2)(x-2)>0 .
\]
The last product is nonnegative since $y\ge1/2$ and $x\le2$.
\item If $x\ge y+1/2$, then $1/2\le y\le3/2$. Expansion around $x=2$ gives
\[
 (2-x)\bigl(18[3-(y-2)^2]+9(y-1)^2(2-x)\bigr)
       +1+(5y-4)^2+(y-1)^2(9y+20)>0 ,
\]
since $2-x\ge0$ and $3-(y-2)^2\ge3/4$.
\end{enumerate}
This proves $\E[C_{\max}]<2$ throughout the closed square.

This pattern can favor the pair map. For example, take $x=y=2$ and
give job $2$ time $12/25$ on machine $0$ before increasing that time
to $1/2$. The probabilities of both moves are $1/15$ for the cyclic
map by~\eqref{eq:n3:7} and $1/5$ for the pair map, where the two
events coincide. With this forced time the four makespans, in table
order, are $1+12/25$, $3$, $3+12/25$, and $2+12/25$.
Thus $\E[C_{\rm cyc}]=252/125>2$, whereas
$\E[C_{\rm pair}]=42/25<2$. The fair mixture averages the two.
 
\textbf{The common destination $(2,2)$.}
This pattern has expected makespan at most two under any coupling. Its expectation is affine in $y_{(2,2,0)}$, with coefficient
\[
 x+y-\max(1,x)-\max(3/2,y)+3/2
   =\min(x,1)+\min(y,3/2)-1\ge0.
\]
It is therefore bounded above by setting $y_{(2,2,0)}=\min(q_{20},q_{21})$.
We use the elementary inequalities
\[
 \frac{t}{1+t^2}\le\frac12,\qquad
 \frac32+\frac{2t-3/2}{1+t^2}\le2,
\]
and the fact that $t/(1+t^2)$ decreases for $t\ge1$. The second
inequality is equivalent to $(t-2)^2\ge0$.

If $x\le y$, the upper bound is
\[
 \frac32+(\max(1,x)-3/2)q_{20}+[x+y-\max(1,x)]q_{21}.
\]
For $x\le1$ this is at most $3/2-q_{20}/2+yq_{21}\le2$; for $x\ge1$
it is at most $3/2+(2x-3/2)q_{20}\le2$, since $yq_{21}\le xq_{20}$.
If $x\ge y$, the upper bound is
\[
 \frac32+(\max(3/2,y)-3/2)q_{21}+[x+y-\max(3/2,y)]q_{20}.
\]
For $y\le3/2$ this is at most $3/2+(2x-3/2)q_{20}\le2$; for
$y\ge3/2$ it is at most $3/2+(2y-3/2)q_{21}\le2$, since $xq_{20}\le yq_{21}$.
This covers all eight original destination choices, both cost orders,
and every endpoint of $[1/2,2]^2$ using the same fair mixture.
\end{proof}

\subsubsection{A chain configuration with one full column}
\begin{lemma}[Chain configuration]\label{lem:n3:chain}
For a normalized closed mask of the following form, each clock map
separately satisfies $\E[C_{\max}]\le2$:

\[
 \begin{pmatrix}1&-&t_{02}\\-&1&t_{12}\\-&t_{21}&1\end{pmatrix},\qquad
 t_{21},t_{02},t_{12}\in[1/2,2],\quad t_{02}\le2t_{12},\quad t_{12}\le2t_{02}.
\]
Rows and columns are indexed by $0,1,2$ in that order, and each
column has its diagonal machine as owner. A dash denotes an ineligible edge.
\end{lemma}
\begin{proof}
\textbf{Assignment costs.}
Job~$0$ stays on machine~$0$. Job~$1$ can move to machine~$2$, and
job~$2$ can go to any machine. The Bounded-Square probabilities for
job~$2$ are
\[
 q_{22}=\frac1{1+t_{02}^{-2}+t_{12}^{-2}},\qquad
 q_{02}=q_{22}/t_{02}^2,\qquad q_{12}=q_{22}/t_{12}^2 .
\]
If job $2$ goes to machine $0$ or $1$, the excess above one unit
is at most $\max(t_{02},t_{21}-1)$ or $\max(t_{12},t_{21}-1)$,
respectively. If it stays on machine $2$, there is an excess only
when job $1$ joins it, giving the assignment $(0,2,2)$. Since
$t_{21}\le2$, for either fixed clock map $\kappa$,
\begin{equation}
 \E[C_\kappa]\le1+\max(t_{02},1)q_{02}+\max(t_{12},1)q_{12}
                       +t_{21}y^\kappa_{(0,2,2)}. \label{eq:n3:9}
\end{equation}
\paragraph{Why the two jobs are unlikely to meet on machine $2$.}
Both maps use one owner clock. For job $2$ to stay, this clock must
be small relative to its two competitors. For job $1$ to move, one of
those competitor clocks must be small relative to the owner clock.
In the cyclic map it is the clock for job $2$ going to machine $0$.
In the pair map it is the clock for job $2$ going to machine $1$.

Condition on job $2$ staying. The owner-clock value has conditional
density $e^{-u/q_{22}}/q_{22}$. To see this, multiply its density
$e^{-u}$ by the probability $e^{-u(t_{02}^{-2}+t_{12}^{-2})}$
that both competitors lose, and divide by $q_{22}$. For fixed $u$,
each competitor clock is conditioned to exceed its own threshold.
By memorylessness, it is that threshold plus a rate-one exponential.
Thus the probability that the shared competitor clock is below
$u/t_{21}^2$, as required for job $1$ to move, is at most
$1-e^{-u/t_{21}^2}$. Job $0$ is forced to machine $0$, so the
conditional probability that job $1$ moves given that job $2$ stays
is $y^\kappa_{(0,2,2)}/q_{22}$. For either map,
\[
 \frac{y^\kappa_{(0,2,2)}}{q_{22}}
 \le\int_0^\infty \frac{e^{-u/q_{22}}}{q_{22}}
                   (1-e^{-u/t_{21}^2})\,\mathrm{d}u
 =\frac{q_{22}}{q_{22}+t_{21}^2}\le q_{21},
\]
where $q_{21}=1/(1+t_{21}^2)$. Hence
$t_{21}y^\kappa_{(0,2,2)}\le q_{22}t_{21}q_{21}\le q_{22}/2$.
\paragraph{Bound the expectation.}
For $1\le t_{02}\le2$, the inequality $(t_{02}-1)/t_{02}^2\le1/4$
is equivalent to $(t_{02}-2)^2\ge0$. When $t_{02}\le1$, we have $(t_{02}-1)_+=0$. The same bounds hold for $t_{12}$. Using
$q_{02}+q_{12}=1-q_{22}$ in~\eqref{eq:n3:9} now gives
\[
\E[C_\kappa]\le1+(1-q_{22})+
 q_{22}\left(\frac{(t_{02}-1)_+}{t_{02}^2}
         +\frac{(t_{12}-1)_+}{t_{12}^2}+\frac12\right)
 \le2 .
\]
\end{proof}

\Needspace{9\baselineskip}
\subsubsection{A star configuration with one full column}
\begin{lemma}[Star configuration]\label{lem:n3:star}
For a normalized closed mask of the following form, the fixed fair
mixture satisfies $\E[C_{\max}]<2$:

\[
 \begin{pmatrix}1&-&y\\x&1&z\\-&-&1\end{pmatrix},\qquad
 x,y,z\in[1/2,2],\quad y\le2z,\quad z\le2y.
\]
Rows and columns are indexed by $0,1,2$ in that order, and each
column has its diagonal machine as owner. A dash denotes an ineligible edge.
\end{lemma}
\begin{proof}
The three variable entries of the displayed matrix are $x=t_{10}$,
$y=t_{02}$, and $z=t_{12}$. Job~$1$ is forced to machine~$1$.
Job~$0$ may move to machine~$1$, and job~$2$ may go to any machine.
The probabilities used below follow directly from~\eqref{eq:n3:1}:
\begin{equation}\label{eq:n3:10}
 q_{10}=\frac{1}{1+x^2},\qquad q_{02}=\frac{1}{1+y^2+y^2/z^2},
 \qquad q_{12}=\frac{y^2}{z^2}q_{02}.
\end{equation}
Enumerating these assignments gives
\[
 C_{\max}=1+x\ind\{A(0)=1\}+z\ind\{A(2)=1\}+y\ind\{A(2)=0\}
              -\min(y,1+x)\ind\{A=(1,1,0)\}.
\]
The last term occurs when jobs~$0$ and~$1$ give machine~$1$ load $1+x$
and job~$2$ gives machine~$0$ load $y$. The makespan is the larger of
these two loads.

\paragraph{The probability of assignment $(1,1,0)$.}
Here a larger probability of $(1,1,0)$ lowers the expected makespan.
In the cyclic map, the moves of job~$0$ to machine~$1$
and job~$2$ to machine~$0$ share an alternative clock. Conditioning
on its value $u$ gives
\[
 y^{\rm cyc}_{(1,1,0)}=\int_0^\infty e^{-u[1+\max(x^2,y^2)+y^2/z^2]}\,\mathrm{d}u
 =\frac1{1+\max(x^2,y^2)+y^2/z^2}.
\]
In the pair map the moves use distinct alternative clocks and the
same owner clock. Condition on the alternative clock $u$ for job~$2$
and the owner clock $v$. The other alternative of job~$2$ must lose,
the owner clock must exceed $y^2u$, and the clock for the move of
job~$0$ must be below $v/x^2$. Thus
\[
 \begin{aligned}
 y^{\rm pair}_{(1,1,0)}
 &=\int_0^\infty e^{-u(1+y^2/z^2)}
       \int_{y^2u}^\infty e^{-v}(1-e^{-v/x^2})\,\mathrm{d}v\,\mathrm{d}u\\
 &=q_{02}-\frac{x^4 q_{02}}{(1+x^2)(x^2+y^2 q_{02})}.
 \end{aligned}
\]
When job $2$ goes to machine $0$, the only other possible assignment
is $(0,1,0)$. The subtracted term is therefore $y^{\rm pair}_{(0,1,0)}$.
On the profile in Section~\ref{n3:allocation}, $x=z=1$ and $y=2$,
so this term is $1/26$. The pair map alone has expected makespan
$473/234>2$ there, and the fair mixture has expected makespan $232/117<2$.
Averaging the two maps in the pointwise makespan formula gives
\begin{equation*}
\begin{aligned}
 \E[C_{\max}]
 &=1+xq_{10}+zq_{12}+yq_{02}\\
 &\quad-\frac{\min(y,1+x)}2
        \bigl(y^{\rm cyc}_{(1,1,0)}+q_{02}-y^{\rm pair}_{(0,1,0)}\bigr).
\end{aligned}
\end{equation*}
The marginals in~\eqref{eq:n3:10} give
$1-q_{02}-2zq_{12}=(z-1)^2q_{12}$. When $y\le1+x$, substituting
this identity in the expectation gives
\[
 \begin{aligned}
 2(2-\E[C_{\max}])={}&\frac{(x-1)^2}{1+x^2}+q_{02}+(z-1)^2q_{12}\\
 &-y\bigl(q_{02}-y^{\rm cyc}_{(1,1,0)}+y^{\rm pair}_{(0,1,0)}\bigr).
 \end{aligned}
\]
We distinguish three cases. The cyclic probability changes formula
at $y=x$, and the load correction $\min(y,1+x)$ changes at $y=1+x$.

\paragraph{When $x\le y\le1+x$.}
If job $2$ moves to machine $0$ under the cyclic map, job $0$ also
moves, so $y^{\rm cyc}_{(1,1,0)}=q_{02}$. Only the pair-map term
$y^{\rm pair}_{(0,1,0)}$ is subtracted in the last display.
For $x\le1$, its formula gives
$y^{\rm pair}_{(0,1,0)}<q_{02}x^2/(1+x^2)\le q_{02}/2$. Since
$y\le2$, the positive term $q_{02}$ alone covers this subtraction.

Now suppose $1\le x\le y\le2$. The cutoff $y\le2z$ gives
$y^2q_{02}\ge y^2/(5+y^2)$. Since also $x^2\ge1$, we obtain
\[
 \frac{x^4}{x^2+y^2q_{02}}
 =x^2-\frac{1}{1/x^2+1/(y^2q_{02})}
 \le x^2-\frac{y^2}{5+2y^2}.
\]
Use this bound in $y^{\rm pair}_{(0,1,0)}$ and use
$1/q_{02}\ge1+y^2$. Dropping the nonnegative $(z-1)^2q_{12}$ gives
\[
 \frac{2(1+x^2)}{q_{02}}(2-\E[C_{\max}])
 \ge 1+(1-y)x^2+(x-1)^2(1+y^2)+\frac{y^3}{5+2y^2}.
\]
Completing the square in $y(x-1)$ and using $2-y\ge0$ bounds the
right side below by $2-y+y^3/(5+2y^2)-(1-1/y)^2$.
Here $2-y+y^3/(5+2y^2)\ge2-y+y^3/13\ge8/13$, since the latter
expression decreases on $[1,2]$. The subtracted square is at most
$1/4$. Thus $\E[C_{\max}]<2$.

\paragraph{When $y<x$.}
Now the cyclic move of job $2$ may occur without the move of job $0$.
Its missing overlap and the pair-map term satisfy
\[
 \frac{q_{02}-y^{\rm cyc}_{(1,1,0)}}{q_{02}}
 =\frac{x^2-y^2}{1+x^2+y^2/z^2}\le\frac{x^2-y^2}{1+x^2},
 \qquad
 \frac{y^{\rm pair}_{(0,1,0)}}{q_{02}}<\frac{x^2}{1+x^2}.
\]
This time retain the slack $(z-1)^2q_{12}$. The two estimates give
\[
 \frac{2(1+x^2)}{q_{02}}(2-\E[C_{\max}])
 \ge \frac{(x-1)^2}{q_{02}}
       +(1+x^2)\bigl[1+y^2(1-1/z)^2\bigr]-y(2x^2-y^2).
\]
Expand $1/q_{02}=1+y^2+y^2/z^2$. Since $1+x^2\ge(x-1)^2$ and
$z^{-2}+(1-1/z)^2\ge1/2$, the right side is at least
\[
 y^3-2x^2y+1+x^2+(x-1)^2(1+\tfrac32y^2).
\]
To see its sign, write $u=x-1\in[-1/2,1]$. Expand in powers of
$y-1$ and use $(y-1)^3\ge-(y-1)^2/2$. This gives the lower bound
\[
 (\tfrac52+\tfrac32u^2)(y-1)^2+(1-4u+u^2)(y-1)+1-2u+\tfrac32u^2 .
\]
Its leading coefficient is positive. The negative of its discriminant is
\[
 (1-u)\bigl[3(3-u)-4u^3\bigr]+4u^4>0 .
\]
Indeed the bracket is at least $6-4=2$, and at $u=1$ the last term
is positive. The quadratic is therefore positive throughout this case.

\paragraph{When $y>1+x$.}
Here $1/2\le x<1$, $y>3/2$, and $y^{\rm cyc}_{(1,1,0)}=q_{02}$.
The load correction is now $1+x$. Since $y\le2z$,
\[
 y^2q_{02}\ge\frac{y^2}{5+y^2}\ge\frac9{29}>\frac14,\qquad
 q_{02}\le\frac1{1+y^2}<\frac12 .
\]
Hence $x^2/(x^2+y^2q_{02})\le4/5$. Substitution in the expectation,
using $2zq_{12}\le1-q_{02}$ and $y\le2$, gives
\[
 2(1+x^2)(2-\E[C_{\max}])
 \ge (1-x)^2-q_{02}\left[(1+x^2)(1-2x)+\tfrac45(1+x)x^2\right].
\]
The bracket is less than $2(1-x)^2$. Indeed, five times their
difference, with the bracket subtracted, is
\[
 (6x+9)(x-2/3)^2+1-2x/3>0 .
\]
Since $2q_{02}<1$, the gap is positive. These three cases prove the
bound for the fair mixture.
\end{proof}

With one full column, one two-edge column, and one owner singleton,
the alternative of the two-edge job leads either to the full column's
owner (the chain) or to the singleton's owner (the star). These two
lemmas finish the $54$ direct cases in Table~\ref{tab:n3-cover}.

\subsection{The finite rational certificate for the other orbits}\label{n3:computational}
The remaining $75$ patterns have too many variable costs for the
preceding calculations. We again express expected makespan in terms
of complete-assignment probabilities. This time we bound them uniformly
on boxes of costs, using the shared clock comparisons.

Two of the patterns have three full columns. Draw an arrow from each
owner to its cheaper alternative. The arrows form either a three-cycle
or a two-cycle with a tail, giving the two all-full orbits. The other
$73$ patterns have mixed supports. For each pattern the initial box is
intersected with the constraints $x\le y\le2x$ on each full column.
For intervals $[\underline x,\overline x]$ and
$[\underline y,\overline y]$, these constraints imply
\[
 \max\{\underline x,\underline y/2\}\le x\le\min\{\overline x,\overline y\},
 \qquad
 \max\{\underline y,\underline x\}\le y\le\min\{\overline y,2\overline x\}.
\]
The verifier tightens the intervals this way and splits boxes at
recorded rational interior points. Both closed children are retained,
so their union covers the parent, including the splitting boundary.
Every branch ends at an empty box or at a box with the bound proved
below.

\subsubsection{Uniform probability and cost bounds on a box}
Fix a box $\underline t\le t\le\overline t$ in one support mask,
and fix one clock map $\kappa$. Job $j$ has owner $j$. A supported
assignment $A$ occurs exactly when the score of its winner is smallest
in every column. Up to ties of probability zero, these comparisons are
\begin{equation}
 t_{A(j),j}^{2}E_{\kappa(A(j),j)}\le t_{ij}^{2}E_{\kappa(i,j)}
 \qquad(j=0,1,2,\ i\in S_j\setminus\{A(j)\}). \label{eq:n3:12}
\end{equation}
All these comparisons use a single set of clock variables. In
particular a label repeated across two columns denotes the same
variable in their two inequalities. Replacing each winner time by its
lower endpoint and each competitor time by its upper endpoint enlarges
this event. Its probability gives an upper bound $y_A^+$ on
$y_A^\kappa(t)$ for every profile in the box. The opposite replacements
give a lower bound $y_A^-$. These two bounds are specific to $\kappa$.
The one-job bounds are the same for both maps.
For each supported edge $(i,j)$, the inverse-square rule gives
$q_{ij}^-\le q_{ij}\le q_{ij}^+$, where
\begin{equation}\label{eq:n3:13}
 q_{ij}^-=\frac{\overline t_{ij}^{-2}}
 {\overline t_{ij}^{-2}+\sum_{k\in S_j\setminus\{i\}}\underline t_{kj}^{-2}},\qquad
 q_{ij}^+=\frac{\underline t_{ij}^{-2}}
 {\underline t_{ij}^{-2}+\sum_{k\in S_j\setminus\{i\}}\overline t_{kj}^{-2}}.
\end{equation}
For an unsupported edge, set $q_{ij}^-=q_{ij}^+=0$.
The times at the upper endpoints also give a bound on each makespan,
\begin{equation}
 C_{\max}(t,A)\le C_A^+=\max_i\sum_{j:A(j)=i}\overline t_{ij}. \label{eq:n3:14}
\end{equation}
For every profile in the box the distribution of this clock map
satisfies
\begin{equation}\label{eq:n3:15}
 y_A^-\le y_A^\kappa(t)\le y_A^+,\qquad \sum_A y_A^\kappa(t)=1,\qquad
 q_{ij}^-\le\sum_{A:A(j)=i}y_A^\kappa(t)\le q_{ij}^+.
\end{equation}
Hence maximizing the average of $C_A^+$ over all distributions
satisfying these bounds can only increase $\E[C_\kappa]$.
A rational dual witness bounds that maximum for the entire box.

\subsubsection{Rational dual witnesses}
Use nonnegative rational weights
$\lambda_{ij},\mu_{ij}$ for the upper and lower bounds on the $q_{ij}$,
and $\alpha_A,\beta_A$ for the upper and lower joint-probability
bounds of the fixed map. Let $\nu$ be a free rational weight for the
total probability constraint.
If, for every supported assignment,
\begin{equation}
 \nu+\sum_j(\lambda_{A(j),j}-\mu_{A(j),j})+\alpha_A-\beta_A\ge C_A^+,               \label{eq:n3:16}
\end{equation}
then weak duality gives the uniform upper bound
\begin{equation}
 U_\kappa=\nu+\sum_{i,j}\bigl(\lambda_{ij}q_{ij}^+-\mu_{ij}q_{ij}^-\bigr)
          +\sum_A\bigl(\alpha_A y_A^+-\beta_A y_A^-\bigr). \label{eq:n3:17}
\end{equation}
Indeed, multiply~\eqref{eq:n3:16} by $y_A^\kappa(t)$ and sum over
supported assignments. For each $i,j$, the terms with $A(j)=i$ sum
to $q_{ij}(t)$, giving
\[
 \begin{aligned}
 \E[C_\kappa]&\le\sum_A C_A^+y_A^\kappa(t)\\
 &\le\nu+\sum_{i,j}(\lambda_{ij}-\mu_{ij})q_{ij}(t)
        +\sum_A(\alpha_A-\beta_A)y_A^\kappa(t)\le U_\kappa .
 \end{aligned}
\]
The last inequality uses upper bounds for quantities with positive
weights and lower bounds for those being subtracted.

For each map the verifier checks that the proposed weights are
nonnegative and chooses rational multipliers with denominators at most
$10^8$ (Appendix~\ref{supp:n3replay}). Using these multipliers, it sets
\[
 \nu=\max_A\left[
 C_A^+-\sum_j(\lambda_{A(j),j}-\mu_{A(j),j})-\alpha_A+\beta_A
 \right]
\]
to enforce~\eqref{eq:n3:16}. It then evaluates~\eqref{eq:n3:17} by
exact arithmetic. A box is accepted when $U_{\rm cyc}+U_{\rm pair}\le4$.
This is the test for a single fair coin. One map's bound can exceed
two if the other is far enough below it. The verified weights need
only yield this inequality.

\subsubsection{Exact probabilities from simplex volumes}
To evaluate a bound $y_A^-$ or $y_A^+$, use the rational endpoint
times in~\eqref{eq:n3:12}. The resulting inequalities are homogeneous
and linear in the clocks. Delete unused clocks. If there are no
comparisons the event has probability one. Otherwise, normalizing the
remaining independent exponentials by their sum gives a uniform point
on a simplex. Thus the required probability is the volume of the
polytope cut out by the comparisons, divided by the volume of the whole
simplex. Appendix~\ref{supp:n3replay} gives the density calculation.

There are at most three active clocks in the cyclic map and four in
the pair map. The verifier intersects the comparison and nonnegativity
hyperplanes with the simplex, enumerates the feasible vertices, and
evaluates the relative volume by exact rational determinants. With
one or two active clocks it suffices to check a point or an interval.
In dimension two it orders polygon vertices and uses the shoelace
formula. In dimension three it triangulates the distinct polygonal
facets about an interior point. Lower-dimensional intersections have
zero relative volume in a positive-dimensional simplex.

\subsection{Finite verification and conclusion}\label{n3:reproducibility}
The rational verification of the remaining $75$ support orbits uses the
following finite cover:
\begin{center}
\begin{tabular}{lrrr}
\toprule
Group & Orbits & Tree records & Exact terminal boxes\\\midrule
All full support &2&65,380&32,691\\
Remaining mixed &73&116,127&58,100\\\midrule
Computational total &75&181,507&90,791\\\bottomrule
\end{tabular}
\end{center}
For every root, the two closed children at each split cover their
parent, and every branch ends at an empty or accepted box. Each of the
$90{,}791$ terminal boxes satisfies the exact dual inequalities
\eqref{eq:n3:16}--\eqref{eq:n3:17} for both clock maps, with
$U_{\rm cyc}+U_{\rm pair}\le4$. The verification uses rational
arithmetic throughout. Thus its conclusion is uniform over every real
profile in the cover, including boundary profiles.

The closed cover and its terminal witnesses establish~\eqref{eq:n3:2}
on these $75$ orbits. The direct bounds establish it on the other $54$,
proving Theorem~\ref{thm:n3normalized}. Appendix~\ref{supp:n3replay}
describes the replay and Appendix~\ref{supp:reproduction} gives the
code and instructions.

\begin{proof}[Proof of Theorem~\ref{thm:n3upper}]
Apply Theorem~\ref{thm:n3normalized} to each completed representative
triple in Section~\ref{n3:transfer}. Averaging bounds the original
expected makespan by $2T=2\OPT(t)$. Zero-minimum jobs contribute no
load, including when the optimum is zero. Section~\ref{n3:payments}
gives truthfulness in expectation and finite expected payments and
utilities. The lower bound in Theorem~\ref{thm:lower} at $n=3$ now
gives $\rho(3)=2$.
\end{proof}

\Needspace{12\baselineskip}
\section{Exact implementation}\label{sec:implementation}
On rational reports the slot matrix is rational, so the general
mechanism can draw a matching by sampling an integer. For two and three
machines, we must also find an auxiliary schedule and sample the
corresponding joint assignment law. Approximating that schedule keeps
each job's marginals and changes the makespan bound by at most the
approximation factor (Lemma~\ref{lem:ownertransfer}). For all three
algorithms, a payment integral can be irrational on rational reports.
We realize it as the mean of a rational bonus.

\subsection{Polynomial bit implementation of slot rounding}
\begin{proposition}\label{prop:slots-efficient}
For rational input, the Bounded-Square slot mechanism can be sampled exactly in expected time polynomial in the total input length, including a variable number of machines. It produces rational payments whose expectations satisfy the exact payment identity. Its approximation ratios from Theorem~\ref{thm:generalupper} are unchanged.
\end{proposition}
\begin{proof}
\textbf{A common rational grid.}
The empty instance has its empty assignment and zero payments. Otherwise
compute each $q_{ij}$ by rational arithmetic. Each column requires at
most $n$ inverse squares to sum and normalize, so these probabilities
have polynomial bit length. Let $D$ be the product of their denominators.
Its bit length is at most the sum of the denominator bit lengths.
The sorted interval endpoints, their overlaps with slots, and the
dummy entries filling slot deficits are all multiples of $1/D$.
There are $K=\sum_i\lceil\sum_jq_{ij}\rceil\le m+n$ slots since
$\sum_{i,j}q_{ij}=m$. Scaling the completed matrix by $D$ gives integer
entries and row and column sums $D$.

\paragraph{Decompose the slot matrix.}
Use the augmenting-path algorithm of
Section~\ref{upper:marginal-preserving-correlated-rounding}. Scaling
preserves the support, so at each step it selects the same matching
as on the original matrix. Subtract the smallest matched entry from
every edge of that matching. The integer subtracted, divided by $D$,
is its probability weight. Each step eliminates a positive entry, so
at most $K^2$ steps manipulate integers between zero and $D$.

\paragraph{Sample the allocation and payments.}
The integers subtracted sum to $D$. Use them as the sizes of consecutive
blocks in $\{0,\ldots,D-1\}$, one per matching. To select a uniform
integer, draw $\lceil\log_2D\rceil$ unbiased bits and repeat if the
value is at least $D$. Each draw succeeds with probability
$D/2^{\lceil\log_2D\rceil}>1/2$. For $D=1$ the sole block is selected
directly. Otherwise the accepted integer selects its block. Delete
its dummy jobs to obtain the assignment.

Fix machine $i$ and job $j$, and hold the other entries fixed.
The probability $q(s)=q_{ij}(s,t_{-ij})$ is nonincreasing in $s$,
can be evaluated exactly in polynomial time at rational $s$, and
vanishes above $B=2\min_{k\ne i}t_{kj}$. Lemma~\ref{lem:paymentbits} therefore
implements the per-job payment in expected polynomial time.
\end{proof}

\subsection{Approximate auxiliary schedules preserve incentives}
\begin{lemma}[Using an approximate auxiliary schedule]\label{lem:ownertransfer}
After removing zero-minimum jobs, use any auxiliary assignment
$A^{\mathrm{aux}}$ of makespan $T_{\mathrm{aux}}$ to choose the owners
before drawing the shared seed. If the two- or three-machine construction
has coupling bound $R$ on normalized instances, then its expected
makespan is at most $R T_{\mathrm{aux}}$. Its approximation factor is
thus at most $R$ times that of the auxiliary schedule. Its assignment
probabilities $q_{ij}$ and its truthfulness guarantee are the same for
every such choice of owners.
\end{lemma}
\begin{proof}
If no positive-minimum jobs remain, all loads are zero and the claim
is immediate. Otherwise \(T_{\mathrm{aux}}>0\).
Apply Lemma~\ref{lem:representative} to this auxiliary schedule.
Each normalized real representative has owner entry one. For two machines,
a pair's nominated-owner assignment has makespan one and hence optimum
at most one. For its matrix $P$, Lemma~\ref{n2:frechet} gives expected
makespan $C_f(P)\le R\OPT(P)\le R$.
A single representative is covered by \eqref{n2:one-job-bound}, and
two dummies cost zero. For three machines, the dummy completion in
Section~\ref{n3:transfer} and Theorem~\ref{thm:n3normalized} give $R=2$.
Consequently \eqref{eq:representative-bound} yields $\E[C_{\max}]\le R T_{\mathrm{aux}}$
in either case. Optimality of the auxiliary schedule was never used.

For either construction, each $q_{ij}$ is independent of the owner labels.
The choice of owners changes correlations but preserves each machine's
expected processing cost and expected payment on every report profile.
This proves the incentive assertion. Restoring zero-minimum jobs at
zero cost adds no load.
\end{proof}

\subsection{A rational owner approximation scheme}
Fix the number of machines and a rational $0<\varepsilon<1$.
The input length includes the binary encodings of the reports and of
$\varepsilon$. Remove zero-minimum jobs and retain a zero-cost owner
for each. If no jobs remain, return those owners. Otherwise, let $m$
be the number of remaining jobs and put $\underline T=\max_j\min_i t_{ij}>0$.
Assigning each job to a cheapest machine gives the upper bound in
\[
 \underline T\le\OPT\le\sum_j\min_i t_{ij}\le m\underline T.
\]
Set $K_\varepsilon=\lceil3m/\varepsilon\rceil$. Test the makespan guesses
\[
 T_k=\underline T(1+\varepsilon/3)^k\qquad(k=0,1,\ldots)
\]
in increasing order. For each $k$, use the rounded processing times
$\widetilde t_{ij}=\lfloor K_\varepsilon t_{ij}/T_k\rfloor$ and the following dynamic program.
\begin{enumerate}
\item A state records the rounded machine loads in $\{0,\ldots,K_\varepsilon\}^n$.
Start with the zero vector before processing any job.
\item For the next job $j$, try each machine $i$ and add
$\widetilde t_{ij}$ to coordinate $i$. Retain the new state only if
that coordinate is at most $K_\varepsilon$, and store one predecessor and machine
choice for each reachable state.
\item Accept the guess if a state is reachable after all jobs. At the
first accepted guess, reconstruct the auxiliary assignment from the
stored predecessors.
\end{enumerate}
We prove two properties of this test: every $T_k\ge\OPT$ is accepted,
and every accepted $T_k$ produces a schedule of makespan at most
$T_k(1+\varepsilon/3)$.

\begin{lemma}\label{lem:ownerfptas}
The reconstructed auxiliary schedule has makespan at most $(1+\varepsilon)\OPT$.
For each fixed $n$, the running time is polynomial in the combined binary
encoding length of the reports and $\varepsilon$, and in $1/\varepsilon$.
\end{lemma}
\begin{proof}
\textbf{Feasibility and rounding loss.}
If an assignment $A$ has makespan at most $T_k$, then each rounded load satisfies
\[
 \sum_{j:A(j)=i}\widetilde t_{ij}
 =\sum_{j:A(j)=i}\left\lfloor\frac{K_\varepsilon t_{ij}}{T_k}\right\rfloor
 \le\frac{K_\varepsilon}{T_k}\sum_{j:A(j)=i}t_{ij}\le K_\varepsilon.
\]
Thus every $T_k\ge\OPT$ is accepted. Conversely, let $A^{\mathrm{aux}}$ be
an assignment returned by a successful test. Since
$t_{ij}<(T_k/K_\varepsilon)(\widetilde t_{ij}+1)$, every machine $i$
that receives a job satisfies
\[
 \sum_{j:A^{\mathrm{aux}}(j)=i}t_{ij}
 <\frac{T_k}{K_\varepsilon}\sum_{j:A^{\mathrm{aux}}(j)=i}(\widetilde t_{ij}+1)
 \le\frac{T_k}{K_\varepsilon}(K_\varepsilon+m)\le T_k(1+\varepsilon/3).
\]
An empty machine has load zero. This proves the two test properties.

\paragraph{Approximation guarantee.}
Let $k$ now be the first accepted index. If $k=0$, then
$T_0=\underline T\le\OPT$. If $k>0$, the preceding guess
$T_{k-1}=T_k/(1+\varepsilon/3)$ was rejected, so $T_{k-1}<\OPT$.
In either case,
$T_k\le(1+\varepsilon/3)\OPT$. Combining the search and rounding losses gives
\[
 C_{\max}(t,A^{\mathrm{aux}})
 \le T_k(1+\varepsilon/3)
 \le(1+\varepsilon/3)^2\OPT<(1+\varepsilon)\OPT,
\]
where the last inequality follows from
$2\varepsilon/3+\varepsilon^2/9\le7\varepsilon/9<\varepsilon$.

\paragraph{Bit complexity.}
A guess reaches $m\underline T\ge\OPT$ within
$O(1+\varepsilon^{-1}\log(m+1))$ iterations. At each guess, the dynamic
program has at most $(K_\varepsilon+1)^n$ states per layer and at most $n$ transitions
per state. At the $k$th guess, the numerator and denominator lengths of
$T_k=\underline T(1+\varepsilon/3)^k$ are at most those of $\underline T$ plus $k$ times the
encoding length of $1+\varepsilon/3$. Each rounded cost is computed by
exact integer division. Since $K_\varepsilon=O(m/\varepsilon)$, the bit complexity
is polynomial for fixed $n$ with the stated dependence on $\varepsilon$.
\end{proof}

\subsection{A piecewise rational probability function for two machines}
For the efficient two-machine mechanism we choose a probability
function $f$ that can be evaluated exactly on rational reports.
It is specified by $172$ rational knots and their values,
\[
1/10=r_0<\cdots<r_{171}=1,
\qquad 1=v_0\ge\cdots\ge v_{171}=1/2.
\]
Set $f(r)=1$ below $r_0$. On $[r_j,r_{j+1}]$, interpolate affinely in $1/r$:
\[
 f(r)=\frac{r_j(r_{j+1}-r)}{r(r_{j+1}-r_j)}v_j
      +\frac{r_{j+1}(r-r_j)}{r(r_{j+1}-r_j)}v_{j+1}.
\]
Equivalently, $f(r)=a_j+b_j/r$ on this interval, where
\[
 a_j=\frac{r_{j+1}v_{j+1}-r_jv_j}{r_{j+1}-r_j},\qquad
 b_j=\frac{r_jr_{j+1}(v_j-v_{j+1})}{r_{j+1}-r_j}.
\]
For $r>1$, use $f(r)=1-f(1/r)$. This defines a continuous
nonincreasing reciprocal rule, with rational evaluation at rational
arguments and $f(r)=0$ for $r\ge10$.

\begin{lemma}[Bound for the piecewise rational function]\label{lem:rationalfallback}
The specified rule satisfies
\[
\sup_{0<s\le t}\{1+sf(s)-(1+s-t)f(t)\}
\le\frac{752998007}{500000000}.
\]
\end{lemma}
\begin{proof}
On $0<s\le t\le1$, the expression is at most
\[
1+s-\frac12(1+s-t)=\frac12(1+s+t)\le\frac32.
\]
For $t\ge10$, we have $f(t)=0$, so the objective is $1+sf(s)$.
If $s\le10$, this value already occurs at $(s,10)$, and if $s>10$
it equals one. It remains to consider $0<s\le t\le10$ with $t\ge1$.

On each piece $i$ of $(0,1]$, write $f(x)=a_i+b_i/x$.
For the constant piece below $r_0$, take $a_i=1$ and $b_i=0$.
When $x\ge1$ and $1/x$ lies in piece $i$, reciprocity gives
$f(x)=1-a_i-b_i x$. Substitution in
$\Phi_f(s,t)=1+sf(s)-(1+s-t)f(t)$ gives two cases. For
$s\le1\le t$, take $i$ to be the piece containing $s$ and $j$ the
one containing $1/t$. Then
\[
 \Phi_f(s,t)=b_i+a_j+(a_i+a_j-1)s+(1-a_j+b_j)t+b_jst-b_jt^2.
\]
For $1\le s\le t$, use the pieces $i,j$ containing $1/s,1/t$,
respectively. Now
\[
 \Phi_f(s,t)=a_j+(a_j-a_i)s+(1-a_j+b_j)t-b_is^2+b_jst-b_jt^2.
\]
The first case partitions into rectangles. The second also gives
rectangles when $1/s$ and $1/t$ lie in different pieces, and triangles
cut out by $s\le t$ when they lie in the same piece. The formulas agree
on shared boundaries and extend continuously to $s=0$.

On each closed polygon, the verifier evaluates the quadratic at the
vertices, the stationary points along the edges, and any isolated
stationary point in the interior. If an edge restriction is constant,
its endpoints attain the same value. If an interior maximum lies on
a line of stationary points, that line reaches the boundary with the
same value. Appendix~\ref{supp:rational} gives the rational formulas
for these evaluations.

The largest value found on the polygons exceeds $3/2$. It is attained
at the positive rational pair $s\approx1.10782$, $t\approx2.69110$.
The verifier evaluates $\Phi_f$ again at the exact pair using the knot
values directly. In the two-job profile with columns $(1,s)$ and
$(t,1)$, both off-diagonal entries exceed one, so the diagonal
assignment is optimal with makespan one. This profile attains the
worst-case ratio in~\eqref{n2:scalar-objective}, namely
\[
 R(f)=\frac{6742585099439191488032872130310743621}
           {4477159992806878068875810000000000000}
       <\frac{752998007}{500000000}.\qedhere
\]
\end{proof}

\Needspace{12\baselineskip}
\subsection{Exact sampling of payments}
Archer et al.~\cite[Section~4]{APTT03} estimate a payment integral without
bias by running the allocation rule at a uniformly sampled bid.
For single-parameter agents, Babaioff, Kleinberg, and Slivkins~\cite{BKS10}
give a payment reduction that evaluates a monotone allocation rule once
at randomized bids. Here we keep the sampled assignment and evaluate
the per-job probability at rational reports to draw a nonnegative bonus
whose mean equals the tail integral in~\eqref{eq:local-payments}.

\begin{lemma}[Monotone-integral Bernoulli sampling]\label{lem:paymentbits}
Let $q:[0,\infty)\to[0,1]$ be nonincreasing, vanish above a known rational $B\ge0$, and be evaluable exactly at rational arguments in polynomial bit time. Given a rational report $s\ge0$ and an allocation indicator $X\in\{0,1\}$ with $\Prb(X=1)=q(s)$, one can generate a nonnegative rational payment in expected polynomial bit time whose expectation is $sq(s)+\int_s^\infty q(u)\operatorname{d}u$. The payment is at least $sX$ in every outcome.
\end{lemma}
\begin{proof}
\textbf{An unbiased payment with rational outcomes.}
First pay $sX$, covering the reported cost of the assignment. If
$s\ge B$, the tail integral is zero and this completes the payment.
For $s<B$, the remaining integral is the area under $q$ on $[s,B]$.
Draw $U,V$ independently and uniformly from $[0,1]$, independently
of the assignment. Pay a bonus of $B-s$ when the point
$(s+(B-s)U,V)$ lies under the graph. The total payment is
\begin{equation}\label{eq:randompayment}
\widehat\pi=sX+(B-s)\ind\{V\le q(s+(B-s)U)\}.
\end{equation}
The point is uniform on $[s,B]\times[0,1]$, so the probability of a
bonus is $(B-s)^{-1}\int_s^B q(x)\operatorname{d}x$.
Multiplying by the bonus amount $B-s$ gives the required tail integral.
Thus $\E[\widehat\pi]=sq(s)+\int_s^\infty q(x)\operatorname{d}x$, and the
realized payment belongs to $\{0,s,B-s,B\}$.

\paragraph{Determine the payment event by rational comparisons.}
Reveal $k$ bits of each uniform. With $h=2^{-k}$, let the resulting
intervals be $U\in[u,u+h]$ and $V\in[v,v+h]$. Monotonicity gives
\[
 q(s+(B-s)(u+h))\le q(s+(B-s)U)\le q(s+(B-s)u).
\]
Thus the indicator in~\eqref{eq:randompayment} is one if
$v+h\le q(s+(B-s)(u+h))$, and zero if $v>q(s+(B-s)u)$.
Both comparisons involve rational numbers. If neither condition holds,
reveal another bit of each uniform and repeat. Outside a null set of
boundary events, this procedure returns the indicator's exact value.

\paragraph{Bound the refinement depth.}
In horizontal bin $\ell$ of width $h=2^{-k}$, the graph descends from
$q(s+(B-s)\ell h)$ to $q(s+(B-s)(\ell+1)h)$. Every unresolved
vertical bin meets the interval between these heights. Their union
extends by at most $h$ beyond each endpoint, so its height is at most
the drop in $q$ plus $2h$. Multiplying by the horizontal width and
summing over the bins gives an unresolved area of at most
\[
 h\sum_{\ell=0}^{2^k-1}\bigl(q(s+(B-s)\ell h)-q(s+(B-s)(\ell+1)h)+2h\bigr)
 =h\bigl(q(s)-q(B)\bigr)+2h\le3h.
\]
Since $(U,V)$ is uniform on the unit square, the probability of needing
level $k+1$ is at most $3\cdot2^{-k}$. The dyadic endpoints at that level
use $O(k+1)$ bits, so forming the two rational arguments of $q$ and
evaluating $q$ takes time polynomial in the input length and $k+1$.
Summing these costs against the geometric tail gives expected polynomial
bit time and almost-sure termination, including when $q$ jumps.
\end{proof}

For $q(s)=q_{ij}(s,t_{-ij})$ under the piecewise rational rule on
two machines, take $B=10t_{kj}$,
where $k$ is the other machine. For the Bounded-Square rule on $n\ge2$
machines, take $B=2\min_{k\ne i}t_{kj}$. If the minimum opponent time
is zero, then $B=0$ and $q(x)=0$ for $x>0$. In either rule the value
at zero does not affect the integral. All endpoint evaluations in the
sampler use rational arithmetic.

\subsection{Exact sampling with shared randomness}
\paragraph{Two machines: one uniform seed.}
One uniform $U$ determines all positive-minimum jobs. A job owned by
machine one goes to machine one when $U\le q_{1j}$, and a job owned by
machine two goes there when $U\ge1-q_{1j}$. At rational reports these
thresholds are rational. After $k$ bits, the interval containing $U$
has length $2^{-k}$. Apart from ties of probability zero, a decision
remains open only if a threshold lies inside this interval. Each of
the at most $m$ thresholds lies inside at most one such interval, so
the probability that another bit is needed is at most $\min(1,m2^{-k})$.
Thus $O(\log(m+1))$ refinement levels suffice in expectation, and
the geometric tail gives expected polynomial bit complexity.

\paragraph{Three machines: a simplex direction.}
For three machines, conditional on the global fair coin, there are $d=3$ or $d=4$ independent exponential clocks. Normalize them by their sum. The direction $Y$ is uniform on the standard $(d-1)$-simplex: the change of variables $E_i=zY_i$ gives density $e^{-z}z^{d-1}$, whose integral is constant in $Y$. Equivalently, $Y$ consists of the $d$ spacings formed by $d-1$ sorted independent uniforms on $[0,1]$. Their ordered density is $(d-1)!$ and the spacing map has determinant one in the first $d-1$ coordinates.

The allocation depends only on this direction, since a common scaling
of the clocks changes no winner. We can therefore decide every winner
from binary prefixes of the same uniforms. Reveal $k$ bits of each
uniform and let $h=2^{-k}$. Each order statistic lies between its sorted lower endpoint and that endpoint plus $h$, so each spacing lies in a rational interval of width at most $2h$. Multiply these bounds by the rational squared costs and certify each job's winner when one score's upper bound is smaller than every competing lower bound. Shared prefixes are refined until all jobs are decided.

Consider machines $i,i'$ competing for the same positive-minimum
job $j$. Let $a,b$ be their squared processing times divided by the
larger of the two squares, so $\max(a,b)=1$. Their clock labels
$\ell=\kappa(i,j)$ and $\ell'=\kappa(i',j)$ are distinct, and their
normalized score difference is $aY_\ell-bY_{\ell'}$. An unresolved comparison implies
$|aY_\ell-bY_{\ell'}|\le4h$. For $d=3,4$, the joint density of
these two simplex coordinates is
\[
(d-1)(d-2)(1-u-v)^{d-3},\qquad u,v\ge0,\quad u+v\le1,
\]
which is at most six. Suppose $b=1$; the case $a=1$ follows by
interchanging the coordinates. For every fixed $u$, the allowed interval
of $v$ in the strip $|au-v|\le4h$ has length at most $8h$. Hence
\[
 \Prb\bigl(|aY_\ell-bY_{\ell'}|\le4h\bigr)
 \le6\int_0^1
       \operatorname{length}\{v\in[0,1-u]:|au-v|\le4h\}\operatorname{d}u
 \le48h.
\]
At most three comparisons per job give
\[
\Prb(\text{some job unresolved at level }k)
\le\min(1,144m2^{-k}).
\]
Ties have probability zero. There are $O(\log(m+1))$ expected levels and polynomially bounded moments; each level requires polynomially many rational operations on polynomial-bit numbers. Zero-minimum jobs are sampled uniformly over their zero-cost machines by exact finite rational sampling.

\begin{proof}[Proof of Theorem~\ref{thm:efficient}]
Proposition~\ref{prop:slots-efficient} proves the variable-machine claim
with the exact approximation ratios in Theorem~\ref{thm:generalupper}.
For two and three machines, Lemma~\ref{lem:ownerfptas} gives owners of
makespan $T_{\mathrm{aux}}\le(1+\varepsilon)\OPT$.
Put $R=752998007/500000000$ for $n=2$ and $R=2$ for $n=3$.
Lemma~\ref{lem:ownertransfer}, the two-machine bound in
Lemma~\ref{lem:rationalfallback}, and the three-machine bound in
Theorem~\ref{thm:n3normalized} then give
\[
 \E[C_{\max}]\le R T_{\mathrm{aux}}\le R(1+\varepsilon)\OPT.
\]
The common-seed samplers realize these joint laws. For the payments,
apply Lemma~\ref{lem:paymentbits} to each machine--job pair using
independent payment bits. The resulting expectations satisfy
\eqref{eq:local-payments}, so truthful reporting maximizes expected
utility. At rational bids $b$, $\widehat P_i\ge L_i(b,A)$ in every
outcome, and the total expected bit cost is polynomial in the stated
parameters. The allocation and expected-payment formulas apply to all finite
nonnegative real reports. When $\OPT=0$, every job goes to a zero-cost machine.
The empty instance has the empty assignment.
\end{proof}


\Needspace{5\baselineskip}
\section*{AI methodology}
The main proof was discovered by ChatGPT.
The author checked and revised the proofs and takes
responsibility for the correctness and originality of the final
manuscript, including its references.

\bibliographystyle{alpha}
\bibliography{references}
\appendix
\section{Computer-assisted verification}\label{sec:supplement}
The computer-assisted parts establish three finite sets of inequalities:
the analytic bounds used to determine the two-machine optimum, the
three-machine inequality on the remaining normalized support patterns,
and the approximation guarantee of the piecewise rational two-machine
rule. The reductions from these inequalities to mechanisms, the
truthfulness arguments, and the matching lower bounds are proved in the
main text. In particular, the proofs for $n\ge7$ require none of the
computations described here.

\begin{center}
\begin{tabular}{>{\raggedright\arraybackslash}p{0.32\linewidth}>{\raggedright\arraybackslash}p{0.59\linewidth}}
\toprule
Claim & Finite input and acceptance test\\
\midrule
The analytic value $R_*$ &
Degree-$24$ rational polynomial proposals and endpoint brackets;
contraction, shape, and boundary inequalities from
Theorem~\ref{n2:verification}.\\[4pt]
The three-machine factor $2$ &
Closed subdivision trees and rational dual weights for $75$ roots;
complete coverage and exact acceptance of all $90{,}791$ terminal boxes.\\[4pt]
The upper bound $752998007/500000000$ &
The $172$ rational knots and values of
Lemma~\ref{lem:rationalfallback}. The exact maximum is computed and
attained at a checked pair of ratios.\\
\bottomrule
\end{tabular}
\end{center}

\subsection{Exact arithmetic and certificate verification}\label{supp:arithmetic}
Each computation has two parts: a finite proposed certificate and a
verifier that checks the inequalities needed in the proof. The verifier
uses integer and rational arithmetic for all acceptance decisions.
For example, with positive denominators, $a/b\le c/d$ is checked as
$ad\le bc$. A terminating decimal denotes its exact rational value.
Square roots are enclosed between rational endpoints. For a
nonnegative rational $u/v$, with integers $u\ge0$ and $v\ge1$, choose
an integer denominator $D\ge1$ and compute the integer square root
$k=\lfloor\sqrt{\lfloor uD^2/v\rfloor}\rfloor$. By its definition,
\[
 k^2\le\frac{uD^2}{v}<(k+1)^2,
 \qquad\text{hence}\qquad
 \frac{k}{D}\le\sqrt{\frac uv}<\frac{k+1}{D}.
\]
Thus both endpoints are obtained by integer arithmetic. The lower
endpoint is exact precisely when $k^2v=uD^2$. Division of intervals is
used only after verifying that the denominator interval excludes zero.

The certificate inputs are embedded in the single Python file described
in Section~\ref{supp:reproduction}. They specify the objects being
checked, but their proposed bounds are not assumed. The verifier
reconstructs the interval bounds and probability integrals and checks
complete domain coverage. Numerical search is not part of this
acceptance procedure. Stored success reports and sampled profiles are
not used to establish any of the three conclusions.

\subsection{The two-machine analytic curve}\label{supp:analytic}
Recall the analytic curve
$\gamma_r(z)=(a_r(z),G_r(z),v_r(z))$ from Section~\ref{n2:section}.
For each $r\in [r_-,r_+]$, it satisfies
$\gamma_r(z)=B_r(\gamma_r(\lambda_r z))$ and the initial conditions
\eqref{n2:jet}. The finite input gives a polynomial approximation $Q_r$
valid throughout $[r_-,r_+]$, two more accurate approximations at $r_-,r_+$,
and rational brackets. Each polynomial has three coordinates of degree
at most $24$. Its coefficients of degrees $2$ through $24$ are fixed
rationals; the constant and linear coefficients are prescribed by
\eqref{n2:jet} and enclosed uniformly in $r$. The purpose of the check
is to prove that an exact analytic solution exists near these polynomials
and has the properties required in the main proof.

\Needspace{5\baselineskip}
\paragraph{Existence from a checked residual.}
Write $(\mathcal T_r\gamma)(z)=B_r(\gamma(\lambda_r z))$ and use the
maximum of the three coordinate supremum norms on $|z|\le1/2$.
The verifier bounds the denominators and Jacobian of $B_r$ on the
inputs arising from $\|\gamma-Q_r\|_\infty\le10^{-7}$, with
\eqref{n2:jet} imposed. To bound the residual $\mathcal T_rQ_r-Q_r$,
it forms the full polynomial numerators in
\eqref{n2:residual-numerators}. The triangle inequality gives
\[
 \sup_{|z|\le1/2}\left|\sum_k u_kz^k\right|
 \le\sum_k|u_k|2^{-k}.
\]
Dividing these numerator bounds by positive lower bounds for the
denominator moduli proves
$\|\mathcal T_rQ_r-Q_r\|_\infty<7.756\cdot10^{-10}$.
The Jacobian bound and Schwarz's lemma, as in \eqref{n2:schwarz}, give
\[
 \|\mathcal T_r\gamma-\mathcal T_r\widetilde\gamma\|_\infty
 \le0.703885\|\gamma-\widetilde\gamma\|_\infty.
\]
In particular, every curve in the ball satisfies
\[
 \|\mathcal T_r\gamma-Q_r\|_\infty
 \le0.703885\|\gamma-Q_r\|_\infty+7.756\cdot10^{-10}<10^{-7}.
\]
These two estimates give the analytic solution in the ball.
At $z=0$ the same Jacobian bound excludes a first differing Taylor
coefficient of order $k\ge2$ in any other solution with the prescribed
initial data, as shown in Lemma~\ref{n2:curve-certificate}.

\paragraph{Shape and boundary conditions.}
The verifier first reconstructs the shape coefficients from the
polynomial and the initial conditions~\eqref{n2:jet}. Cauchy's estimate
\eqref{n2:cauchy-error} then turns the uniform analytic error into
bounds on the first two derivatives. The verifier subdivides
$I_z=[-437/5000,211/500]$ into rational intervals, evaluates the polynomial
and derivative enclosures by Horner's rule, and applies the chain-rule
formulas \eqref{n2:coordinate-derivatives}. Every accepted interval must
satisfy the shape inequalities and any low- or high-interval conditions
that meet its enclosed range of $a_r(z)$. The accepted intervals cover
all of $I_z$. The zero $b$ and the points where $a=1$ and $a=9/10$ are
handled by the bracketing and slope tests in the main proof.

The two endpoint polynomials use radius-$10^{-20}$ balls. The verifier
checks that each is contained in the uniform ball, so uniqueness
identifies their fixed points with the same family. For each bracket
$[z_-,z_+]$, it verifies $a_r(z_-)<1<a_r(z_+)$ and encloses
$G_r([z_-,z_+])-1/2$ with the analytic error included. The opposite signs
in \eqref{n2:boundary-signs} establish a boundary root. The mathematical
optimality argument proves uniqueness; exact evaluation of $\Psi$ at
$r_-$ and $r_+$ gives \eqref{n2:decimal-bound}.

For pairs with both ratios at least one, the companion module
\texttt{verify\_high\_high\_constants.py} checks the bounds on
$-F_{xx}$, $-F_{yy}$, and $|F_{xy}|$ in~\eqref{n2:hessian-margin}
using exact fractions.

\subsection{The three-machine inequality}\label{supp:n3replay}
The verifier enumerates the $9^3=729$ labeled support and minimum-order
patterns and checks their reduction to $129$ orbits. The $54$ direct
cases are covered by Section~\ref{n3:analytic}. The other $75$ roots
consist of the two full-support orbits and $73$ mixed-support orbits.
The encoded trees contain $90{,}716$ splits and $90{,}791$ terminal boxes,
for $181{,}507$ records in total. The verifier also checks that relabeling
machines and owners only relabels the clocks of either component map. Before checking the boxes, the replay uses
\texttt{law} in \texttt{three/rational\_cone.py} to reproduce the two
worked profiles in Sections~\ref{n3:allocation} and~\ref{n3:analytic}.
For each map it checks the indicated joint probability and sums the
makespans over all assignments.

\paragraph{A closed cover, reconstructed from the roots.}
At a node, the verifier intersects the inherited coordinate intervals
with the root's ordering and eligibility constraints. For example,
suppose a full column has two variable costs $x\le y\le2x$, with
$x\in[\underline x,\overline x]$ and
$y\in[\underline y,\overline y]$. The inequalities $y/2\le x\le y$
and $x\le y\le2x$ imply, respectively,
\[
 \max\{\underline x,\underline y/2\}\le x\le\min\{\overline x,\overline y\},\qquad \max\{\underline y,\underline x\}\le y\le\min\{\overline y,2\overline x\}.
\]
Updating these bounds preserves every feasible point.
A split at a rational interior point $c$ replaces an interval
$[\underline x,\overline x]$ by $[\underline x,c]$ and
$[c,\overline x]$, whose union is the original interval. The children are reconstructed from the
parent; their recorded identifiers must be distinct and unused.
A terminal record must match its reconstructed box exactly. An empty
node can be discarded only when exact tightening makes its bounds
inconsistent. Acceptance requires every root to be exhausted, with no
unprocessed node. These checks prove coverage, including the common
boundaries of adjacent boxes. As a redundant check, the verifier sums
the exact clipped terminal volumes and compares them with the root
volumes; volume equality alone is not the coverage argument.

\paragraph{A uniform bound on each box.}
For a fixed support and clock map $\kappa$, the verifier computes the
$q_{ij}^-,q_{ij}^+$ of \eqref{eq:n3:13}, the assignment-probability
bounds $y_A^-,y_A^+$ on $y_A^\kappa(t)$, and the cost bounds $C_A^+$.
For a fixed profile, each assignment event is defined by homogeneous
linear inequalities in the clocks, as in \eqref{eq:n3:12}. Relabel the
$d$ active clocks as $E_1,\ldots,E_d$. They are independent
exponentials of rate one, and their normalized vector satisfies
\[
 \left(\frac{E_1}{\sum_{k=1}^dE_k},\ldots,
       \frac{E_d}{\sum_{k=1}^dE_k}\right)
 \quad\hbox{is uniform on}\quad
 \left\{x\in\R_{\ge0}^d:\sum_{k=1}^d x_k=1\right\}.
\]
Indeed, the change of variables $z=\sum_i E_i$ and $x_i=E_i/z$
gives joint density $e^{-z}z^{d-1}$. Integrating over $z>0$ leaves a
constant density in the direction $x$.
For the rational endpoint costs used by the verifier, the assignment
inequalities define a rational polytope in this simplex. Its volume,
divided by the volume of the simplex, is the required probability.
Here $d\le4$. The verifier enumerates feasible
vertices, orders polygon vertices by exact cross products, and computes
areas or triangulated facet volumes by rational determinants. Unused
clocks can be deleted. Degenerate intersections have probability zero.

For each map, the verifier reads each proposed weight as an exact
rational number and checks that it is nonnegative. It then chooses a
nearest rational with denominator at most $10^8$ for each weight.
These numbers are the $\lambda_{ij},\mu_{ij},\alpha_A,\beta_A$ used in
the calculation. Once they are fixed, it computes the free multiplier
\[
 \nu=\max_A\left\{C_A^+
 -\sum_j(\lambda_{A(j),j}-\mu_{A(j),j})-\alpha_A+\beta_A\right\}.
\]
This is the least $\nu$ satisfying~\eqref{eq:n3:16} for every
assignment on the box. Substituting it in~\eqref{eq:n3:17} gives
$U_{\rm cyc}$ and $U_{\rm pair}$ for the two maps. Only terms with
nonzero multipliers require probability integrals. The box passes
when $(U_{\rm cyc}+U_{\rm pair})/2\le2$, the bound for the single
fair coin that selects the map for all jobs. Once every box passes,
Section~\ref{n3:transfer} extends the bound from normalized triples
to arbitrary job counts.

\subsection{The piecewise rational two-machine rule}\label{supp:rational}
The finite input gives the $172$ knots and values in
Lemma~\ref{lem:rationalfallback}. The verifier checks the knot order,
the range and monotonicity of the values, and both endpoints. It
reconstructs $f(x)=a_j+b_j/x$ on each knot interval and uses reciprocity
above one. For $t\ge1$ the two expansions in that lemma give the
quadratic on each rectangle or triangle that needs to be checked.

On each compact polygon, the verifier writes the objective in the original
ratio variables $s,t$:
\[
 Q(s,t)=c_{00}+c_{10}s+c_{01}t+c_{20}s^2+c_{11}st+c_{02}t^2.
\]
The subscripts record the powers of $s$ and $t$. In the Python function
\texttt{value}, the coefficient tuple has exactly this order.
The maximum occurs at a vertex, at a stationary point on an edge, or
at an interior stationary point.

For an edge from $(s_0,t_0)$ to $(s_1,t_1)$, evaluate
$Q(s_0+\tau(s_1-s_0),t_0+\tau(t_1-t_0))$ for $0\le\tau\le1$.
Its derivative is
\[
\begin{aligned}
 &(s_1-s_0)(c_{10}+2c_{20}s_0+c_{11}t_0)
 +(t_1-t_0)(c_{01}+c_{11}s_0+2c_{02}t_0)\\
 &\qquad+2\tau\bigl(c_{20}(s_1-s_0)^2
       +c_{11}(s_1-s_0)(t_1-t_0)+c_{02}(t_1-t_0)^2\bigr).
\end{aligned}
\]
The code calls its constant term \texttt{edge\_linear} and half its
$\tau$ coefficient \texttt{edge\_quadratic}. If the latter is nonzero,
it checks $\tau=-\texttt{edge\_linear}/(2\,\texttt{edge\_quadratic})$
when $0<\tau<1$. Otherwise the restriction is linear or constant,
so its endpoints suffice.

At an interior stationary point, the two gradient equations are
\[
 2c_{20}s+c_{11}t=-c_{10},\qquad c_{11}s+2c_{02}t=-c_{01}.
\]
If $4c_{20}c_{02}-c_{11}^2\ne0$, their unique solution is
\[
 s=\frac{c_{11}c_{01}-2c_{02}c_{10}}{4c_{20}c_{02}-c_{11}^2},\qquad
 t=\frac{c_{11}c_{10}-2c_{20}c_{01}}{4c_{20}c_{02}-c_{11}^2}.
\]
It is checked only when it lies in the polygon. If the determinant is
zero, an interior maximum is constant along a null direction of the
quadratic part reaching the boundary; the edge candidates therefore
suffice. The low--low and unbounded regions have the bounds proved in
Lemma~\ref{lem:rationalfallback}. The verifier makes $178{,}506$
candidate evaluations across $44{,}118$ polygons, counting a shared
boundary on each polygon that contains it. As a separate attainment
check, it interpolates $f$ directly from the knots at the supplied
pair of ratios. Its report gives the pair, the exact value $R(f)$,
and the difference between $752998007/500000000$ and $R(f)$.

\Needspace{7\baselineskip}
\subsection{Code and reproduction}\label{supp:reproduction}
The arXiv ancillary file \texttt{verify\_scheduling.py} contains the
readable modules and all certificate inputs. From its directory, run
the following command with Python~3.11 or newer and assertions enabled.
Only the standard library is required.
\begin{quote}\ttfamily
\verb|python3 verify_scheduling.py --workers 2|
\end{quote}
A full successful run writes \texttt{PASS} for all three checks to
\texttt{verification\_run.json}. The supplied
\texttt{VERIFICATION\_REPORT.json} records the run for this release.
The worker count controls the three-machine calculation. One worker
also suffices. With \texttt{--only rational}, for example, the program
checks just the piecewise rational rule and reports \texttt{PARTIAL\_PASS}
on success.

To inspect the modules and inputs, extract them into an empty directory:
\begin{quote}\ttfamily
\verb|python3 verify_scheduling.py --extract-only readable_source|
\end{quote}
The accompanying \texttt{NOTATION\_MAP.txt} relates the Python names
to the formulas. For example, \texttt{assignment[j]} is the winner
$A(j)$, while the owner on a normalized triple is
$A^{\mathrm{aux}}(j)=j$. The repository is also available at
\begin{quote}\small
\url{https://github.com/liyunpeng1991/truthful-scheduling-verifier}.
\end{quote}

\end{document}